\documentclass[aps,pre,twocolumn,superscriptaddress,showpacs,showkeys,amsmath,amssymb,floatfix
]{revtex4-2}

\usepackage{epsfig,amsmath,amssymb,xcolor,bm}
\usepackage{physics}
\usepackage[T1]{fontenc}
\usepackage[colorlinks = true,linkcolor = red,citecolor = magenta]{hyperref}
\usepackage[sort&compress]{natbib}
\usepackage{orcidlink}
\usepackage[normalem]{ulem}

\begin{document}

\title{Flat bands in the $J_1$-$J_2$-$J_3$ XXZ sawtooth chain}

\author{Vadim Ohanyan\,\orcidlink{0000-0002-7810-7321}}
  \affiliation{Laboratory of Theoretical Physics
          Yerevan State University,
         1 Alex Manoogian Str., 0025 Yerevan, Armenia}
  \affiliation{CANDLE, Synchrotron Research Institute, 31 Acharyan Str., 0040 Yerevan, Armenia}

\author{Johannes Richter\,\orcidlink{0000-0002-5630-3786}}
  \affiliation{Institut f\"{u}r Physik, Universit\"{a}t Magdeburg, P.O. Box 4120, D-39016 Magdeburg, Germany}
  \affiliation{Max-Planck-Institut f\"{u}r Physik Komplexer Systeme, N\"{o}thnitzer Stra{\ss}e 38, D-01187 Dresden, Germany}

\author{Michael Sekania\,\orcidlink{0000-0002-4270-0238}}
  \affiliation{Rechenzentrum, University of Augsburg, 86135 Augsburg, Germany}
  \affiliation{Andronikashvili Institute of Physics, Javakhishvili Tbilisi State University, Tamarashvili str. 6, 0177 Tbilisi, Georgia}

\author{Lucas Giambattista\,\orcidlink{0009-0004-5595-1193}}
  \affiliation{Theoretical Physics III, Center for Electronic Correlations and Magnetism, Institute of Physics, University of Augsburg, 86135 Augsburg, Germany}

\author{Alexei Andreanov\,\orcidlink{0000-0002-3033-0452}}
  \affiliation{Center for Theoretical Physics of Complex Systems, Institute for Basic Science, Daejeon 34126, Republic of Korea}
  \affiliation{Basic Science Program, Korea University of Science and Technology, Daejeon 34113, Korea}

\author{Marcus Kollar\,\orcidlink{0000-0003-2118-9490}}
  \affiliation{Theoretical Physics III, Center for Electronic Correlations and Magnetism, Institute of Physics, University of Augsburg, 86135 Augsburg, Germany}

\date{\today}



\begin{abstract}
  We consider a generalization of the XXZ model on the sawtooth spin chain with Dzyaloshinskii-Moriya interactions in which all exchange constants (symmetric, antisymmetric, and axial anisotropy) are different for the three different bonds of each triangle.
  We derive and resolve algebraic constraints on the exchange constants ensuring the appearance of a flat band in the one-magnon spectrum.
  The properties of the corresponding flat magnon bands and localized magnon states are analyzed.
  We further construct the mapping of the flat-band conditions for the Dzyaloshinskii-Moriya constants onto the Katsura-Nagaosa-Balatsky parameters.
  Based on the mapping, the possibility of the electric-field-driven flat bands with the aid of the magnetoelectric coupling is examined.
\end{abstract}

\maketitle

\section{Introduction}
\label{sec1}

Frustrated quantum magnetism has been the subject of intensive research over the past several decades.
Frustrated spin systems exhibit a wide range of unusual and intriguing phenomena~\cite{fru1}.
One of the most notable and at the same time simplest geometrically frustrated spin models is the one-dimensional chain of corner-sharing triangles, commonly referred to as the sawtooth chain.
The sawtooth chain is well known for its remarkable properties, such as an exact twofold degenerate dimerized ground states and quantum spin-\(1/2\) kink-antikink excitations in the case of uniform antiferromagnetic coupling~\cite{kubo93, nak96, sen96, hao11}.
The valence-bond order of the dimerized ground states, however, can be destroyed by a uniform out-of-plane Dzyaloshinskii-Moriya (DM) interaction.
It was argued in Ref.~\onlinecite{hao11} that there is a critical value of the DM interaction at which the spin gap closes and the ground state becomes a Luttinger liquid.
Another important and straightforward modification is a sawtooth chain with non-uniform couplings, one for the basal line, \(J_1\), and another one for the zigzag part of the chain, \(J_2\).
The ground states and excitations of this \(J_1\)-\(J_2\) sawtooth chain depend on the ratio \(J_1/J_2\) and have a rich phase diagram,
i.e., a zero-field ground state can exhibit quasi-N{\'e}el long-range order, dimerized valence bond order or more complicated spiral configurations~\cite{blu03,ton04,kab05,jia15,paul19, FAF1, FAF1a, FAF2, FAF3, FAF_field, rau25, schnack23a}.

\begin{figure}[tb]
\begin{center}
  \includegraphics[width=75.5mm]{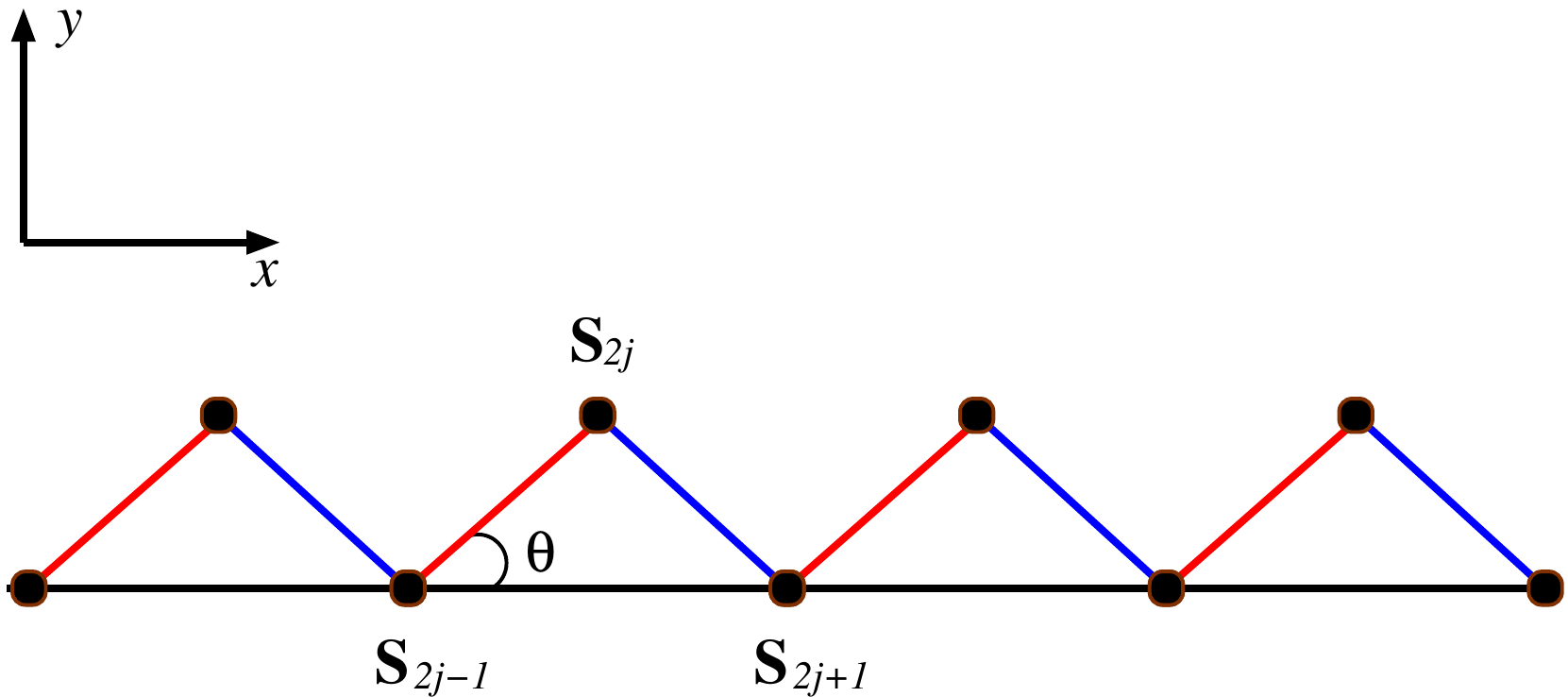}
  \caption{
    (Color online) Sawtooth chain with different couplings on each bond of the triangle.
    The filled circles show the lattice sites occupied by spins.
    The exchange coupling, the XXZ anisotropy and DM interaction along the basal line (black) are \(J_1\), \(\Delta_1\) and \(D_1\), respectively.
    Left (red) and right (blue) bonds along the zigzag line feature the parameters $J_2$, $\Delta_2$ and $D_2$ (red) and $J_3$, $\Delta_3$ and $D_3$ (blue), respectively.
  }
  \label{fig1}
\end{center}
\end{figure}

Chains with a sawtooth arrangement of sites are also among the simplest examples of exactly solvable classical spin models containing finite clusters of quantum spins~\cite{oha09, bel10,bel13}.
A particularly important feature arises in the \(J_1\)-\(J_2\) sawtooth model under certain fine-tuning conditions.
One branch of the one-magnon excitations then becomes flat and localized magnon states appear in the spectrum~\cite{schul02, rich04, rich05, zhi04,der04,der06,der07,rich08,der15,met20, der20, CMP20, ace20, SC_KNB, schnack23a, schnack23b, schnack24, schnack25}.
The flat bands for magnon excitations have been found in various lattice spin models in \(d=1,2,3\) dimensions~\cite{der06,der07}.
A number of unconventional phenomena directly linked to the flat magnon band emerge in magnetic fields, such as a magnetization jump at the saturation field~\cite{schul02,rich04,rich05,zhi04, der15}, magnon crystals in \(d=2\)~\cite{zhi04,zhi05,prl2020},
a magnetic-field-driven spin-Peierls instability~\cite{spin-peierls}, a finite residual entropy at the saturation field~\cite{zhi04,der06,rich04,zhi05,der04},
anomalously enhanced magnetocaloric~\cite{der06,zhi05,zhi04h} and electrocaloric~\cite{SC_KNB} effects and an additional low-temperature maximum of the specific heat signaling the appearance of an additional low-energy scale~\cite{der06,rich04}.

Localized magnon states appear in the spectrum of the \(J_1\)-\(J_2\) sawtooth chain at certain exact relations between the exchange couplings~\cite{schul02}.
Interestingly, both the purely antiferromagnetic (AFM) sawtooth chain (\(J_1,J_2>0\)) as well as mixed AFM \(J_1>0\) and ferromagnetic (F) \(J_2<0\) sawtooth chains admit the localized magnons in their spectrum.
For the AFM sawtooth chain, $J_1,J_2>0$, the lowest band of one-magnon excitations above the fully polarized state becomes dispersionless at the flat-band point $J_2=2J_1$~\cite{schul02, zhi04, der15, der06, rich04, der04, met20, der20}.
This flat magnon band corresponds to multimagnon (flat-band) states in the magnetization sectors $N/4 \leq S^z < N/2$, which constitute the ground states within these sectors.
At $S^z = N/4$, corresponding to half of the saturation magnetization, the system exhibits a wide magnetization plateau, with the ground state forming a magnon-crystal configuration~\cite{schul02}.
All flat-band states are linearly independent~\cite{sch06} and their number grows exponentially with the system size $N$, leading to an extensive degeneracy.
These states dominate the low-temperature thermodynamics in magnetic fields close to the saturation~\cite{schul02, zhi04, der15, der06, rich04, der04, met20, der20}.
A second flat-band scenario arises in the sawtooth Heisenberg chain with ferromagnetic exchange between apical and basal spins ($J_2 < 0$) and antiferromagnetic coupling along the basal line ($J_1 > 0$)~\cite{FAF1, FAF3, FAF_field, ton04, kab05}.
In this case, the lowest one-magnon band becomes dispersionless at $J_2=-2J_1$~\cite{FAF1}.
Notably, the localized states exist here in all magnetization sectors $0 \leq S^z < N/2$, and again they are the lowest states in the
corresponding magnetization sectors.
A striking difference to the AFM case is that flat-band physics takes place at zero magnetic field.

The main challenge for the experimental realization of the flat-band physics in magnetic materials comes from the constraints imposed on the parameters of the Heisenberg Hamiltonian.
Although a number of minerals and synthesized compounds possess magnetic lattices which can be approximated by a sawtooth chain, none of them is close to the exact flat-band condition $J_2 = \pm 2 J_1$.
This makes experimental observation of any physical effect related to localized magnons challenging.
For example, a half-magnetization plateau has been reported in the mineral atacamite, Cu$_2$Cl(OH)$_3$; however it was mainly attributed to the weak interchain coupling~\cite{ata21, ata25, ata25b}.
Further examples of magnetic compounds with sawtooth chain geometry of exchange bonds are the euchroite Cu$_2$(AsO$_4$)(OH)$\cdot$3H$_2$O~\cite{euch},
sawtooth Fe$^{3+}$ chains in Rb$_2$Fe$_2$O(AsO$_4$)$_2$~\cite{Fe3+} and Fe$_2$O(SeO$_3$)$_2$~\cite{Fe3+2},
a family of olivine-type compounds Mn$_2$SiS$_{4-x}$Se$_x$ ($x=0\ldots4$)~\cite{oliv1} and Fe$_2$SiSe$_4$~\cite{oliv2}, the chromium compound BeCr$_2$O$_4$~\cite{spir}.
The last two compounds feature spiral spin textures.
Molecular magnets constitute another class of magnetic materials with sawtooth chain geometry, such as the spin ring Mo$_{75}$V$_{20}$~\cite{MoV} and the mixed-spin Fe$_{10}$Gd$_{10}$ molecule~\cite{FeGd}.

Recently an approach based on electric-field-driven flat bands was proposed to overcome the fine-tuning constraint in sawtooth chains~\cite{SC_KNB}.
The central result of Ref.~\cite{SC_KNB} is that for the \(J_1\)-\(J_2\) Heisenberg sawtooth chain,
the Katsura-Nagaosa-Balatsky (KNB) mechanism of spin-driven ferroelectricity~\cite{KNB1, KNB2, Sol21, Sol25} turns flat the lower band of the one-magnon spectrum by fine-tuning the electric field magnitude to
\begin{gather}
  \label{eq:E_FB}
  E_{\mathrm{FB}} = \pm\frac{\sqrt{4J_1^2-J_2^2}}{\sin\theta},
\end{gather}
provided the electric field is parallel to the basal line of the sawtooth chain.
Here \(\theta\) is the angle between basal-basal and basal-apical bonds (see Fig.~\ref{fig1}).
Thus, for any values of the exchange coupling satisfying \(-2J_1\leq J_2 \leq 2J_1\) the hard flat-band constraint \(J_2=\pm 2J_1\) is replaced by the appropriate tuning of the electric field.

Magnetoelectrics are a class of multiferroic materials~\cite{khom} which can exhibit simultaneous magnetic and dielectric order~\cite{MEE1, MEE2, Tok10, Tok14}.
The most straightforward manifestation of the magnetoelectric effect in such  materials is the explicit dependence of the magnetization (dielectric polarization) on the electric (magnetic) field.
The microscopic link between spin texture in lattice spin models and local polarization, as proposed by the KNB theory~\cite{KNB1, KNB2, Sol21, Sol25}, is the emergence of dielectric polarization for non-collinear configuration of spins at the edges of a lattice bond:
\begin{gather}
  \label{eq:KNB}
  \bm{P}_{ij} = \gamma_{ij} \bm{e}_{ij}\times\bm{S}_i\times\bm{S}_j.
\end{gather}
Here \(\bm{e}_{ij}\) is the unit vector pointing from site \(i\) to site \(j\), and \(\gamma_{ij}\) is a material-dependent constant, ensuring that the KNB mechanism expresses the magnetoelectric polarization in terms of spin-operators.
While this expression strongly depends on the topology of lattice bonds, for a simple linear chain parallel to \(x\)-axis with uniform bonds, the components of the polarization are given by~\cite{bro13}
\begin{align}
  \label{eq:KNB_lin}
  P^x & = 0, \\
  P^y & = \gamma\sum_{j=1}^N\left(S_J^yS_{j+1}^x-S_{j}^x S_{j+1}^y\right) \notag, \\
  P^z & = \gamma\sum_{j=1}^N\left(S_J^zS_{j+1}^x-S_{j}^x S_{j+1}^z\right) \notag.
\end{align}
Thus, for an electric field in the $(x,y)$ plane, the spin Hamiltonian acquires an additional term, describing interaction between the KNB polarization and the field,
\begin{gather}
  \label{eq:KNB_Ham_lin}
  -\bm{E}\cdot\bm{P} = E_y\sum_{j=1}^N\left(S_J^yS_{j+1}^x-S_{j}^x S_{j+1}^y\right),
\end{gather}
where the coefficients $\gamma_{ij}$ are absorbed into the definitions of the electric-field components.  Consequently, hereafter $E_x$ and $E_y$ are expressed in energy units. However, when the same chain is physically folded in the \((x, y)\) plane into a regular zigzag geometry, the expression for the polarization becomes more complicated, demonstrating a staggered structure~\cite{baran18}:
\begin{align}
  \label{eq:KNB_Ham_zigzag}
  -\bm{E}\cdot\bm{P} = & \sum_{j=1}^N\left(E_y\cos\theta+({-}1)^j E_x\sin\theta\right)\times \notag \\
  & \left(S_{j}^xS_{j+1}^y {-} S_{j}^yS_{j+1}^x\right).
\end{align}
Several theoretical studies on various aspects of one-dimensional magnetoelectric spin chains with KNB mechanism were published~\cite{SC_KNB, bro13, thakur18, XYZ, oha20, mench15, sznajd18, sznajd19, oles, stre20, cabra19, baran18, bar21}.
The arrangement of sites in a sawtooth chain includes both cases of the KNB polarization, Eqs.~\eqref{eq:KNB_Ham_lin} and~\eqref{eq:KNB_Ham_zigzag}.
The linear-chain KNB polarization corresponds to the basal line of the sawtooth chain, while  the zigzag part features the staggered structure.
Thus the emergent model considered in Ref.~\onlinecite{SC_KNB} is a $J_1$-$J_2$ sawtooth chain with an additional DM interaction, which is different for each side of triangle.
The values of the DM interaction constants are determined in turn by the external electric field magnitude $E$ as well as its angle with the basal line $\phi$:
\begin{align}
  \label{eq:DDD}
  D_{2j-1, 2j+1}&\equiv D_1=aE\sin\phi,\\
  D_{2j-1, 2j}&\equiv D_2=E\sin\left(\phi-\theta\right), \notag \\
  D_{2j, 2j+1}&\equiv D_3=E\sin\left(\phi+\theta\right). \notag
\end{align}
Here \(\theta\) is the same angle as in Eq.~\eqref{eq:E_FB} and the constant $a$ reflects the possible quantum-chemical non-equivalence between basal and lateral bonds~\cite{SC_KNB}.
In Ref.~\cite{SC_KNB} two special values of the angle between the electric field and the lattice bonds, $\phi=0$ and $\phi=\theta$, were considered.
As mentioned above, the most remarkable result is the possibility of driving sawtooth chain into the flat-band regime by tuning the magnitude of the electric field to the value given in Eq.~\eqref{eq:E_FB} without any additional constraints.
The case of \(\phi=\theta\), however, requires an additional relation between \(J_1\) and \(J_2\).

In this paper we present a detailed analysis of the flat-band scenarios for the generalized model of the sawtooth chain with three exchange couplings, \(J_1, J_2, J_3\)~\cite{J1J2J3, euch}, one for each side of the triangle,
and the three DM parameters, \(D_1, D_2, D_3\) which are in general not given by the KNB formula~\eqref{eq:DDD}.
For each bond of the triangle its own axial anisotropy $\Delta_a$, $a=1,2,3$ is also added.
We derive the flat-band constraints, which turn out to be given by two rational functions of $J_1, J_2, J_3, D_1, D_2, D_3$ and $\Delta_1$.
Interestingly, the coefficients of axial anisotropy corresponding to the zigzag part of the chain, \(\Delta_2, \Delta_3\), do not enter the flat-band constraints.
There are many ways of handling the solutions of the flat-band conditions, as they contain seven parameters.
Based on the mathematical structure of the solutions we consider several distinct classes:
\begin{itemize}
  \item Solutions expressing a pair of symmetric exchange constants and corresponding DM parameters from the left or right side of the triangle in terms of other parameters, $J_{3,2}=J_{3,2}\left(J_1, J_{2,3}, \Delta_1, D_1, D_2, D_3\right)$ and $D_{3,2}=D_{3,2}\left(J_1, J_{2,3}, \Delta_1, D_1, D_2, D_3\right)$.
  \item Solutions for the \(J_1\)-\(J_2\)-\(J_3\) sawtooth chain without the DM interaction.
  \item Solutions expressing the DM constants in terms of symmetric exchange parameters and axial anisotropy \(\Delta_1\), \(D_a=D_a\left(J_1, J_2, J_3, \Delta_1 \right)\) with additional constraint for the symmetric couplings \(J_1, J_2, J_3\).
  \item Solutions expressing the DM constants in terms of symmetric exchange parameters and axial anisotropy \(\Delta_1\), \(D_a=D_a\left(J_1, J_2, J_3, \Delta_1 \right)\), \(a=1, 2, 3\), without additional constraints.
  A special case is a one-parametric solution for the uniform couplings (\(J_1=J_2=J_3=1\)).
\end{itemize}
We also analyze the compatibility of the solutions with the KNB mechanism, and also present a generalization of the results obtained in Ref.~\onlinecite{SC_KNB} to the case of the XXZ sawtooth chain with three couplings \(J_1, J_2, J_3\).

The paper is organized as follows.
Section~\ref{sec:2} introduces the model.
In Section~\ref{sec:3}, we derive the flat-band constraints from the one-magnon spectrum and resolve them, analyze the general properties of flat bands and construct localized states of magnons.
In Section~\ref{sec:4}, we connect the KNB mechanism to given DM parameters.
In the next Section~\ref{sec:5} we interpret results in terms of the KNB mechanism.
Section~\ref{sec:6} provides a linear spin-wave theory treatment of the model.
We conclude with a summary of the main results in the final section.

\section{Model Hamiltonian for a general sawtooth Heisenberg chain}
\label{sec:2}

We focus on the frustrated quantum \(J_1\)-\(J_2\)-\(J_3\) XXZ sawtooth chain with arbitrary spin-\(S\) and three DM couplings, i.e., all exchange couplings can be different,
namely for the basal line ($J_1$ and $D_1$) and for the left ($J_2$ and $D_2$) and right ($J_3$ and $D_3$) bonds of the zigzag part, see Fig.~\ref{fig1}.
Furthermore we suppose that  all DM vectors point along the $z$-axis.
This gives the following general Hamiltonian:
\begin{widetext}
\begin{align}
  \label{ham_gen}
  \mathcal{H} &= \sum_{j=1}^{N/2}\left(\frac{J_1+iD_1}{2}S_{2j-1}^{+}S_{2j+1}^{-}+\frac{J_1-iD_1}{2}S_{2j-1}^{-}S_{2j+1}^{+}+J_1\Delta_1 S_{2j-1}^{z}S_{2j+1}^{z}\right)\\
  &+ \sum_{j=1}^{N/2}\left(\frac{J_2+iD_2}{2}S_{2j-1}^{+}S_{2j}^{-}+\frac{J_2-iD_2}{2}S_{2j-1}^{-}S_{2j}^{+}+J_2\Delta_2 S_{2j-1}^{z}S_{2j}^{z}\right)\notag\\
  &+ \sum_{j=1}^{N/2}\left(\frac{J_3+iD_3}{2}S_{2j}^{+}S_{2j+1}^{-}+\frac{J_3-iD_3}{2}S_{2j}^{-}S_{2j+1}^{+}+J_3\Delta_3 S_{2j}^{z}S_{2j+1}^{z}\right)-B\sum_{j=1}^N S_j^z, \notag
\end{align}
\end{widetext}
where $S_j^{\alpha}$, $\alpha=x, y, z$ and $S_j^{\pm}=S_j^x \pm i S_j^y$ stand for the standard $SU(2)$ generators with $[S_j^{\alpha},\; S_{j^{\prime}}^{\beta}]=\delta_{jj^{\prime}}\epsilon^{\alpha\beta\gamma} S_{j}^{\gamma}$.
We consider generic values for the DM parameters and do not assume that the parameters of DM interaction to be generated by the KNB mechanism.
However, later we discuss the feasibility of electric-field induced flat bands for some types of flat-band solutions, generalizing the results of Ref.~\onlinecite{SC_KNB}.

\section{Flat bands and localized magnon states}
\label{sec:3}

The sawtooth chain with two couplings, one for the basal line and another one for the zigzag bonds, is well known as a paradigmatic example of a frustrated spin model exhibiting localized magnons or flat magnon bands~\cite{schul02,zhi04,der04,der06,der07,rich08,FAF1,der15,met20,der20,CMP20,ace20,SC_KNB,schnack23a,schnack23b, schnack24}.
For the chain with arbitrary spin-\(S\) the flat-band constraint has the following form in Ref.~\onlinecite{schul02}:
\begin{gather}
  \label{FB12}
  J_2=\pm\sqrt{2\left(1+\Delta_1\right)}J_1.
\end{gather}
However, in previous works only three particular cases have been identified and studied in detail,
\begin{align}
  \label{FB_old}
  J_2 &= \pm 2 J_1, \Delta_1=1 \\
  J_2 &= J_1, \Delta_1=-1/2. \notag
\end{align}
All these flat-band scenarios share multiple features:
a magnetization plateau at half-magnetization caused by condensation of localized magnons, a sharp jump to saturation, large degeneracy at the quantum critical point, etc.
Furthermore the cases $J_2=-2J_1>0$, $\Delta_1=1$~\cite{FAF1} and $J_2=J_1>0$, $\Delta_1=-1/2$~\cite{schul02,der06} are remarkable for an extended version of the localized magnon ground state manifold,
as they admit also special overlapping localized magnons configurations~\cite{der20}.
A central aspect of localized magnon physics involves the constraints which the exchange couplings must satisfy in order to develop flat magnon bands.
In this section, we will derive and solve the flat-band constraints for the general \(J_1\)-\(J_2\)-\(J_3\) sawtooth chain with DM terms for which the DM vectors point in the $z$-direction.
In this general case the model has seven adjustable parameters, $J_1, J_2, J_3, D_1, D_2, D_3$ and $\Delta_1$, providing broad opportunities for flat-band scenarios.
The most direct way to derive flat-band constraints is by analyzing the spectrum of one-magnon states around the fully polarized high-field state \(\ket{0}=\ket{\uparrow\uparrow\ldots\uparrow}\) with energy $E_{\mathrm{FM}}=\frac{N}{2}S^2\sum_{a=1}^3 J_a \Delta_a-NSB$.
Assuming periodic boundary conditions for the spins, $S_{N+1}^{\alpha}=S_1^{\alpha}$ (where $N$ is even), one can construct low-lying excitations around the polarized reference state, i.e., one-magnon states.
Since the chain has two sites per unit cell, the one-magnon eigenstates are given by
\begin{gather}
  \ket{1_k}=\sum_{l=0,1}a_l\sum_{j=1}^{N/2}e^{ijk}S_{2j+l}^{-}\ket{0},
\end{gather}
where $k$ is the quasi-momentum of the one-magnon excitation taking $N/2$ values between $-\pi$ and $\pi$.
For the generalized model given by the Hamiltonian~\eqref{ham_gen} the two branches of the one-magnon spectrum are:
\begin{widetext}
\begin{gather}
  \label{1m_sp_gen}
  \varepsilon^{\pm}_1(k)=B-\Gamma+S\left[\rho_1\cos(k-\phi_1)\pm\sqrt{\left(J_1\Delta_1-\rho_1\cos(k-\phi_1)\right)^2+2\rho_2\rho_3\cos(k-\phi_2-\phi_3)+\rho_2^2+\rho_3^2}\right],
\end{gather}
\end{widetext}
where
\begin{align}
  \label{eq:notations}
  && J_a+i D_a =\rho_a e^{i \phi_a},\; \rho_a=\sqrt{J_a^2+D_a^2}, \\
  && \phi_a=\arctan\frac{D_a}{J_a}, \qquad \Gamma=S \sum_{a=1}^3J_a\Delta_a. \notag
\end{align}
For a flat band to emerge, the cosine terms under the square root in Eq.~\eqref{1m_sp_gen} must have the same $k$ dependence, i.e., may differ at most by a sign $\sigma=\pm 1$.
This implies that $\phi_1$ and $\phi_2+\phi_3$ must either be equal (for $\sigma=+1$) or differ by $\pi$  (for $\sigma=-1$).
Next the square root in~\eqref{1m_sp_gen} must simplify to $|\rho_1\cos(k-\phi_1)|$.
We then arrive at the following conditions:
\begin{align}
  \label{FB_pr}
  \phi_1 &= \phi_2+\phi_3\pm \pi \delta_{\sigma, -1}, \\
  \rho_1^2 & \left(\rho_2^2+\rho_3^2\right)+2\sigma J_1\Delta_1\rho_1\rho_2\rho_3-\rho_2^2\rho_3^2=0, \notag \\
  \sigma &= \left\{-1, 1\right\}. \notag
\end{align}
These constraints generalize the flat-band condition~\eqref{FB12}.
Interestingly, neither $\Delta_2$ nor $\Delta_3$ appear in the above.
Also, Eqs.~\eqref{FB_pr} are symmetric under the interchange $\rho_2\longleftrightarrow \rho_3$.
Since the constraints are independent of the spin value, $S$, we set $S=1/2$ without loss of generality.
The one-magnon spectrum (\ref{1m_sp_gen}), after substitution of the flat-band constraints in Eq.~\eqref{FB_pr}, takes the following form:
\begin{align}
  \label{eq:one-m_spec}
  \varepsilon^{+}_1(k) &= B - E_0 + \rho_1\cos\left(k-\phi_1\right), \\
  \varepsilon^{-}_1(k) &= B - B_0, \notag
\end{align}
where
\begin{align}
  E_0 &= \frac{1}{2} \left(2J_1\Delta_1+J_2\Delta_2+J_3\Delta_3-\frac{\sigma \rho_2 \rho_3}{\rho_1}\right), \notag \\
  B_0 &= \frac{1}{2} \left(J_2\Delta_2+J_3\Delta_3+\frac{\sigma \rho_2 \rho_3}{\rho_1}\right).
\end{align}
The branch corresponding to $\varepsilon^{-}_1(k)$ is flat.
However, to observe all features of the localized magnons at zero temperature, the flat branch of the one-magnon spectrum must always be below the dispersive one.
Therefore, an additional condition (see Appendix~\ref{app:derivation}) should be added to the flat band constraints~\eqref{FB_pr},
\begin{align}
  \label{sigma}
  \sigma & =\text{sign}\left(1+\frac{D_1D_2}{J_1J_2}\right)\equiv \text{sign}\left(1+\frac{D_1D_3}{J_1J_3}\right) \notag \\
  & \equiv \text{sign}\left(1-\frac{D_2D_3}{J_2J_3}\right)=1,
\end{align}
in order for the flat band to be the lower one.

The dispersionless magnonic band has localized excitations, and as is known from previous works, localized magnons in the sawtooth chain are superpositions of flipped spins on three neighbouring sites between two teeth, labeled by $2j, 2j+1$ and $2j+2$~\cite{der20}:
\begin{gather}
  \label{loc_st}
  \ket{1_j} = a\ket{2j} + b\ket{2j+1} + c\ket{2j+2}
\end{gather}
where $\ket{j} = S_j^-\ket{0}$.
In our case the coefficients of the superposition can be chosen in the following way:
\begin{gather}
  \label{lm_st_gen}
  \ket{1_j} = \frac{\rho_1}{\rho_2} e^{i\left( \phi_2-\phi_1\right)} \ket{2j} - \ket{2j+1} + \frac{\rho_1}{\rho_3} e^{i\left( \phi_1-\phi_3\right)} \ket{2j+2}, \notag \\
\end{gather}
provided all $\rho_a$ and $\phi_a$ satisfy the relations~\eqref{FB_pr}.

\subsection{Solution with respect to $\rho_a$}

\begin{figure}[tb]
\begin{center}
  \includegraphics[width=75.5mm]{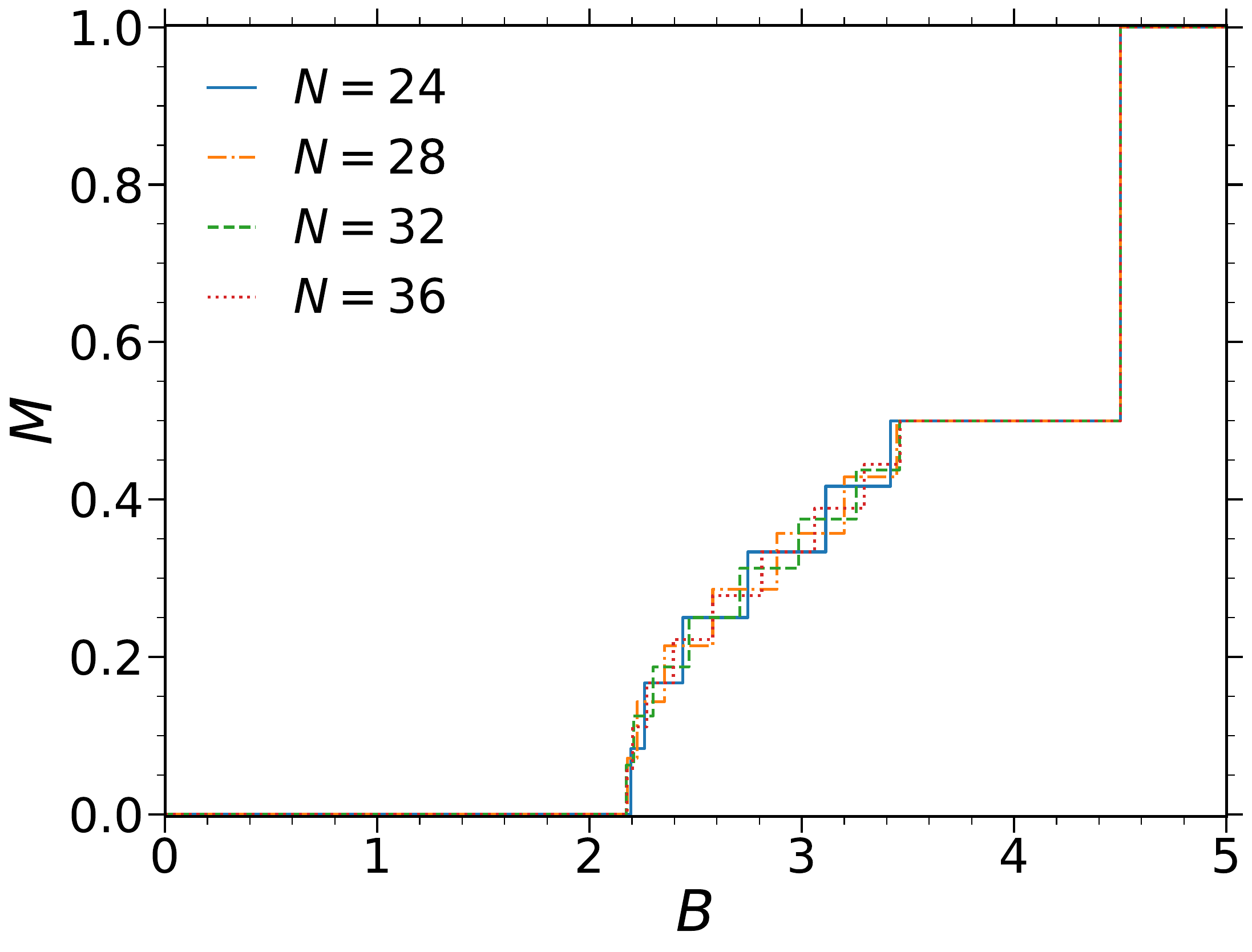}
  \caption{
    (Color online) Zero-temperature exact diagonalization magnetization plots for $J_1=1, J_2=1, J_3=3, D_1=1, D_2=2, D_3=-1$ and for $\Delta_1=\Delta_2=\Delta_3=1$.
  }
  \label{fig2}
\end{center}
\end{figure}

\begin{figure}[tb]
\begin{center}
  \includegraphics[width=75.5mm]{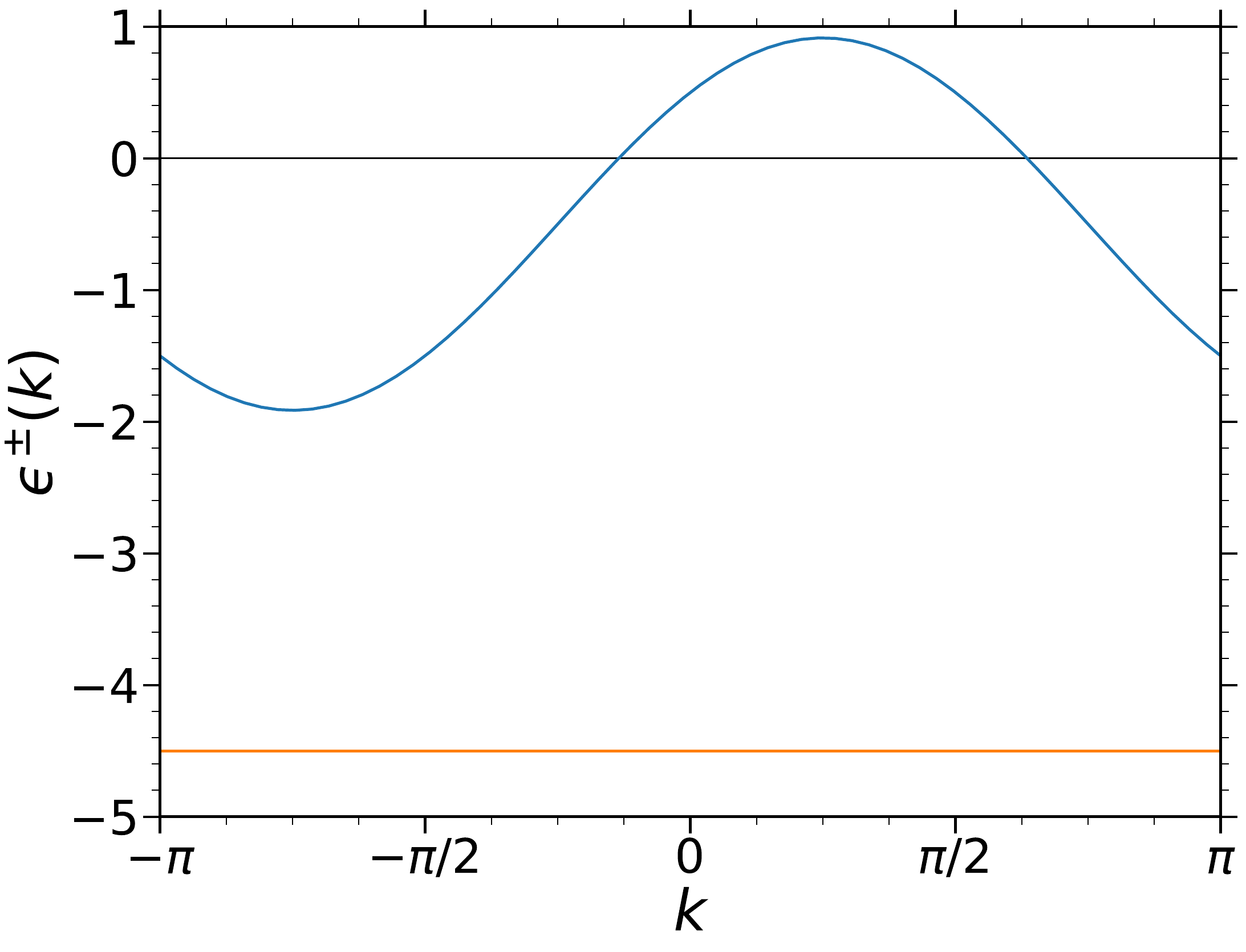}
  \caption{
    (Color online) The one-magnon spectrum for $J_1=1, J_2=1, J_3=3, D_1=1, D_2=2, D_3=-1$ and for $\Delta_1=\Delta_2=\Delta_3=1$ with the lower flat band.
  }
  \label{fig3}
\end{center}
\end{figure}

The equations~\eqref{FB_pr} can be directly solved with respect to $\rho_2$ ($\rho_3$), provided $\rho_3^2 \geq \rho_1^2-J_1^2 \Delta_1^2$ ($\rho_2^2 \geq \rho_1^2-J_1^2 \Delta_1^2$) and $\rho_2\neq\rho_1$ ($\rho_3\neq\rho_1$).
As the flat-band constraints are symmetric with respect to permutations of $\rho_2$ and $\rho_3$, we present here only one of the cases.
Taking into account the condition $\sigma=1$, the solution for the positive quantity $\rho_3$ is given by:
\begin{gather}
  \label{rho3}
  \rho_3=\frac{\pm\rho_1\rho_2}{\sqrt{J_1^2\Delta_1^2-\rho_1^2+\rho_2^2}\mp  J_1\Delta_1}.
\end{gather}
Then the localized states take the following form:
\begin{align}
  \label{Lk}
  \ket{1_j} &= \rho_1 e^{-i\phi_1} \ket{2j} - \rho_2 e^{-i\phi_2} \ket{2j+1} \mp \rho^{\pm} \ket{2j+2}, \notag \\
  \rho^{\pm} &= \sqrt{J_1^2\Delta_1^2+\rho_2^2-\rho_1^2} \mp J_1\Delta_1,
\end{align}
It is then straightforward to use the flat-band constraint to express $J_3$ and $D_3$ in terms of the other parameters.
Multiplying both sides of the relation~\eqref{rho3} by $e^{i\phi_3}= e^{i\left(\phi_1-\phi_2\right)}$ we get two relations:
\begin{align}
  \label{J3D3}
  && J_3=\frac{\pm\left(J_1J_2+D_1D_2\right)}{\sqrt{J_1^2\Delta_1^2-\rho_1^2+\rho_2^2}\mp J_1\Delta_1}, \\
  && D_3=\frac{\pm\left(J_2D_1-J_1D_2\right)}{\sqrt{J_1^2\Delta_1^2-\rho_1^2+\rho_2^2}\mp J_1\Delta_1}. \notag
\end{align}
These relations automatically satisfy the phase condition from Eq.~\eqref{FB_pr},
\begin{gather}
  \label{D/J}
  \frac{D_3}{J_3}=\frac{\frac{D_1}{J_1}-\frac{D_2}{J_2}}{1+\frac{D_1 D_2}{J_1 J_2}}.
\end{gather}
One must also check the phase shift condition~\eqref{sigma}.
As this particular solution expresses $J_3$ and $D_3$ through other parameters, it is conveniently expressed as
\begin{gather}
  \frac{D_1D_2}{J_1J_2}>-1.
\end{gather}
However, as $\rho_3>0$, not all values of $J_3$ and $D_3$ given by Eq.~\eqref{J3D3} result in flat bands.
For the solution of Eqs.~\eqref{J3D3} with the upper signs one gets
\begin{align}
  \label{FB_cond_0_pl}
  & J_1\Delta_1<0, \;\; \text{all solutions are acceptable} \\
  & J_1\Delta_1>0, \;\; \text{lower flat band only for} \;\rho_2^2 > \rho_1^2\; \notag
\end{align}
Taking the lower signs in Eqs.~\eqref{J3D3} leads to the following constraints:
\begin{align}
  \label{FB_cond_0_mi}
  & J_1\Delta_1<0, \;\;\text{lower flat band only for} \;\rho_2^2 < \rho_1^2\; \notag \\
  & J_1\Delta_1>0, \;\;  \text{no flat band}
\end{align}
Hence if $\rho_2^2<\rho_1^2$ there is an extra constraint:
\begin{gather}
  \label{rho_2_range}
  \rho_1^2-J_1^2\Delta_1^2 \leq \rho_2^2 < \rho_1^2.
\end{gather}
The one-magnon spectrum~\eqref{eq:one-m_spec} in this case has
\begin{align}
  \label{EB}
  E_0^{\pm} &= \frac{1}{2}\left(2J_1\Delta_1+J_2\Delta_2\pm \frac{\left(J_1J_2+D_1D_2\right)\Delta_3-\rho_2^2}{\sqrt{J_1^2\Delta_1^2-\rho_1^2+\rho_2^2}\mp  J_1\Delta_1}\right),\notag\\
  B_0^{\pm} &= \frac{1}{2}\left(J_2\Delta_2\pm \frac{\left(J_1J_2+D_1D_2\right)\Delta_3+\rho_2^2}{\sqrt{J_1^2\Delta_1^2-\rho_1^2+\rho_2^2}\mp J_1\Delta_1}\right),
\end{align}
for the upper and lower sign choice in Eq.~\eqref{rho3}, respectively.
We present in Fig.~\ref{fig2} the zero-temperature exact diagonalization magnetization plots for $\Delta_1=\Delta_2=\Delta_3=1$, $J_1=J_2=1$, $D_1=1, D_2=2$,
while the values of the other parameters are obtained from Eq.~\eqref{J3D3}, namely $J_3=3$ and $D_3=-1$.
The plateau at $M=1/2$ and the jump to saturation are observed exactly at the value of the saturation field $B_0=9/2$ from Eq.~\eqref{EB}.
The corresponding one-magnon spectrum is shown in Fig.~\ref{fig3}.

\subsection{Flat band in case of no DM interactions}

To connect with the other well-known cases of flat bands in sawtooth chain, it is useful to consider separately $J_1$-$J_2$-$J_3$ sawtooth chain without DM interaction.
The flat-band condition~\eqref{FB_pr} simplifies to
\begin{gather}
  \label{fb_0}
  J_1^2\left(J_2^2+J_3^2+2\Delta_1 J_2 J_3\right)=J_2^2 J_3^2
\end{gather}
$J_3=J_2$ gives the well-known solution~\eqref{FB_old}.
Also Eq.~\eqref{fb_0} can be resolved with respect to either $J_3 (J_2)$ or $J_1$.
The former case can be obtained by setting all $D_a=0$ in Eqs.~(\ref{rho3})-(\ref{Lk}).
The latter leads to the following flat-band condition
\begin{gather}
  \label{fb_01}
  J_1=\frac{\pm J_2 J_3}{\sqrt{J_2^2+J_3^2+2\Delta_1 J_2 J_3}}
\end{gather}
The case of a minus sign corresponds to an upper flat band.
Thus the one-magnon spectrum in this case takes the following form:
\begin{align}
  \label{spec_01}
  \varepsilon_1(k) &= B-E_0+\frac{J_2J_3}{\sqrt R}\cos k, \\
  \varepsilon_2(k) &= B-B_0, \notag\\
  E_0 &= \frac{1}{2}\left(\frac{2\Delta_1J_2J_3}{\sqrt R}+J_2\Delta_2+J_3\Delta_3- \sqrt R\right), \notag\\
  B_0 &= \frac{1}{2} \left(J_2\Delta_2+J_3\Delta_3+\sqrt{R}\right), \notag \\
  R &= J_2^2+J_3^2+2\Delta_1J_2J_3. \notag
\end{align}
Since \(R>0\), an additional constraint $J_2^2+J_3^2+2\Delta_1J_2J_3>0$ is necessary.
The corresponding unnormalized localized state is
\begin{gather}
  \label{locm_01}
  \ket{1_j} = J_3\ket{2j} - \sqrt R \ket{2j+1} + J_2 \ket{2j+2}.
\end{gather}
Therefore in addition to the well-known flat-band case \(J_2=J_3\)~\eqref{fb_0}~\cite{schul02,zhi04,der04,der06,der07,rich08,FAF1,der15,met20,der20,CMP20,ace20,SC_KNB,schnack23a,schnack23b, schnack24},
the sawtooth chain with three couplings and no DM terms admits another simplified flat-band case, corresponding to the model with only two exchange constants, which can be obtained by setting either $J_2 = \mu J_1$, or $J_3=\mu J_1$, $\mu=\pm 1$. Then, the solution of (\ref{fb_0}) gives
\begin{gather}
  \label{fb_simp}
  J_3=-\frac{\mu J_1}{2\Delta_1}\;\; \text{or}\;\; J_2=-\frac{\mu J_1}{2\Delta_1}.
\end{gather}
The one-magnon spectrum has the following form (for the choice $J_2=\mu J_1$):
\begin{align}
  \label{spec_J}
  \varepsilon_1(k) &= B-E_0^{\mu}+J_1\cos k, \\
  \varepsilon_2(k) &= B-B_0^{\mu}, \notag \\
  E_0^{\mu} &= \frac {J_1}{2}\left(2\Delta_1+\mu\Delta_2+\frac{1-\mu \Delta_3}{2\Delta_1}\right), \notag \\
  B_0^{\mu} &= \frac {J_1}{2} \left(\mu\Delta_2-\frac{1+\mu \Delta_3}{2 \Delta_1}\right), \notag
\end{align}
The flat band is the lowest one for $J_1\Delta_1<0$.
The unnormalized localized magnons are:
\begin{gather}
  \ket{1_j} = \ket{2j} - \mu \ket{2j+1} -2\Delta_1 \ket{2j+2}.
\end{gather}
Setting $\Delta_1=-1/2$ and $\mu=1$ leads to the well-known flat band in the sawtooth chain with uniform couplings $J_1=J_2=J_3=1$~\cite{der20} with the saturation field $B_0=\frac 32$.
The choice $\mu=-1$, on the other hand, gives two interesting cases: $J_2=J_3=-J_1=-1$ and $\Delta_1=-1/2$, as well as $J_1=J_3=-J_2=-1$ and $\Delta_1=1/2$.
However, for the former case the value of the saturation field is negative: for these values of the parameters the sawtooth chain already has a ferromagnetic ground state.
The latter case is more remarkable, as here the saturation field, $B_0=\frac{1}{2}$ is reduced compared to the case of uniform $J_a=J$.

\subsection{Solutions in terms of $D_1$, $D_2$ and $D_3$}

The most remarkable feature of the flat-band constraints~\eqref{FB_pr} is the possibility of resolving them with respect to $D_1, D_2$ and $D_3$.
This is particularly important in the light of the models with KNB mechanism.
We provide the interpretation of the results in terms of electric-field-driven flat band later.
In the present subsection we analyze the generic  solutions, expressing $D_1, D_2$ and $D_3$ in terms of given $J_1, J_2, J_3$ and $\Delta_1$.
Let us define $d_a$ as
\begin{gather}
  \rho_a=|J_a|\sqrt{1+d_a^2}.
\end{gather}
Then the first condition of Eq.~\eqref{FB_pr} becomes
\begin{gather}
  \label{fb_ddd}
  d_1=\frac{d_2+d_3}{1-d_2d_3}.
\end{gather}
Substituting this relation into the second equation of~\eqref{FB_pr} and taking into account that $\text{sign}\left(1-d_2d_3\right)=\text{sign}\left(1+d_1d_2\right)=\sigma$ one arrives at the following equation:
\begin{align}
  \label{FB_d2d3}
  && J_1^2\left[J_2^2(1+d_2^2)+J_3^2(1+d_3^2)+2\xi \Delta_1 J_2J_3(1-d_2d_3)\right] \notag \\
  && =J_2^2J_3^2\left(1-d_2d_3\right)^2, \notag \\
  && \xi=\mathrm{sign}\left(J_1J_2J_3\right).
\end{align}
Thus, formally one can conclude that the values of $d_2\equiv x$ and $d_3\equiv y$ corresponding to the flat band lie on a quartic planar curve~\cite{AlGem,cruc}:
\begin{gather}
  \label{quartic}
  x^2y^2-A^2x^2-B^2y^2-2Cxy-D=0,
\end{gather}
where
\begin{align}
  A & = \left|\frac{J_1}{J_3}\right|, \quad B=\left|\frac{J_1}{J_2}\right|, \quad C = 1-\xi\Delta_1\frac{J_1^2}{J_2J_3}, \notag \\
  \label{eq:DR_def}
  D & = \frac{J_1^2\widetilde{R}}{J_2^2J_3^2}-1, \quad \widetilde{R} = J_2^2+J_3^2+2\xi\Delta_1 J_2J_3.
\end{align}
Using the terminology of algebraic geometry the set of solutions $(D_1, D_2, D_3)$ corresponding to the flat band is given by
\begin{widetext}
\begin{gather}
\begin{pmatrix}
  D_1 \\
  D_2 \\
  D_3
\end{pmatrix}=
\left\{
\begin{pmatrix}
  J_1\frac{d_2+d_3}{1-d_2d_3} \\
  J_2 d_2 \\
  J_3d_3
\end{pmatrix}
\left|\begin{pmatrix}
  d_2 \\
  d_3
\end{pmatrix}
\in \mathcal{C}
\left(\left|\frac{J_1}{J_3}\right|, \left|\frac{J_1}{J_2}\right|,1-\xi\Delta_1\frac{J_1^2}{J_2J_3},\frac{J_1^2\widetilde{R}}{J_2^2J_3^2}-1 \right) \right.\right\}
\end{gather}
\end{widetext}
where
\begin{gather}
  \mathcal{C}\left(A, B, C, D\right)= \\
  \left\{
  \begin{pmatrix}
    x \\
    y
  \end{pmatrix}
  \in\mathbb{R}^2\left|x^2y^2-A^2x^2-B^2y^2-2Cxy-D=0\right.\right\} \notag
\end{gather}

The case of a homogeneous quartic curve, $D=0$, is worth a separate consideration.
It includes the $D_a=0$ flat-band constraint from Eq.~\eqref{fb_0}, provided $\text{sign}(J_1)=\text{sign}(J_2J_3)$.
However, it turns out that there is a whole family of $D_a\neq 0$ solutions, corresponding to the additional constraint given in Eq.~\eqref{fb_0}.
The set of solutions for $D=0$ can be parametrized by an angle $\varphi$ in the following way~\cite{cruc}:
\begin{align}
  x & =\frac{r\left(\varphi\right)}{\cos\varphi}, \qquad y = \frac{r\left(\varphi\right)}{\sin\varphi},  \\
  r(\varphi) & = \sqrt{A^2\sin^2\varphi+C \sin2\varphi+B^2\cos^2\varphi}, \notag \\
  & 0<\varphi <2\pi, \; \varphi \neq \pm \pi/2, \pm\pi. \notag
\end{align}
Despite the simplification of the geometry of the plane quartic curve, the solutions of the flat-band constraint are quite complicated as the condition $D=0$ implies an additional constraint for the symmetric exchange constants and $\Delta_1$, which we resolve with respect to $J_1$:
\begin{align}
  J_1=\mu\frac{J_2J_3}{\sqrt{\widetilde{R}}}, \qquad \mu=\left\{-1,1\right\}
\end{align}
Interestingly, the sign factor \(\xi\) in \(\widetilde{R}\), Eq.~\eqref{eq:DR_def}, can be replaced by \(\mu\) in this particular case, as $\mathrm{sign}(J_1)=\mu\,\text{sign}(J_2J_3)$, thus making the definition of $J_1$ unambiguous,
\begin{align}
  J_1 &= \mu\frac{J_2J_3}{\sqrt{R}_{\mu}}, \\
  R_{\mu} &= J_2^2+J_3^2+2\mu\Delta_1 J_2 J_3. \notag
\end{align}
Then, the solutions in terms of the DM interaction parameters can be written in the following form~\cite{AlGem,cruc}:
\begin{align}
  D_1 &= \frac{\mu J_2 J_3\left(\text{sign}\left(\cos\varphi\right)\sqrt{X\left(\varphi\right)}+\text{sign}\left(\sin\varphi\right)\sqrt{Y\left(\varphi\right)}\right)}
{\left|R_{\mu}\right|-\text{sign}\left(\tan\varphi\right)\sqrt{X\left(\varphi\right)Y\left(\varphi\right)}}, \notag \\
  D_2 &= J_2\,\text{sign}\left(\cos\varphi\right)\sqrt{\frac{X\left(\varphi\right)}{R_{\mu}}}, \\
  D_3 &= J_3\,\text{sign}\left(\sin\varphi\right)\sqrt{\frac{Y\left(\varphi\right)}{R_{\mu}}}, \notag \\
  & \text{sign}\left(\tan\varphi\right)\frac{\sqrt{X\left(\varphi\right)Y\left(\varphi\right)}}{\left|R_{\mu}\right|}<1, \notag \\
  X(\varphi) &= J_2^2\tan^2\varphi+2\left(R_{\mu}-\mu\Delta_1J_2J_3\right)\tan\varphi+J_3^2, \notag \\
  Y(\varphi) &= J_3^2\cot^2\varphi+2\left(R_{\mu}-\mu\Delta_1J_2J_3\right)\cot\varphi+J_2^2. \notag
\end{align}

Another class of solutions is possible for the case $D\neq 0$.
However, in this case one has to fix the value of one of the DM interaction parameters $D_2$ or $D_3$ in addition to the values of $J_a$ and $\Delta_1$.
These flat bands correspond to the formal solution of Eq.~\eqref{quartic} with respect to either $x$ or $y$:
\begin{align}
  && x_{\pm}=\frac{Cy\pm\sqrt{C^2y^2+\left(y^2-A^2\right)\left(B^2y^2+D\right)}}{y^2-A^2}, \notag \\
  && y\neq\pm A,
\end{align}
or
\begin{align}
  && y_{\pm}=\frac{Cx\pm\sqrt{C^2x^2+\left(x^2-B^2\right)\left(A^2x^2+D\right)}}{x^2-B^2}, \notag \\
  && x\neq\pm B.
\end{align}
Thus this case of flat-band constraints drives the spectrum into the flat-band mode by adjusting $D_1$ and $D_2$ or $D_3$.
Let us write down the corresponding solution for the case of fixed $D_2$:
\begin{widetext}
\begin{align}
  \label{D1D3}
  D_1 &= J_1\frac{\left[J_1^2 \left(J_3+\xi J_2\Delta_1\right)-J_3\left(J_2^2+D_2^2\right)\right]D_2\mp J_2\text{sign}(J_3)\sqrt{Q}}{J_1^2\left(J_2 J_3-\xi D_2^2\Delta_1 \right)\pm \text{sign}(J_3) D_2 \sqrt{Q}} \\
  D_3 &= \frac{\left(J_2 J_3-\xi J_1^2\Delta_1 \right)D_2\pm \text{sign}(J_3)\sqrt{Q}}{D_2^2-J_1^2} \notag \\
  Q &= J_1^2\left(D_2^4+\left[J_2^2+J_3^2-J_1^2\left(1-\Delta_1^2\right)\right]D_2^2+J_2^2J_3^2-J_1^2\widetilde{R}\right) \notag
\end{align}
\end{widetext}
Degenerate solutions of the flat-band constraint~\eqref{quartic}, when $x=\pm B$ or $y=\pm A$, lead to the following expressions for DM parameters:
\begin{align}
  \label{DDD_1}
  D_1 &= -\mu\,\text{sign}(J_1 J_2) \frac{J_2\left(J_1^2\widetilde{J}^2-J_2^2J_3^2\right)+V}{J_1^2\widetilde{J}^2+J_2^2J_3^2}, \notag \\
  D_2 &= \mu\,\text{sign}(J_2) |J_1|, \\
  D_3 &= -\mu\,\frac{J_1^2\left(J_1^2+\widetilde{R}\right)-J_2^2J_3^2}{2\,\text{sign}(J_2)|J_1|\left(J_2J_3-\Delta_1J_1^2\right)}, \notag \\
  V &= 2J_1^2J_3\left(\xi\Delta_1\left(J_1^2+J_2^2\right)-J_2J_3\right), \notag \\
  \widetilde{J}^2 &= J_1^2+J_2^2+J_3^2, \notag \\
  \mu &= \left\{-1, 1\right\}, \notag
\end{align}
for the $x=\pm B$ case, while the solution for $y=\pm A$ is obtained by replacing $D_2\leftrightarrow D_3$ and $J_2 \leftrightarrow J_3$.
As before, the additional condition for the flat band to be the lower one is necessary here:
\begin{gather}
  \frac{J_1^2\left(J_1^2+\widetilde{R}\right)-J_2^2J_3^2}{2J_2J_3\left(J_2J_3-\Delta_1J_1^2\right)} > -1.
\end{gather}

\begin{figure}[tb]
\begin{center}
  \includegraphics[width=75.5mm]{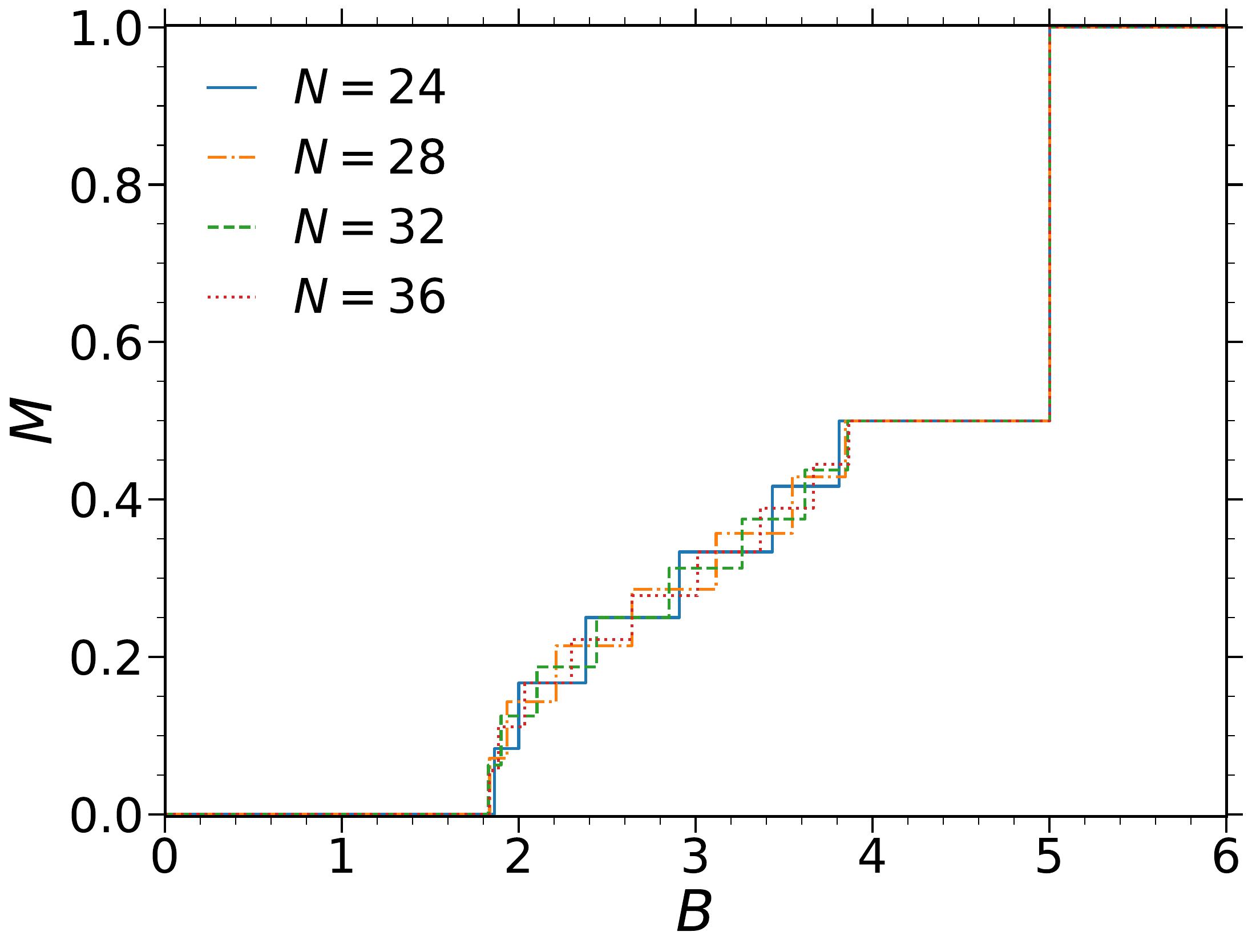}
  \caption{
    (Color online) Zero-temperature exact diagonalization magnetization plots for $J_1=1, J_2=2, J_3=3, \Delta_1=\Delta_2=\Delta_3=1$ and for the values of DM interactions obtained according to Eq.~(\ref{DDD_1}), $ D_1= D_2=D_3=1$.
  }
  \label{fig4}
\end{center}
\end{figure}
A particularly interesting case of these solutions is illustrated in Fig.~\ref{fig4}, when all DM constants are equal to unity and $J_1=1, J_2=2, J_3=3$ with $\Delta_1=\Delta_2=\Delta_3=1$.
Many other flat-band scenarios are possible if one puts additional constraints into the system parameters.
For instance, one can introduce the ratio of $d_3$ and $d_2$, substituting into the general flat-band condition~\eqref{FB_d2d3} $y=\alpha x$, leading to $D_3=\alpha \frac{J_3}{J_2} D_2$.
In this case one can express all three DM couplings in terms of the other Hamiltonian parameters in the following way:
\begin{align}
  \label{sol_y=ax}
  D_1 &= \mp J_1\frac{2J^2\alpha\left(1+\alpha\right)\sqrt{\frac{1}{\alpha}+\frac{ W_{\mu}(\alpha)}{2J^2 \alpha^2}}}{ W_{\mu}(\alpha)},\\
  D_2 &= \pm J_2\sqrt{\frac{1}{\alpha}+\frac{ W_{\mu}(\alpha)}{2J^2 \alpha^2}}, \notag \\
  D_3 &= \pm J_3\alpha\sqrt{\frac{1}{\alpha}+\frac{ W_{\mu}(\alpha)}{2J^2\alpha^2}}, \notag
\end{align}
where
\begin{align}
  \label{sol_y=ax_2}
  J &= \frac{J_2J_3}{J_1}, \\
  W_{\mu}(\alpha) &= R(\alpha)+\mu\sqrt{R^2(\alpha)+4 J^2 \alpha (1+\alpha)\left(J_2^2+\alpha J_3^2\right)}, \notag \\
  R(\alpha) &= J_2^2+\alpha^2 J_3^2-2\alpha \xi \Delta_1 J_2 J_3, \notag \\
  \mu &= \left\{-1,1\right\}. \notag
\end{align}
Taking into account square roots and condition for proper order of the bands one can obtain additional constraint for the function $W_{\mu}(\alpha)$:
\begin{align}
  \label{W_const}
  && \alpha >0, \;\;\; W_{\mu}(\alpha)\in \left(-2J^2\alpha, \, 0\right), \\
  && \alpha<0, \;\;\; W_{\mu}(\alpha)\in \left(0, \, -2J^2\alpha\right).  \notag
\end{align}
As $W_{1}(\alpha)$ is more likely to be positive and $W_{-1}(\alpha)$ to be negative, it is easier to find flat bands with $\mathrm{sign}(\mu\alpha)=-1$, although flat bands in other cases are still possible.

\subsection{The case of sawtooth chain with uniform exchange coupling, $J_1=J_2=J_3$}

To obtain unambiguous situations in which the analysis of the signs of $\alpha$ and $\mu$ is easier, and also to consider an important particular case of the sawtooth chain with homogeneous symmetric exchange, let us consider separately the values of DM parameters given in Eq.~\eqref{sol_y=ax_2} for $J_1=J_2=J_3=J=1$ and $\Delta_1=\pm 1$,
\begin{align}
  \label{sol_y=ax_2b}
  D_1 &= \mp \frac{2\alpha\left(1+\alpha\right)\sqrt{\frac{1}{\alpha}+\frac{ X_{\mu}(\alpha)}{2 \alpha^2}}}{ X_{\mu}(\alpha)},\\
  D_2 &= \pm \sqrt{\frac{1}{\alpha}+\frac{ X_{\mu}(\alpha)}{2 \alpha^2}}, \notag \\
  D_3 &= \pm \alpha\sqrt{\frac{1}{\alpha}+\frac{ X_{\mu}(\alpha)}{2\alpha^2}}, \notag
\end{align}
where
\begin{align}
  \label{Xmu}
  X_{\mu}(\alpha)=\\
  \left\{
  \begin{array}{ll}
    (1-\alpha)^2+\mu\sqrt{\left(1-\alpha\right)^4+4\alpha \left(1+\alpha\right)^2}, \; \Delta_1=1 \\
    \left(1+\alpha\right)^2+\mu\sqrt{\left(1+\alpha\right)^4+4\alpha \left(1+\alpha\right)^2}, \; \Delta_1=-1
  \end{array}\right.  \notag
\end{align}
The positivity of the expression under the square root imposes an additional constraint restricting the possible values of $\alpha$ for given values of $\mu$.
For $\Delta_1=1$ only $\mu=1$ and $\alpha<0$ are acceptable, however, so that the square root in Eq.~\eqref{Xmu} is always positive.
For the case $\Delta_1=-1$ both signs of $\mu$ are compatible with the flat bands value, however, $\alpha$ still must be negative, with an additional forbidden range of the values of $\alpha$ appearing, for which the expression under the square root is negative:
\begin{gather}
  \label{al_range}
  \alpha \in \left(-3-2\sqrt 2, -3+2\sqrt 2\right).
\end{gather}
Thus, for the sawtooth chain with isotropic and uniform symmetric exchange couplings there are infinitely many sets of DM interactions that lead to a flat band.
We note that the expressions for the DM constants given in Eqs.~(\ref{sol_y=ax_2})-(\ref{Xmu}) possess an interesting symmetry with respect to inversion of the negative numerical coefficient $\alpha$:
\begin{align}
  \alpha &\rightarrow\frac{1}{\alpha}, \\
  D_1 &\rightarrow -D_1, \notag \\
  D_2 &\rightarrow -D_3, \notag \\
  D_3 &\rightarrow - D_2. \notag
\end{align}

\begin{figure}[tb]
\begin{center}
\includegraphics[width=75.5mm]{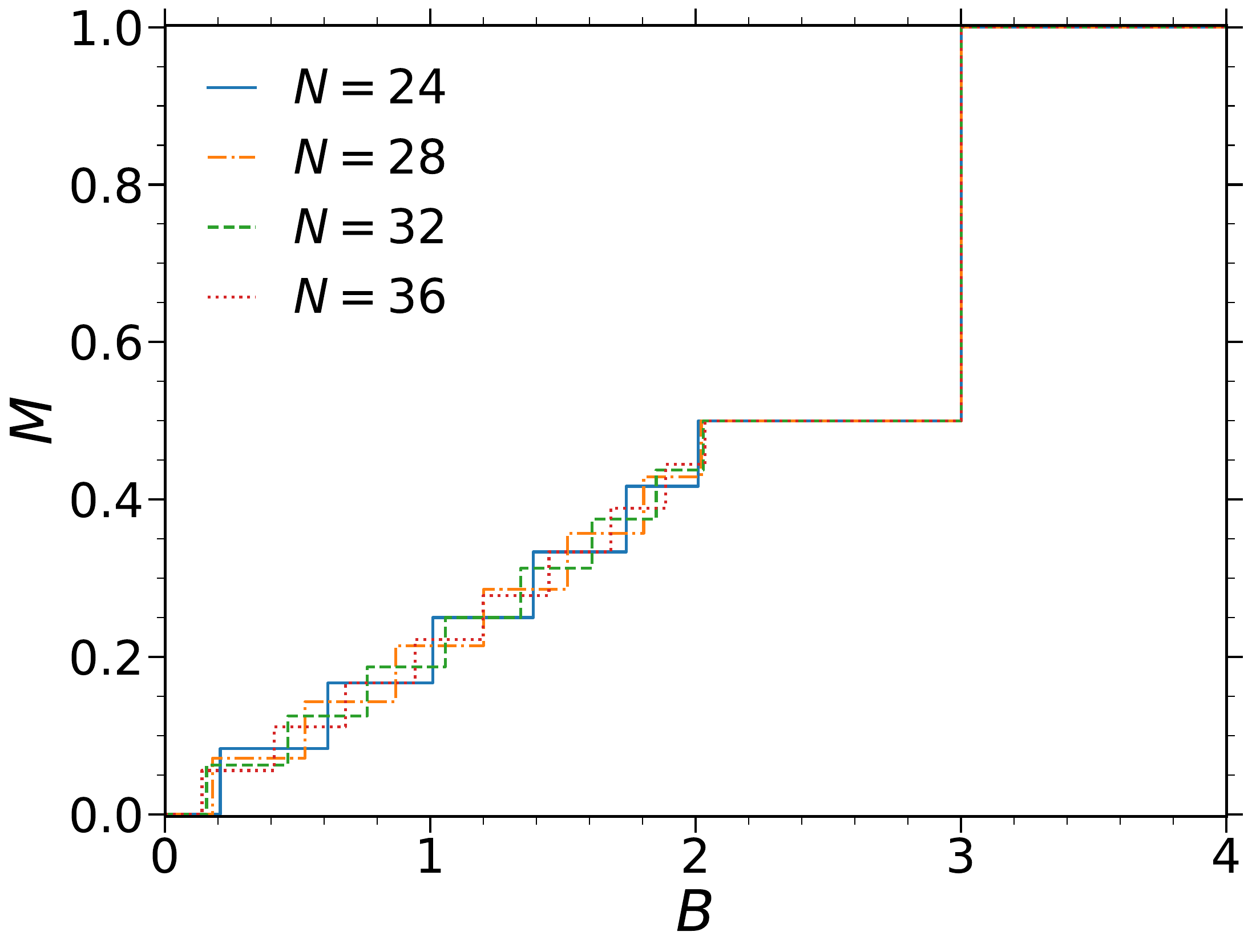}
  \caption{
    (Color online)
    Zero-temperature magnetization curves for a finite system with uniform exchange couplings, $J_1=J_2=J_3=1$ and $\Delta_1=1$ with the values of DM interaction corresponding to the first line of the Table~\ref{Tab:1}, $D_1=0, D_2=\sqrt 3, D_3=-\sqrt 3$.
    DM interactions can be implemented via electric-field-driven flat band~\cite{SC_KNB}.
    Note the absence of the plateau at \(M=0\).
  }
  \label{fig5}
\end{center}
\end{figure}

\begin{figure}[tb]
\begin{center}
  \includegraphics[width=75.5mm]{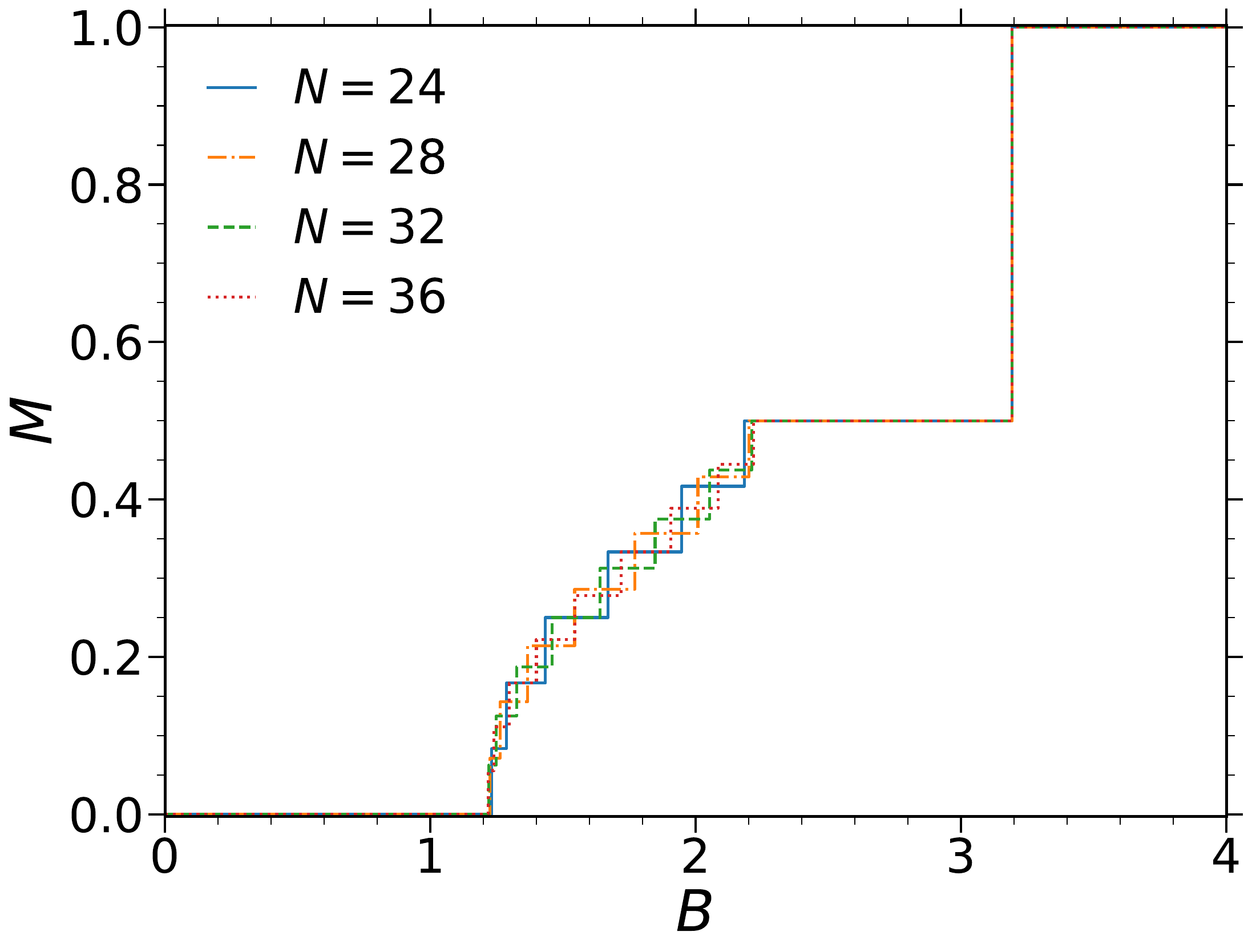}
  \caption{
    (Color online)
    Zero-temperature magnetization curves for a finite system with uniform exchange couplings, $J_1=J_2=J_3=1$ and $\Delta_1=1$, corresponding to the second line of the Table~\ref{Tab:1}, i.e., $\alpha=-2$, $\mu=1$.
  }
  \label{fig6}
\end{center}
\end{figure}

\begin{figure}[tb]
\begin{center}
  \includegraphics[width=75.5mm]{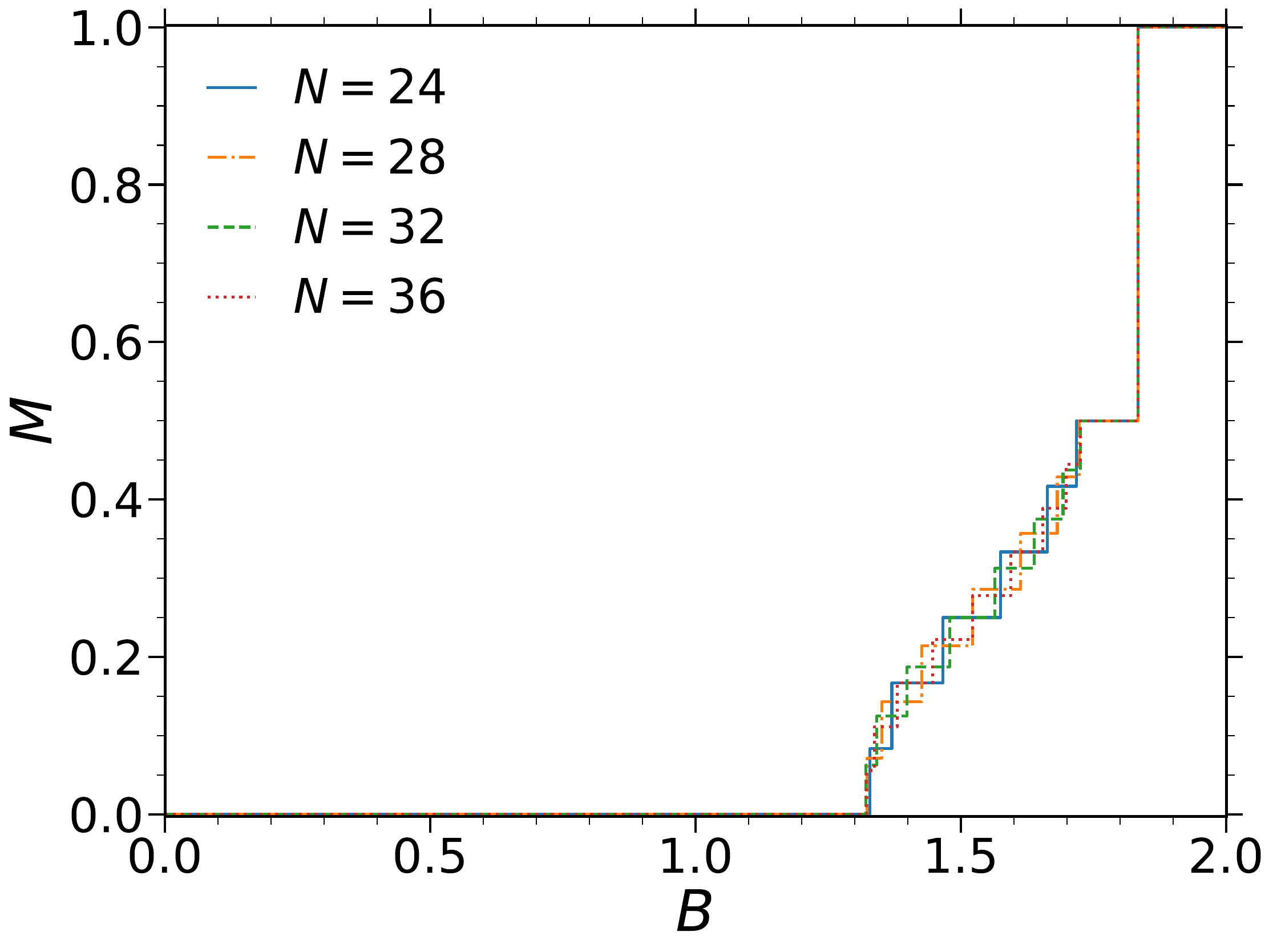}
  \caption{
    (Color online)
    Zero-temperature magnetization plots for a finite system with uniform exchange couplings, $J_1=J_2=J_3=1$ and ferromagnetic axial anisotropy, $\Delta_1=-1$.
    The values of the DM parameters can be found in the first line of the Table \ref{Tab:2}, which corresponds to $\alpha=-6$, $\mu=-1$ leading to $D_1=-1$, $D_2=1/3$ and $D_3=-2$.
    Here one can see substantial extension of the plateau at zero magnetization and reduction of the localized magnons crystal region, the $M=1/2$ plateau.
  }
  \label{fig7}
\end{center}
\end{figure}

Examples of the solutions are presented in Tables~\ref{Tab:1}
and~\ref{Tab:2} for the cases of $\Delta_1=1$ and $\Delta_1=-1$,
respectively.  As an illustration of the flat bands cases for uniform
exchange couplings, $J_1=J_2=J_3=1$ for $\Delta_1=\pm 1$,
magnetization plots are presented in Fig.~\ref{fig5}-\ref{fig7}.  The
case presented in Fig.~\ref{fig5} is a remarkable, as it can be
obtained by an electric field in the sawtooth model with KNB
mechanism, namely when the electric field direction is parallel to the
basal line of the chain~\cite{SC_KNB}.  It is also the simplest case
of the solutions, given in Eq.~(\ref{sol_y=ax_2}) for $\alpha=-1$ and
$\mu=1$.  The corresponding value of the electric field magnitude is
given by $E=\pm\frac{\sqrt 3}{\sin\theta}$.  The corresponding
magnetization plot lacks the plateau at $M=0$.  The next two cases,
shown in Figs.~\ref{fig6} and~\ref{fig7}, illustrate the magnetization
curves for two values of $\alpha$ from Tables~\ref{Tab:1}
and~\ref{Tab:2}.

\begin{table*}[tb!]
  \caption{
    Examples of the values of the sets of DM parameters leading to the flat band for uniform coupling, $J_1=J_2=J_3=1$ and $\Delta_1=1$, obtained from Eqs.~(\ref{sol_y=ax_2}-\ref{Xmu}) for different values of $\alpha$.
    Only $\mu=1$ is possible in this case.
  }
  \begin{ruledtabular}
  \begin{tabular}{ccccc}
    $\alpha$    & $\mu$ & $D_1$ & $D_2$ & $D_3$ \vspace{0.4em} \\
    \hline
    $-1$                         & $1$ & $0$  & $\sqrt 3$ & $-\sqrt 3$ \vspace{0.2em} \\
    $-2$                         & $1$ & $\displaystyle -\frac{\sqrt{2\left(5+\sqrt{73}\right)}}{9+\sqrt{73}}$      & $ \displaystyle\frac{1}{2} \sqrt{\frac{1}{2}\left(5+\sqrt{73}\right)}$ & $ \displaystyle-\sqrt{\frac{1}{2}\left(5+\sqrt{73}\right)}$             \vspace{0.2em} \\
    $-3$                         & $1$ & $\displaystyle -\frac{\sqrt{5+2\sqrt{13}}}{4+\sqrt{13}}$                   & $ \displaystyle\frac{1}{3}\sqrt{5+2\sqrt{13}}$                         & $ \displaystyle-\sqrt{5+2\sqrt{13}}$                                 \vspace{0.2em} \\
    $-4$                         & $1$ & $\displaystyle -\frac{3\sqrt{2\left(17+\sqrt{481}\right)}}{25+\sqrt{481}}$ & $ \displaystyle\frac 14 \sqrt{\frac{1}{2}\left(17+\sqrt{481}\right)}$  & $ \displaystyle-\sqrt{\frac{1}{2}\left(17+\sqrt{481}\right)}$           \vspace{0.2em} \\
    $-5$                         & $1$ & $\displaystyle -\frac{2\sqrt{13+2\sqrt{61}}}{9+\sqrt{61}}$                 & $ \displaystyle\frac 15 \sqrt{13+2\sqrt{61}}$                          & $ \displaystyle-\sqrt{13+2\sqrt{61}}$                                \vspace{0.2em} \\
    $\displaystyle -\frac{1}{2}$ & $1$ & $\displaystyle\frac{\sqrt{2\left(5+\sqrt{73}\right)}}{9+\sqrt{73}}$        & $ \displaystyle\sqrt{\frac{1}{2}\left(5+\sqrt{73}\right)}$             & $ \displaystyle-\frac{1}{2} \sqrt{\frac{1}{2}\left(5+\sqrt{73}\right)}$ \vspace{0.2em} \\
  \end{tabular}
  \label{Tab:1}
  \end{ruledtabular}
\end{table*}

\begin{table*}[tb!]
  \caption{
    Examples of the values of the sets of DM parameters leading to the flat band for uniform coupling, $J_1=J_2=J_3=1$ and $\Delta_1=-1$, obtained from Eqs.~(\ref{sol_y=ax_2}-\ref{Xmu}) for different values of $\alpha$.
    Both signs of $\mu$ are acceptable, however, there are forbidden values of $\alpha$ given in Eq.~\eqref{al_range}.
  }
  \begin{ruledtabular}
  \begin{tabular}{ccccc}
    $\alpha$                      & $\mu$ & $D_1$ & $D_2$ & $D_3$ \vspace{0.4em} \\
    \hline
    $-6$                          & $-1$ & $-1$                                                                     & $ \displaystyle\frac{1}{3}$ & $-2$ \vspace{0.2em} \\
    $-6$                          & $1$  & $-1$                                                                     & $ \displaystyle\frac{1}{2}$ & $-3$ \vspace{0.2em} \\
    $\displaystyle -\frac{1}{10}$ & $1$  & $ \displaystyle\frac{\sqrt{2\left(61+9\sqrt{41}\right)}}{9+\sqrt{41}}$   & $ \displaystyle\sqrt{\frac{1}{2}\left(61+9\sqrt{41}\right)}$             & $ \displaystyle-\frac{1}{10}\sqrt{\frac{1}{2}\left(61+9\sqrt{41}\right)}$ \vspace{0.2em} \\
    $\displaystyle -\frac{1}{10}$ & $-1$ & $ \displaystyle\frac{\sqrt{2\left(61-9\sqrt{41}\right)}}{9-\sqrt{41}}$   & $ \displaystyle\sqrt{\frac{1}{2}\left(61-9\sqrt{41}\right)}$             & $ \displaystyle-\frac{1}{10}\sqrt{\frac{1}{2}\left(61-9\sqrt{41}\right)}$ \vspace{0.2em} \\
    $-10$                         & $1$  & $ \displaystyle-\frac{\sqrt{2\left(61+9\sqrt{41}\right)}}{9+\sqrt{41}}$  & $ \displaystyle\frac{1}{10}\sqrt{\frac{1}{2}\left(61+9\sqrt{41}\right)}$ & $ \displaystyle-\sqrt{\frac{1}{2}\left(61+9\sqrt{41}\right)}$             \vspace{0.2em} \\
    $-8$                          & $1$  & $ \displaystyle-\frac{\sqrt{66+14\sqrt{17}}}{7+\sqrt{17}}$               & $ \displaystyle\frac 18 \sqrt{\frac{1}{2}\left(33+7\sqrt{17}\right)}$    & $ \displaystyle-\sqrt{\frac{1}{2}\left(33+7\sqrt{17}\right)}$             \vspace{0.24em} \\
  \end{tabular}
  \label{Tab:2}
  \end{ruledtabular}
\end{table*}

\section{Particular configurations of $D_a$ and comparison with the electric-field-driven flat bands scenario from Ref.~\onlinecite{SC_KNB}}
\label{sec:4}

\subsection{$D_1=0, D_3/J_3=-D_2/J_2$}

Let us first focus on the simplest case $D_1=0, D_3/J_3=-D_2/J_2$.
The value of the phase shift  under this assumption is always zero, since $\sigma=1$.
Therefore, the flat band conditions~\eqref{FB_pr} now read:
\begin{align}
  \label{D_fb_1}
  J_1^2\widetilde{R} &=J_2^2 J_3^2\left(1+d^2\right)\\
  d &\equiv \frac{D_2}{J_2}=-\frac{D_3}{J_3}. \notag
\end{align}
These equations can be solved with respect to $J_1$, giving the solution similar to those given in Eqs.~(\ref{fb_01})-(\ref{locm_01}).
However, it is more instructive to look for the solution with respect to $d$, which is related to the case of an electric-field-driven flat band in the sawtooth chain with KNB mechanism~\cite{SC_KNB}.
This solution is also a special case of Eq.~\eqref{sol_y=ax} with $\alpha=-1$ and $\mu=1$.
Thus, here we again have an example of the flat band induced by tuning of the DM terms for arbitrary values of the symmetric exchange couplings:
\begin{align}
  \label{D2D3_1}
  D_1 &= 0, \\
  D_2 &= \pm J_2\sqrt{\widetilde{R}/J^2 -1}, \notag \\
  D_3 &= \mp J_3\sqrt{\widetilde{R}/J^2 -1}. \notag
\end{align}
This solution requires an additional inequality between the model exchange constants:
\begin{gather}
  \label{ineq1}
  J_1^2 \widetilde{R}>J_2^2 J_3^2.
\end{gather}
The one-magnon spectrum in this case is independent of the sign choice in Eqs.~\eqref{D2D3_1}:
\begin{align}
  \epsilon^{+}(k) &= B - E_0 + J_1\cos k, \\
  \epsilon^{-}(k) &= B - B_0, \notag
\end{align}
with
\begin{align}
  E_0 &= \frac{1}{2}\left[J_2\Delta_2+J_3\Delta_3-\frac{\left(J_2^2+J_3^2\right)}{J}\right], \\
  B_{0} &= \frac{1}{2}\left[2J_1\Delta_1+J_2\Delta_2+J_3\Delta_3+\frac{\left(J_2^2+J_3^2\right)}{J}\right]. \notag
\end{align}

The corresponding one-magnon localized states are conveniently presented in the following form:
\begin{align}
  \ket{l_j} &= J_3 \ket{2j} - \sqrt{\widetilde{R}} e^{-i\phi} \ket{2j+1} + J_2 \ket{2j+2}, \notag \\
  \phi &= \pm\arctan\left(\sqrt{\widetilde{R}/J^2-1}\right).
\end{align}
The DM terms in the Hamiltonian make it possible to attain the flat-band regime even in the case of uniform exchange couplings, $J_1=J_2=J_3>0$:
\begin{align}
  D_1 &= 0, \\
  D_2 &= \pm J_1\sqrt{1+2\Delta_1}, \notag \\
  D_3 &= \mp J_1\sqrt{1+2\Delta_1}, \notag
\end{align}
which for $\Delta_1=1$ gives the results of Eqs.~(\ref{sol_y=ax_2}-\ref{Xmu}) with $\alpha=-1$ and $\mu=1$.
The parameters of the one-magnon spectrum read
\begin{align}
  E_0 &= \frac {J_1}{2}\left(\Delta_2+\Delta_3-2\right) \\
  B_{0} &= \frac {J_1}{2}\left(2\Delta_1+\Delta_2+\Delta_3+2\right). \notag
\end{align}
The localized magnon states become
\begin{align}
  \label{lm_D2D3_J}
  \ket{l_j} &= \ket{2j} - \sqrt{2(1+\Delta_1)}e^{-i\phi} \ket{2j+1} + \ket{2j+2}, \notag \\
  \phi &= \pm\arctan\left(\sqrt{1+2\Delta_1}\right).
\end{align}
In this scenario the flat band is due to the DM interaction terms: one needs a fine tuning of the DM parameters, depending on the values of the exchange constants and the axial anisotropy along the basal line only.
The values of $J_1$, $J_2$, $J_3$ and $\Delta_1$ can be arbitrary, provided the inequality of Eq.~(\ref{ineq1}) is satisfied.
Moreover, the DM terms are controlled in this case by a single parameter, $d=D_2/J_2=-D_3/J_3$, which has a staggered structure along the zigzag part of the chain in the case of $J_2=J_3$.
On the other hand, the DM interaction for the basal line must be zero.
We conclude that the $J_1$-$J_2$ XXZ sawtooth chain with staggered DM interaction along the zigzag bonds can be driven to a flat band regime by fine-tuning a single DM parameter.

Previously Ref.~\onlinecite{SC_KNB} discussed a flat band induced by an electric field in the $J_1$-$J_2$ sawtooth chain ($J_3=J_2$) by means of the KNB mechanism.
This flat band requires the direction of electric field to be along the basal line, so that $D_1=0$, $D_2=-D_3$, with $D_2=\pm E \sin\theta$ for direction of the electric field in the positive and negative direction of $x$-axis, respectively.
Tuning the magnitude of electric field one can achieve the flat band at $E=\pm\frac{\sqrt{4J_1^2-J_2^2}}{\sin\theta}$.
However, the form of the DM constants considered in this subsection is incompatible with the expressions given by the KNB mechanism, unless $J_3=J_2$.
Thus, we conclude that an electric-field-driven flat band of Ref.~\cite{SC_KNB} can only be obtained in the case of uniform exchange coupling along zigzag part of the chain ($J_3=J_2$).
The generalization of this result for the case of $\Delta_1 \neq 1$ and negative $J_1$ reads
\begin{gather}
  E = \pm\frac{\sqrt{2J_1^2(1+\text{sign}(J_1)\Delta_1)-J_2^2}}{\sin\theta}.
\end{gather}
Parameters of the one-magnon spectrum~\eqref{eq:one-m_spec} are the following in this case:
\begin{align}
  \label{E_scen1}
  E_0 &= J_2\Delta_2-J_1, \\
  B_{0} &= J_1\left(1+\Delta_1\right)+J_2\Delta_2, \notag \\
  \phi &= \pm\arctan\left(\frac{\sqrt{2J_1^2(1+\text{sign}(J_1)\Delta_1)-J_2^2}}{J_2}\right), \notag
\end{align}
The localized magnon states are given by
\begin{align}
  \ket{l_j} = J_1 \ket{2j} - J_2 e^{-i\phi} \ket{2j+1} + J_1 \ket{2j+2}.
\end{align}

\subsection{$D_1/J_1=D_2/J_2, D_3=0$ }

The next case with non-trivial DM terms is $D_1/J_1=D_2/J_2, D_3=0$.
By virtue of the symmetry under interchange of $J_2 (D_2)$ and $J_3 (D_3)$, we put $D_3=0$ without loss of generality.
Then the flat band condition reads
\begin{align}
  J_1^2 \left(\widetilde{R}+J_2^2d^2\right) &= J_2^2J_3^2, \notag \\
  d\equiv\frac{D_1}{J_1} &= \frac{D_2}{J_2}.
\end{align}
Alternatively, one can just set $y=0$ in Eqs.~\eqref{quartic} and solve the resulting trivial equation with respect to $x$.
Again, here we can treat the DM terms as given and adjust some of the couplings, say $J_1$, which leads to the flat band formation at
\begin{gather}
  J_1^2=\frac{J_2^2 J_3^2}{J_3^2+(1+d^2)J_2^2+2 J_2 J_3\Delta_1},
\end{gather}
However it is more interesting to tune the DM interaction parameters for the given set of exchange constants so as to drive the system into flat band regime:
\begin{align}
  D_1 &= \pm\xi J_3\sqrt{1-\widetilde{R}/J^2}, \notag \\
  D_2 &= \pm\left|J\right|\sqrt{1-\widetilde{R}/J^2}, \notag \\
  D_3 &= 0, \notag \\
  J_1^2 \widetilde{R} &< J_2^2J_3^2.
\end{align}
The one-magnon spectrum in this case has the following form:
\begin{align}
  \label{spec_0pp}
  \varepsilon^{+}_1(k) &= B - E_0 + \frac{1}{\left|J_2\right|}\sqrt{J_2^2\left(J_1^2-J_3^2\right)+J_1^2 \widetilde{R}}\cos (k-\phi_1), \notag \\
  \varepsilon^{-}_1(k) &= B - B_{0}, \notag \\
  E_0 &= \frac{1}{2}\left(2J_1\Delta_1+J_2\Delta_2+J_3\Delta_3-\frac{|J_2|J_3}{|J_1|}\right), \\
  B_0 &= \frac{1}{2} \left(J_2\Delta_2+J_3\Delta_3+\frac{|J_2|J_3}{|J_1|}\right), \notag \\
  \phi_1 &= \pm\arctan\left(\frac{\xi J_3\sqrt{1-\widetilde{R}/J^2}}{J_1}\right). \notag
\end{align}
The localized magnon state is given by the following superposition:
\begin{align}
  \ket{l_j} &=& \\
  & J_3\left(|J_1| \ket{2j} - |J_2| \ket{2j+1}\right) + We^{i\phi} \ket{2j+2}, \notag \\
  W &= \sqrt{J_2^2\left(J_1^2-J_3^2\right)+J_1^2 \widetilde{R}}. \notag
\end{align}
This flat band has interesting properties.
Unlike the previous case, it does not exist for uniform coupling, $J_1=J_2=J_3$.
It is also related to the electric-field induced flat band of Ref.~\onlinecite{SC_KNB}.
In the context of the KNB mechanism the DM interaction parameters are not independent but are related by the electric field magnitude and angle as well as the bond angle.
From the condition $D_3=0$ the angle of the electric field is fixed to be $\phi=-\theta$ or $-\theta\pm\pi$, and then $D_1/J_1=D_2/J_2$ leads to an additional constraint for the  exchange couplings:
\begin{gather}
  \label{J2alpha}
  J_2=-\frac{2 \cos\theta}{\alpha}J_1.
\end{gather}
This electric-field-driven flat band was obtained in Ref.~\onlinecite{SC_KNB} for the case $J_3=J_2$ and $\Delta_1=1$.
In the KNB case, when the electric field is parallel to a lattice bond, the corresponding effective DM constant vanishes (see Eq.~\eqref{eq:DDD}).
For $D_2=0$ or $D_3=0$ one finds $\phi=\theta, \theta\pm\pi$ or $\phi=-\theta, \theta\pm\pi$ respectively.
When $D_2=0$ the additional constraint for the exchange couplings takes the following form:
\begin{gather}
  \label{J3alpha}
  J_3=\frac{2 \cos\theta}{\alpha}J_1.
\end{gather}
We can now generalize these results for the case of a non-symmetric sawtooth chain with $J_3\neq J_2$ and anisotropic coupling.
This flat band in $J_1$-$J_2$-$J_3$ XXZ sawtooth chain with KNB mechanism can be realized when the electric field points along the zigzag bonds, provided the additional constraint~\eqref{J2alpha} is satisfied and the magnitude of the electric field is equal to
\begin{align}
  E &=& \\
  && \pm\frac{\sqrt{\frac{4\cos\theta}{\alpha}\left(\frac{\cos\theta}{\alpha}\left(J_3^2-J_1^2\right)+\xi \Delta_1 J_1J_3\right)-J_3^2}}{\sin2\theta} \notag
\end{align}
In the case that the electric field points along the left links of the zigzag part of the sawtooth chain, i.e., $\phi=\theta$ or $\phi=\theta\pm\pi$, the constraint~\eqref{J3alpha} yields
\begin{align}
  E &=& \\
    && \pm\frac{\sqrt{\frac{4\cos\theta}{\alpha}\left(\frac{\cos\theta}{\alpha}\left(J_2^2-J_1^2\right)-\xi \Delta_1 J_1J_2\right)-J_2^2}}{\sin2\theta} \notag
\end{align}
For the case of a symmetric sawtooth chain, $J_2=J_3$, but with non-isotropic interaction in the basal line, $\Delta_1 \neq 1$, and possibly negative values of $J_1$, one finds
\begin{gather}
  E=\pm\frac{2\sqrt{\cos^2\theta-\alpha^2\frac{1+\text{sign}(J_1)\Delta_1}{2}}}{\alpha^2\sin\theta}\left|J_1\right|.
\end{gather}

\section{Mapping between the general DM and KNB cases: more possibilities of electric-field-driven flat bands}
\label{sec:5}

In the previous section we established a link between particular solutions of the flat band constraints~\eqref{FB_pr} and the KNB term generated by an electric field.
Those solutions feature one vanishing DM interaction, \(D_1\) or \(D_3(D_2)\).
A natural connection with the KNB interaction when electric field is parallel to the one of the sawtooth chain bonds was demonstrated.
However, do the cases considered in Ref.~\onlinecite{SC_KNB} exhaust all possible electric-field-driven flat band scenarios in the sawtooth chain model with KNB mechanism?
The answer is negative.
Here we work out explicitly the mapping between the KNB case and our solution given in terms of $D_1, D_2$ and $D_3$.
For any given set of DM parameters, \(D_1, D_2, D_3\), that satisfy the flat-band constraints~\eqref{FB_pr},
it is possible to derive an angle \(\phi\) and magnitude \(E\) of the in-plane electric field, and the bond angle \(\theta\) of the chain which corresponds to the values of $D_1, D_2$ and $D_3$ with only an additional inequality on their values.
Using the expressions for $D_1, D_2$ and $D_3$ from the KNB formulas~\eqref{eq:DDD} one can express the parameters of the KNB model via the corresponding DM interaction couplings
\begin{align}
  \label{eq:phi_E}
  \phi &= \arctan\left(\frac{D_3+D_2}{D_3-D_2}\tan\theta\right), \\
  E^2 &= \frac 14\left\{\left(D_3+D_2\right)^2\frac{1}{\cos^2\theta}+\left(D_3-D_2\right)^2\frac{1}{\sin^2\theta}\right\}. \notag
\end{align}
The bond angle \(\theta\) is constrained by Eq.~\eqref{eq:DDD} and should be expressed via  \(D_1, D_2, D_3\) to provide one-to-one correspondence between two sets of three variables, \(\left(D_1, D_2, D_3\right)\) and \(\left(E, \phi, \theta\right)\):
\begin{gather}
  \label{eq:theta}
  \tan^2\theta=\frac{4D_1^2-a^2\left(D_2+D_3\right)^2}{a^2\left(D_2+D_3\right)^2},
\end{gather}
This allows us to write:
\begin{align}
  \label{eq:phi_E_2}
  \phi &= \arctan\left(\frac{\text{sign}(D_3+D_2)}{|a|(D_3-D_2)}\sqrt{4D_1^2-a^2(D_2+D_3)^2}\right), \notag\\
  E &= \pm\frac{2|D_1|}{|a|}\sqrt{\frac{D_1^2-a^2D_2D_3}{4D_1^2-a^2(D_2+D_3)^2}}.
\end{align}
Thus, for any solution given in the previous section the corresponding electric-field-driven flat band within the KNB model can be constructed by the fine-tuned choice of the electric field direction and magnitude, provided
\begin{gather}
  \label{eq:4D1^2}
  4D_1^2-a^2(D_2+D_3)^2>0.
\end{gather}
Interestingly, the microscopic coefficient of non-uniformity of the KNB mechanism realization, \(a\) enters as a parameter in the mapping from general \(D\) to the KNB case.
For each given set of $D$ parameters a one-parameter family of $\phi(a), E(a)$ and $\theta(a)$ exists.
Let us illustrate these relations for some solutions presented in previous sections.
For the flat band case illustrated in Fig.~\ref{fig2} one has \(J_1=1, J_2=1, J_3=1\) and \(D_1=1, D_2=2, D_3=-1\), which correspond to the following parameters of the electric field and lattice geometry:
\begin{align}
  \label{eq:Fig2}
  \theta &= \arctan\frac{\sqrt{4-a^2}}{|a|}, \\
  \phi &= -\arctan\frac{\sqrt{4-a^2}}{3|a|}, \notag \\
  E &= \pm\frac{2}{|a|}\sqrt{\frac{1+2a^2}{4-a^2}}, \quad |a| < 2. \notag
\end{align}
For the particular value \(a=1\) (i.e., for a homogeneous KNB setting)
we get
\begin{align}
  \theta = \frac{\pi}{3} \qquad \phi = -\frac{\pi}{6}, \qquad E = \pm 2\notag
\end{align}
The case presented in Fig.~\ref{fig4}, $J_1=1, J_2=2, J_3=3$ and $D_1=D_2=D_3=1$, which is interesting by itself, leads to a remarkable configuration of the electric field, orthogonal to the basal line of the sawtooth chain, $\phi=\frac{\pi}{2}$.
However, the KNB interpretation is only possible for $a<1$, leading to
\begin{align}
  \theta = \arctan\frac{\sqrt{1-a^2}}{|a|}, \qquad E = \pm\frac{1}{|a|}. \notag
\end{align}
Most solutions presented in the Tables~\ref{Tab:1} and~\ref{Tab:2} do not satisfy the condition~\eqref{eq:4D1^2}: it can be shown that KNB interpretation is only possible for large negative values of $\alpha$.
The case from Table~\ref{Tab:2} with $\alpha=-6$ which is illustrated in Fig.~\ref{fig7} ($J_1=J_2=J_3=1$, $D_1=-1, D_2=1/3, D_3=-2$) is one of the exceptions:
\begin{align}
  \theta &= \arctan\frac{\sqrt{36-25a^2}}{5|a|},\\
  \phi &= \arctan\frac{\sqrt{36-25a^2}}{7|a|},\notag \\
  E &= \pm\frac{2}{|a|}\sqrt{\frac{9+6a^2}{36-25a^2}}. \notag
\end{align}

\section{Linear spin wave theory, tight-binding model view on flat bands}
\label{sec:6}

The most common many-body quantum models where localized states are found and well understood are tight-binding models, or lattice models with mobile non-interacting particles~\cite{der15, ber13,par13,ley13,ley18,bae23,lee24,mall25}.
The first indication regarding the possible dispersionless single-particle eigenstates of $2D$ tight-binding Hamiltonian was given in the 1980s~\cite{suth86}.
The study of special topologies of the lattice then led to the discovery of localization-induced ferromagnetism in Hubbard models at half-filling for finite on-site repulsion $U$~\cite{mielke,tas92,mielke93,tas98, Moes10, Moes12}.
Further recent developments of these ideas provided the description of the para-ferro transition in the Hubbard model with flat bands in terms of Pauli-correlated percolation~\cite{Moes10,Moes12}.
In recent decades an essential progress was achieved in the experimental realization of the aforementioned models with flat bands in photonic lattices~\cite{Ley18a,nat24,phot25}, cold atoms in optical lattices~\cite{Taie15,Ozawa17,Taie20}, as well as twisted graphene bilayers~\cite{gr23,gr24}.
The Heisenberg model, and more complicated quantum spin Hamiltonians in general, are strongly interacting systems, suggesting that localized states in translationally invariant quantum spin systems are many-body localized states.
Indeed, this is the case for one-dimensional quantum spin models with disorder~\cite{MBL}, however the physics of localized magnons has much in common with this well-known example of localization in tight-binding models.
The standard technique to work with spin Hamiltonians is based on the Holstein-Primakoff~\cite{HP} or Dyson-Maleev~\cite{Dys1, Dys2, Mal} transformation from spin to bosonic operators.
In practice  Heisenberg models may be handled  in terms of bosonic operators by a perturbative expansion of the Hamiltonian in powers of the ratio of average boson number and $2S$, where $S$ is the spin magnitude.
The lowest order of these expansions corresponds to free bosons and formally leads to a tight-binding Hamiltonian, an approach which is referred to as a Linear Spin Wave Theory (LSWT).
It will be instructive to compare our exact one-magnon spectrum of the Heisenberg Hamiltonian and its flat-band constraints with the spectrum of the tight-binding Hamiltonian obtained within the LSWT.
We thus consider the sawtooth chain with Hamiltonian~\eqref{ham_gen} for arbitrary spin $S$ and employ standard LSWT for two sites per unit cell.
Namely, the Holstein-Primakoff transformations with respect to the ferromagnetic  reference state have the following form:
\begin{align}
  S_{2j-1}^- &= a_j^+\sqrt{2S-a_j^+a_j}, \;\;\; S_{2j}^-=b_j^+\sqrt{2S-b_j^+b_j} \notag \\
  S_{2j-1}^+ &= \sqrt{2S-a_j^+a_j}a_j, \;\;\; S_{2j}^+=\sqrt{2S-b_j^+b_j}b_j \notag \\
  S_{2j-1}^z &= S-a_j^+a_j,\;\;\;  S_{2j}^z=S-b_j^+b_j
\end{align}
where operators $a_j$ and $b_j$ satisfy standard bosonic commutation relations,
\begin{gather}
  \left[a_j, a_l^+\right]=\delta_{jl}, \;\;\left[b_j, b_l^+\right]=\delta_{jl}, \;\; \left[a_j, b_l^+\right]=0
\end{gather}
Substituting these expressions into the spin Hamiltonian and keeping only quadratic in bosonic operators terms we arrive at the LSWT approximation.
The corresponding Hamiltonian has the following form:
\begin{align}
  \mathcal{H} &= E_\mathrm{FM} + \sum_{j=1}^{N/2}\left(K_1 a_j a_{j+1}^++K_1^* a_j^+ a_{j+1} + \right. \\
  & + K_2 a_j b_{j}^++K_2^* a_j^+ b_{j} + K_3 b_j a_{j+1}^++K_3^* b_j^+ a_{j+1}+ \\
  & \left. + (\mu-2SJ_1\Delta_1)  a_j^+ a_{j}+\mu  b_j^+ b_{j}\right), \notag
\end{align}
where $E_\mathrm{FM}$ is the energy of the fully polarized (ferromagnetic) ground state defined in Section~\ref{sec:2} and
\begin{align}
  K_a &= S(J_a+i D_a)=S\rho_a e^{i\phi_a}, \;\; a=1,2,3, \notag \\
  \mu &= B - S\left(J_2\Delta_2+J_3\Delta_3\right).
\end{align}
In order to diagonalize the Hamiltonian one should use Fourier transformations for the bosonic operators,
\begin{align}
  a_j &= \frac{1}{\sqrt{N/2}}\sum_{k}e^{-ijk} a_k, \\
  b_j &= \frac{1}{\sqrt{N/2}}\sum_{k}e^{-ijk} b_k, \notag\\
  \left[a_k, a_q^+\right] &= \delta_{kq}, \;\;\left[b_k, b_q^+\right]=\delta_{kq}, \;\; \left[a_k, b_q^+\right]=0, \notag
\end{align}
yielding
\begin{gather}
  \mathcal{H}=E_\mathrm{FM} + \sum_k\left(a_k^+, b_k^+\right) M_k
  \begin{pmatrix}
    a_k \\
    b_k
  \end{pmatrix} \\
  M_k =
  \begin{pmatrix}
    \mu-2SJ_1\Delta_1+2S\rho_1\cos(k+\phi_1), & K_2^*+K_3 e^{ik} \\
    K_2+K_3^*e^{-ik}, & \mu
  \end{pmatrix}
  \notag
\end{gather}
The final diagonalization of the Hamiltonian is then  performed by a Bogoliubov transformation to bosonic operators $c_k$ and $d_k$,
\begin{align}
  \begin{pmatrix}
    a_k \\
    b_k
  \end{pmatrix}
  & =
  \begin{pmatrix}
    \cos\theta_k & e^{i\varphi_k}\sin\theta_k \\
    -e^{-i\varphi_k}\sin\theta_k & \cos\theta_k
  \end{pmatrix}
  \begin{pmatrix}
    c_k \\
    d_k
  \end{pmatrix}, \notag \\
  \tan 2\theta_k &= \frac{\sqrt{\rho_2^2+\rho_3^2+2\rho_2\rho_3\cos\left(k+\phi_2+\phi_3\right)}}{\rho_1 \cos\left(k+\phi_1\right)-J_1\Delta_1}, \notag \\
  \tan\varphi_k &= \frac{\rho_3\sin\left(k+\phi_3\right)-\rho_2\sin\phi_2}{\rho_3\cos\left(k+\phi_3\right)-\rho_2\cos\phi_2},
\end{align}
leading to
\begin{gather}
  \mathcal{H}=E_\mathrm{FM}+\sum_k\left(\epsilon^+_k c_k^+c_k+\epsilon^-_k d_k^+d_k\right).
\end{gather}
Two branches of the spectrum of one-magnon excitations for the spin-$S$ sawtooth chain within the LSWT turn out to be identical to the exact one-magnon spectrum~\eqref{1m_sp_gen},
\begin{widetext}
\begin{gather}
  \label{eq:LSWT_ferro_dis}
  \varepsilon^{\pm}_1(k)=B-\Gamma+S\left[\rho_1\cos(k+\phi_1)\pm\sqrt{\left(J_1\Delta_1-\rho_1\cos(k+\phi_1)\right)^2+2\rho_2\rho_3\cos(k+\phi_2+\phi_3)+\rho_2^2+\rho_3^2}\right],
\end{gather}
\end{widetext}
The only difference is the sign of the phases in the arguments of the cosine functions, which do not substantially affect the flat-band constraints.
Thus the exact one-magnon spectrum and the LSWT approach yield equivalent results egarding their flat-band properties.
We expect that this remarkable principle is universal and model independent, because for any spin Hamiltonian admitting localized magnon states, a corresponding tight-binding model with localized excitations can be constructed using LSWT around the ferromagnetic classical ground state.

\section{Conclusion}

We have presented a detailed analysis of several classes of flat-band scenarios for the generalized model of the famous sawtooth chain with symmetric and DM couplings for each side of the chain triangles. We discussed the realization of the magnonic flat bands by tuning of the electric field for the models with KNB mechanism of magnetoelectricity.
For easier reference, the main results are summarized again in Appendix~\ref{app:summary}.
In general, despite reliable experimental evidence~\cite{Fe3+2, oliv2}, the model of sawtooth chain with three exchange couplings received little theoretical attention so far~\cite{J1J2J3}, nor were the corresponding conditions for a magnonic flat band studied.
We also extended our previous results~\cite{SC_KNB} on the $J_1$-$J_2$ sawtooth chain with the KNB mechanism for an in-plane electric field.

As the KNB mechanism establishes a geometry-dependent link between local spin configurations and the dielectric polarization of the bond between two lattice sites (spin-induced ferroelectricity)~\cite{KNB1, KNB2, Sol21, Sol25} in the form of DM interaction terms, our emergent effective spin model, which includes interactions between the KNB polarization and in-plane electric field, then also includes three constants of the DM interactions.
Generalizing the results of  Ref.~\cite{SC_KNB}, we discussed how the electric field can be used to induce the magnonic flat bands.
In Ref.~\onlinecite{SC_KNB} only two cases were considered, namely an electric field parallel to either the basal line or to the zigzag bond of the lattice.
Here we demonstrated that an equivalent electric field, e.g. direction and magnitude, can be identified for a range of values of \(D_1, D_2, D_3\).
While ground state properties of the sawtooth chain with DM interactions were considered before~\cite{hao11}, their role in the formation of flat bands was not discussed.

Our starting point was the set of general flat-band constraints derived from the exact one-magnon spectrum.
Formally, the constraints form a system of seven variables, $J_1, J_2, J_3, D_1, D_2, D_3$ and $\Delta_1$.
Interestingly, the coefficient of exchange anisotropy $\Delta_1$ from the basal line enters the constraints, while $\Delta_2$ and $\Delta_3$ do not.
In contrast to the well-known case of the $J_1$-$J_2$ sawtooth chain, where a flat band implies a linear relation between $J_1$ and $J_2$, here strongly non-linear equations appear,
which can be resolved in multiple ways depending on which group of parameters one considers as input.
We discussed several classes of solutions expressing pairs of symmetric and antisymmetric exchange couplings, (\(J_a, D_a\)), in terms of other parameters.
We remark that when  all three DM couplings are expressed in terms of $J_1, J_2, J_3$ and $\Delta_1$, some additional relations between the symmetric exchanges are still required.
The mathematical structure of our conditions is of geometrical interest, as the solutions in terms of $D_1, D_2$ and $D_3$ are obtained as points of a planar quartic curve, which helps in understanding the flat band manifold.
Moreover, we note that for a given set of symmetric exchange couplings there is an entire one-parameter family of the solutions $D_a(\alpha)$, $\alpha\in\mathbb{R}$.

Because the magnitude of the conventional spin-orbit interaction induced DM terms in solid magnets is usually at least one order of magnitude smaller than the corresponding symmetric exchange, our solutions of the flat-band constraint in terms of $D_a$ may be beyond physically realizable values of DM constants.
However, this limitation could be lifted if they are induced by an electric field via the KNB mechanism.
Finally we note that in artificial many-body strongly correlated systems, such as ultra-cold atoms in optical lattices, DM interaction can be modeled directly~\cite{DM_OL}.
Recent developments suggest that for magnets that are modeled in Rydberg atom quantum simulators~\cite{Ryd}, the DM constants can even be made larger than the corresponding symmetric exchange.
Regarding electric-field-induced flat bands we note that any
sawtooth-chain material with spiral spin order can be a candidate for
the experimental realization of electric-field control over the
formation of localized magnon states. For example, the multiferroic
sawtooth material BeCr$_2$O$_4$ is quite promising, for which multiple
magnetic transitions and at least two phases with noncollinear
magnetic structure were demonstrated~\cite{BeCr}. As previously
discussed in Ref.~\onlinecite{SC_KNB}, quantitative estimates of the
required electric field strength are highly material-dependent, as the
KNB polarization contains the constant $\gamma$ which depends on
quantum-chemical details of the bond between two magnetic
ions~\cite{KNB1, KNB2, Sol21, Sol25}.  Recent experimental
results~\cite{koc21, ukl25} indicate that the necessary electric field
strengths are feasible in principle, ranging from $10^{-1}$ to
10~MV/m, in order to control the chiral spin texture and to switch
between various spin states.

\section*{Acknowledgements}
We thank Sergej Flach for fruitful discussions and stimulating ideas.
V.O. is grateful to the Center for Electronic Correlations and Magnetism for the hospitality during his visits to the University of Augsburg;
he also acknowledges partial financial support from ANSEF (Grants No. PS-condmatth-2884 and PS-condmatth-3273) and CS RA MESCS (Grants No. 21AG-1C047 and 23AA-1C032).
V.O. and M.K. acknowledge partial funding by Deutsche Forschungsgemeinschaft (DFG, German Research Foundation) -- TRR360 -- 492547816.
This work was supported by the Institute for Basic Science, Project Code (Project No.~IBS-R024-D1).

\section*{Data availability statement}
The data that supports the findings of this article are available upon reasonable request from the authors.

\appendix

\section{Summary of the main flat-band solutions}
\label{app:summary}

As the variety of solutions of the general flat-band constraints (\ref{FB_pr}) is quite diverse,  here we summarize their main classes.
\subsubsection{$D_1=D_2=D_3=0$ case}
This case generalizes the know results for $J_1$-$J_2$-$J_2$ $XXZ$ sawtooth chain, $J_2^2=2 J_1^2\left(1+\Delta_1\right)$ to the  $J_1$-$J_2$-$J_3$ case:
\begin{eqnarray}
J_1^2\left(J_2^2+J_3^2+2\Delta_1 J_2 J_3\right)=J_2^2 J_3^2
\end{eqnarray}
which can be solved either with respect to $J_1$ or with respect to $J_2$ or $J_3$ giving
\begin{eqnarray}
J_1=\frac{\pm J_2 J_3}{\sqrt{J_2^2+J_3^2+2\Delta_1 J_2 J_3}},
\end{eqnarray}
or
\begin{eqnarray}
J_2=\frac{\pm J_1 J_3}{\sqrt{J_3^2-J_1^2\left(1-\Delta_1^2\right)}}
\end{eqnarray}
There is full symmetry with respect to interchange $J_2\leftrightarrow J_3$. Additional constraints necessary for the realization of this flat band scenario can be also obtained demanding the positivity of the expressions under square roots.

\subsubsection{DM-interaction-driven flat bands, solution with respect to the pair of interaction constants $\left(J_{2,3}, D_{2,3}\right)$}

\begin{align}
  && J_3=\frac{\pm\left(J_1J_2+D_1D_2\right)}{\sqrt{J_1^2\Delta_1^2-\rho_1^2+\rho_2^2}\mp J_1\Delta_1}, \\
  && D_3=\frac{\pm\left(J_2D_1-J_1D_2\right)}{\sqrt{J_1^2\Delta_1^2-\rho_1^2+\rho_2^2}\mp J_1\Delta_1}. \notag
\end{align}
Here, additional constraints restricting the possible range of the system parameters corresponding to the flat-band solution are given in Eqs. (\ref{FB_cond_0_pl})-(\ref{rho_2_range}).

\subsubsection{DM-interaction-driven flat bands, solutions in terms of $D_1, D_2$ and $D_3$}

It is possible to present the solution of the general flat-band condition in terms of DM parameters only, when the symmetric exchange constants with the corresponding axial anisotropy are considered as given, and $D_1$, $D_2$ and $D_3$ are expressed in terms of them. Interestingly, $D_2/J_2$ and $D_3/J_3$ in this case are the points of the plane quartic curve, while $D_1/J_1$ is found then according to the phase condition. 
\begin{align}
  D_1 &= \mp J_1\frac{2J^2\alpha\left(1+\alpha\right)\sqrt{\frac{1}{\alpha}+\frac{ W_{\mu}(\alpha)}{2J^2 \alpha^2}}}{ W_{\mu}(\alpha)},\\
  D_2 &= \pm J_2\sqrt{\frac{1}{\alpha}+\frac{ W_{\mu}(\alpha)}{2J^2 \alpha^2}}, \notag \\
  D_3 &= \pm J_3\alpha\sqrt{\frac{1}{\alpha}+\frac{ W_{\mu}(\alpha)}{2J^2\alpha^2}}, \notag
\end{align}
where
\begin{align}
  J &= \frac{J_2J_3}{J_1}, \\
  W_{\mu}(\alpha) &= R(\alpha)+\mu\sqrt{R^2(\alpha)+4 J^2 \alpha (1+\alpha)\left(J_2^2+\alpha J_3^2\right)}, \notag \\
  R(\alpha) &= J_2^2+\alpha^2 J_3^2-2\alpha \xi \Delta_1 J_2 J_3, \notag \\
  \mu &= \left\{-1,1\right\}. \notag
\end{align}
Important particular case corresponding to the case of uniform symmetric exchange, $J_1=J_2=J_3=1$ and $\Delta_1=\pm 1$ is given by
\begin{align}
  D_1 &= \mp \frac{2\alpha\left(1+\alpha\right)\sqrt{\frac{1}{\alpha}+\frac{ X_{\mu}(\alpha)}{2 \alpha^2}}}{ X_{\mu}(\alpha)},\\
  D_2 &= \pm \sqrt{\frac{1}{\alpha}+\frac{ X_{\mu}(\alpha)}{2 \alpha^2}}, \notag \\
  D_3 &= \pm \alpha\sqrt{\frac{1}{\alpha}+\frac{ X_{\mu}(\alpha)}{2\alpha^2}}, \notag
\end{align}
This case can be directly linked to the electric-field induced flat-band in the model with KNB mechanism~\cite{SC_KNB}.

\section{Derivation of Eq.~(\ref{sigma})}
\label{app:derivation}

Here we discuss details of the derivation of Eq.~(\ref{sigma}).  The
correct ordering of the one-magnon branches implies the following
inequality:
\begin{gather}
  \label{eq:ordering}
  \sigma \rho_2 \rho_3>\rho_1\left(J_1\Delta_1 +\rho_1\right).
\end{gather}
An intriguing feature of the band order is its connection to the phase-shift parameter $\sigma$.
We substitute the second equation of the flat-band constraint into inequality~\eqref{eq:ordering},
\begin{gather}
  \label{eq:ordering2}
  \sigma\left(\frac{\rho_2\rho_3}{\rho_1}+\rho_1\left[\frac{\rho_2}{\rho_3}+\frac{\rho_3}{\rho_2}\right]\right)>2\rho_1.
\end{gather}
Since all $\rho_a$ are positive, this inequality cannot be satisfied for $\sigma=-1$.
On the other hand, for $\sigma=1$ the l.h.s is always greater than the r.h.s..
Therefore to use this criterion along with the flat band constraints~\eqref{FB_pr} we need to express $\sigma$ in terms of the exchange parameters.
This can be done by taking the tangent out of the first equation in Eqs.~\eqref{FB_pr}.
With the aid of the trigonometric identity,
\begin{align}
  \arctan x+\arctan y &= \arctan\frac{x+y}{1-xy}\\
  &+ \pi\,\text{sign}\left(x+y\right)\Theta\left(xy-1\right), \notag
\end{align}
where $\Theta(x)$ is the Heaviside step function, the additional constraint for the existence of the lower flat band is
\begin{gather}
  \frac{D_2D_3}{J_2J_3} < 1.
\end{gather}
Thus,
\begin{align}
  \sigma & =\text{sign}\left(1+\frac{D_1D_2}{J_1J_2}\right)\equiv \text{sign}\left(1+\frac{D_1D_3}{J_1J_3}\right) \notag \\
  & \equiv \text{sign}\left(1-\frac{D_2D_3}{J_2J_3}\right)=1.
\end{align}

\providecommand{\urlprefix}{}%

\begin{thebibliography}{109}%
\makeatletter
\providecommand \@ifxundefined [1]{%
 \@ifx{#1\undefined}
}%
\providecommand \@ifnum [1]{%
 \ifnum #1\expandafter \@firstoftwo
 \else \expandafter \@secondoftwo
 \fi
}%
\providecommand \@ifx [1]{%
 \ifx #1\expandafter \@firstoftwo
 \else \expandafter \@secondoftwo
 \fi
}%
\providecommand \natexlab [1]{#1}%
\providecommand \enquote  [1]{``#1''}%
\providecommand \bibnamefont  [1]{#1}%
\providecommand \bibfnamefont [1]{#1}%
\providecommand \citenamefont [1]{#1}%
\providecommand \href@noop [0]{\@secondoftwo}%
\providecommand \href [0]{\begingroup \@sanitize@url \@href}%
\providecommand \@href[1]{\@@startlink{#1}\@@href}%
\providecommand \@@href[1]{\endgroup#1\@@endlink}%
\providecommand \@sanitize@url [0]{\catcode `\\12\catcode `\$12\catcode
  `\&12\catcode `\#12\catcode `\^12\catcode `\_12\catcode `\%12\relax}%
\providecommand \@@startlink[1]{}%
\providecommand \@@endlink[0]{}%
\providecommand \url  [0]{\begingroup\@sanitize@url \@url }%
\providecommand \@url [1]{\endgroup\@href {#1}{\urlprefix }}%
\providecommand \urlprefix  [0]{URL }%
\providecommand \Eprint [0]{\href }%
\providecommand \doibase [0]{https://doi.org/}%
\providecommand \selectlanguage [0]{\@gobble}%
\providecommand \bibinfo  [0]{\@secondoftwo}%
\providecommand \bibfield  [0]{\@secondoftwo}%
\providecommand \translation [1]{[#1]}%
\providecommand \BibitemOpen [0]{}%
\providecommand \bibitemStop [0]{}%
\providecommand \bibitemNoStop [0]{.\EOS\space}%
\providecommand \EOS [0]{\spacefactor3000\relax}%
\providecommand \BibitemShut  [1]{\csname bibitem#1\endcsname}%
\let\auto@bib@innerbib\@empty
\bibitem [{\citenamefont {Lacroix}\ \emph {et~al.}(2011)\citenamefont
  {Lacroix}, \citenamefont {Mendels},\ and\ \citenamefont {Mila}}]{fru1}%
  \BibitemOpen
  \bibinfo {editor} {\bibfnamefont {C.}~\bibnamefont {Lacroix}}, \bibinfo
  {editor} {\bibfnamefont {P.}~\bibnamefont {Mendels}},\ and\ \bibinfo {editor}
  {\bibfnamefont {F.}~\bibnamefont {Mila}},\ eds.,\ \href
  {https://doi.org/10.1007/978-3-642-10589-0} {\emph {\bibinfo {title}
  {Introduction to Frustrated Magnetism: {Materials}, Experiments, Theory}}}\
  (\bibinfo  {publisher} {Springer},\ \bibinfo {address} {Berlin, Heidelberg},\
  \bibinfo {year} {2011})\BibitemShut {NoStop}%
\bibitem [{\citenamefont {Kubo}(1993)}]{kubo93}%
  \BibitemOpen
  \bibfield  {author} {\bibinfo {author} {\bibfnamefont {K.}~\bibnamefont
  {Kubo}},\ }\bibfield  {title} {\emph {\bibinfo {title} {Excited states and
  the thermodynamics of a fully frustrated quantum spin chain}},\ }\href
  {https://doi.org/10.1103/PhysRevB.48.10552} {\bibfield  {journal} {\bibinfo
  {journal} {Phys. Rev. B}\ }\textbf {\bibinfo {volume} {48}},\ \bibinfo
  {pages} {10552} (\bibinfo {year} {1993})}\BibitemShut {NoStop}%
\bibitem [{\citenamefont {Nakamura}\ and\ \citenamefont {Kubo}(1996)}]{nak96}%
  \BibitemOpen
  \bibfield  {author} {\bibinfo {author} {\bibfnamefont {T.}~\bibnamefont
  {Nakamura}}\ and\ \bibinfo {author} {\bibfnamefont {K.}~\bibnamefont
  {Kubo}},\ }\bibfield  {title} {\emph {\bibinfo {title} {Elementary
  excitations in the {$\Delta$} chain}},\ }\href
  {https://doi.org/10.1103/PhysRevB.53.6393} {\bibfield  {journal} {\bibinfo
  {journal} {Phys. Rev. B}\ }\textbf {\bibinfo {volume} {53}},\ \bibinfo
  {pages} {6393} (\bibinfo {year} {1996})}\BibitemShut {NoStop}%
\bibitem [{\citenamefont {Sen}\ \emph {et~al.}(1996)\citenamefont {Sen},
  \citenamefont {Shastry}, \citenamefont {Walsted},\ and\ \citenamefont
  {Cava}}]{sen96}%
  \BibitemOpen
  \bibfield  {author} {\bibinfo {author} {\bibfnamefont {D.}~\bibnamefont
  {Sen}}, \bibinfo {author} {\bibfnamefont {B.~S.}\ \bibnamefont {Shastry}},
  \bibinfo {author} {\bibfnamefont {R.~E.}\ \bibnamefont {Walsted}},\ and\
  \bibinfo {author} {\bibfnamefont {R.~J.}\ \bibnamefont {Cava}},\ }\bibfield
  {title} {\emph {\bibinfo {title} {Quantum solitons in the sawtooth
  lattice}},\ }\href {https://doi.org/10.1103/PhysRevB.53.6401} {\bibfield
  {journal} {\bibinfo  {journal} {Phys. Rev. B}\ }\textbf {\bibinfo {volume}
  {53}},\ \bibinfo {pages} {6401} (\bibinfo {year} {1996})}\BibitemShut
  {NoStop}%
\bibitem [{\citenamefont {Hao}\ \emph {et~al.}(2011)\citenamefont {Hao},
  \citenamefont {Wan}, \citenamefont {Rousochatzakis}, \citenamefont
  {Wildeboer}, \citenamefont {Seidel}, \citenamefont {Mila},\ and\
  \citenamefont {Tchernyshyov}}]{hao11}%
  \BibitemOpen
  \bibfield  {author} {\bibinfo {author} {\bibfnamefont {Z.}~\bibnamefont
  {Hao}}, \bibinfo {author} {\bibfnamefont {Y.}~\bibnamefont {Wan}}, \bibinfo
  {author} {\bibfnamefont {I.}~\bibnamefont {Rousochatzakis}}, \bibinfo
  {author} {\bibfnamefont {J.}~\bibnamefont {Wildeboer}}, \bibinfo {author}
  {\bibfnamefont {A.}~\bibnamefont {Seidel}}, \bibinfo {author} {\bibfnamefont
  {F.}~\bibnamefont {Mila}},\ and\ \bibinfo {author} {\bibfnamefont
  {O.}~\bibnamefont {Tchernyshyov}},\ }\bibfield  {title} {\emph {\bibinfo
  {title} {Destruction of valence-bond order in a {$S$ $=$ $1/2$} sawtooth
  chain with a {Dzyaloshinskii-Moriya} term}},\ }\href
  {https://doi.org/10.1103/PhysRevB.84.094452} {\bibfield  {journal} {\bibinfo
  {journal} {Phys. Rev. B}\ }\textbf {\bibinfo {volume} {84}},\ \bibinfo
  {pages} {094452} (\bibinfo {year} {2011})}\BibitemShut {NoStop}%
\bibitem [{\citenamefont {Blundell}\ and\ \citenamefont
  {Núñez-Regueiro}(2003)}]{blu03}%
  \BibitemOpen
  \bibfield  {author} {\bibinfo {author} {\bibfnamefont {S.~A.}\ \bibnamefont
  {Blundell}}\ and\ \bibinfo {author} {\bibfnamefont {M.~D.}\ \bibnamefont
  {Núñez-Regueiro}},\ }\bibfield  {title} {\emph {\bibinfo {title} {Quantum
  topological excitations: from the sawtooth lattice to the {Heisenberg}
  chain}},\ }\href {https://doi.org/10.1140/epjb/e2003-00054-2} {\bibfield
  {journal} {\bibinfo  {journal} {Eur. Phys. J. B}\ }\textbf {\bibinfo {volume}
  {31}},\ \bibinfo {pages} {453} (\bibinfo {year} {2003})}\BibitemShut
  {NoStop}%
\bibitem [{\citenamefont {Tonegawa}\ and\ \citenamefont
  {Kaburagi}(2004)}]{ton04}%
  \BibitemOpen
  \bibfield  {author} {\bibinfo {author} {\bibfnamefont {T.}~\bibnamefont
  {Tonegawa}}\ and\ \bibinfo {author} {\bibfnamefont {M.}~\bibnamefont
  {Kaburagi}},\ }\bibfield  {title} {\emph {\bibinfo {title} {Ground-state
  properties of an {$S$ $=$ $1/2$} {$\Delta$}-chain with ferro- and
  antiferromagnetic interactions}},\ }\href
  {https://doi.org/10.1016/j.jmmm.2003.11.367} {\bibfield  {journal} {\bibinfo
  {journal} {J. Magn. Magn. Mater.}\ }\textbf {\bibinfo {volume} {272-276}},\
  \bibinfo {pages} {898} (\bibinfo {year} {2004})}\BibitemShut {NoStop}%
\bibitem [{\citenamefont {Kaburagi}\ \emph {et~al.}(2005)\citenamefont
  {Kaburagi}, \citenamefont {Tonegawa},\ and\ \citenamefont {Kang}}]{kab05}%
  \BibitemOpen
  \bibfield  {author} {\bibinfo {author} {\bibfnamefont {M.}~\bibnamefont
  {Kaburagi}}, \bibinfo {author} {\bibfnamefont {T.}~\bibnamefont {Tonegawa}},\
  and\ \bibinfo {author} {\bibfnamefont {M.}~\bibnamefont {Kang}},\ }\bibfield
  {title} {\emph {\bibinfo {title} {Ground state phase diagrams of an
  anisotropic spin-1/2 {$\Delta$}-chain with ferro- and antiferromagnetic
  interactions}},\ }\href {https://doi.org/10.1063/1.1851893} {\bibfield
  {journal} {\bibinfo  {journal} {J. Appl. Phys.}\ }\textbf {\bibinfo {volume}
  {97}},\ \bibinfo {pages} {10B306} (\bibinfo {year} {2005})}\BibitemShut
  {NoStop}%
\bibitem [{\citenamefont {Jiang}\ \emph {et~al.}(2015)\citenamefont {Jiang},
  \citenamefont {Liu}, \citenamefont {Tang}, \citenamefont {Yang},\ and\
  \citenamefont {Sheng}}]{jia15}%
  \BibitemOpen
  \bibfield  {author} {\bibinfo {author} {\bibfnamefont {J.-J.}\ \bibnamefont
  {Jiang}}, \bibinfo {author} {\bibfnamefont {Y.-J.}\ \bibnamefont {Liu}},
  \bibinfo {author} {\bibfnamefont {F.}~\bibnamefont {Tang}}, \bibinfo {author}
  {\bibfnamefont {C.-H.}\ \bibnamefont {Yang}},\ and\ \bibinfo {author}
  {\bibfnamefont {Y.-B.}\ \bibnamefont {Sheng}},\ }\bibfield  {title} {\emph
  {\bibinfo {title} {Analytical and numerical studies of the one-dimensional
  sawtooth chain}},\ }\href {https://doi.org/10.1016/j.physb.2015.01.036}
  {\bibfield  {journal} {\bibinfo  {journal} {Physica B}\ }\textbf {\bibinfo
  {volume} {463}},\ \bibinfo {pages} {30} (\bibinfo {year} {2015})}\BibitemShut
  {NoStop}%
\bibitem [{\citenamefont {Paul}\ and\ \citenamefont {Ghosh}(2019)}]{paul19}%
  \BibitemOpen
  \bibfield  {author} {\bibinfo {author} {\bibfnamefont {S.}~\bibnamefont
  {Paul}}\ and\ \bibinfo {author} {\bibfnamefont {A.~K.}\ \bibnamefont
  {Ghosh}},\ }\bibfield  {title} {\emph {\bibinfo {title} {Magnetic and
  thermodynamic properties of a frustrated sawtooth chain}},\ }\href
  {https://doi.org/10.1140/epjb/e2019-100238-3} {\bibfield  {journal} {\bibinfo
   {journal} {Eur. Phys. J. B}\ }\textbf {\bibinfo {volume} {92}},\ \bibinfo
  {pages} {40} (\bibinfo {year} {2019})}\BibitemShut {NoStop}%
\bibitem [{\citenamefont {Krivnov}\ \emph {et~al.}(2014)\citenamefont
  {Krivnov}, \citenamefont {Dmitriev}, \citenamefont {Nishimoto}, \citenamefont
  {Drechsler},\ and\ \citenamefont {Richter}}]{FAF1}%
  \BibitemOpen
  \bibfield  {author} {\bibinfo {author} {\bibfnamefont {V.~Y.}\ \bibnamefont
  {Krivnov}}, \bibinfo {author} {\bibfnamefont {D.~V.}\ \bibnamefont
  {Dmitriev}}, \bibinfo {author} {\bibfnamefont {S.}~\bibnamefont {Nishimoto}},
  \bibinfo {author} {\bibfnamefont {S.-L.}\ \bibnamefont {Drechsler}},\ and\
  \bibinfo {author} {\bibfnamefont {J.}~\bibnamefont {Richter}},\ }\bibfield
  {title} {\emph {\bibinfo {title} {Ground state of the frustrated
  ferromagnetic spin-1/2 chain}},\ }\href
  {https://doi.org/10.1103/PhysRevB.90.014441} {\bibfield  {journal} {\bibinfo
  {journal} {Phys. Rev. B}\ }\textbf {\bibinfo {volume} {90}},\ \bibinfo
  {pages} {014441} (\bibinfo {year} {2014})}\BibitemShut {NoStop}%
\bibitem [{\citenamefont {Dmitriev}\ and\ \citenamefont
  {Krivnov}(2015)}]{FAF1a}%
  \BibitemOpen
  \bibfield  {author} {\bibinfo {author} {\bibfnamefont {D.~V.}\ \bibnamefont
  {Dmitriev}}\ and\ \bibinfo {author} {\bibfnamefont {V.~Y.}\ \bibnamefont
  {Krivnov}},\ }\bibfield  {title} {\emph {\bibinfo {title} {Thermodynamics of
  a frustrated ferromagnetic spin chain}},\ }\href
  {https://doi.org/10.1103/PhysRevB.92.184422} {\bibfield  {journal} {\bibinfo
  {journal} {Phys. Rev. B}\ }\textbf {\bibinfo {volume} {92}},\ \bibinfo
  {pages} {184422} (\bibinfo {year} {2015})}\BibitemShut {NoStop}%
\bibitem [{\citenamefont {Dmitriev}\ and\ \citenamefont
  {Krivnov}(2016{\natexlab{a}})}]{FAF2}%
  \BibitemOpen
  \bibfield  {author} {\bibinfo {author} {\bibfnamefont {D.~V.}\ \bibnamefont
  {Dmitriev}}\ and\ \bibinfo {author} {\bibfnamefont {V.~Y.}\ \bibnamefont
  {Krivnov}},\ }\bibfield  {title} {\emph {\bibinfo {title} {Low-temperature
  thermodynamics of a frustrated ferromagnetic spin chain}},\ }\href
  {https://doi.org/10.1088/0953-8984/28/50/506002} {\bibfield  {journal}
  {\bibinfo  {journal} {J. Phys.: Condens. Matter}\ }\textbf {\bibinfo {volume}
  {28}},\ \bibinfo {pages} {506002} (\bibinfo {year}
  {2016}{\natexlab{a}})}\BibitemShut {NoStop}%
\bibitem [{\citenamefont {Krivnov}\ \emph {et~al.}(2019)\citenamefont
  {Krivnov}, \citenamefont {Dmitriev}, \citenamefont {Richter},\ and\
  \citenamefont {Schnack}}]{FAF3}%
  \BibitemOpen
  \bibfield  {author} {\bibinfo {author} {\bibfnamefont {V.~Y.}\ \bibnamefont
  {Krivnov}}, \bibinfo {author} {\bibfnamefont {D.~V.}\ \bibnamefont
  {Dmitriev}}, \bibinfo {author} {\bibfnamefont {J.}~\bibnamefont {Richter}},\
  and\ \bibinfo {author} {\bibfnamefont {J.}~\bibnamefont {Schnack}},\
  }\bibfield  {title} {\emph {\bibinfo {title} {Magnetization process of the
  frustrated ferromagnetic spin chain}},\ }\href
  {https://doi.org/10.1103/PhysRevB.99.094410} {\bibfield  {journal} {\bibinfo
  {journal} {Phys. Rev. B}\ }\textbf {\bibinfo {volume} {99}},\ \bibinfo
  {pages} {094410} (\bibinfo {year} {2019})}\BibitemShut {NoStop}%
\bibitem [{\citenamefont {Dmitriev}\ \emph {et~al.}(2020)\citenamefont
  {Dmitriev}, \citenamefont {Krivnov}, \citenamefont {Schnack},\ and\
  \citenamefont {Richter}}]{FAF_field}%
  \BibitemOpen
  \bibfield  {author} {\bibinfo {author} {\bibfnamefont {D.~V.}\ \bibnamefont
  {Dmitriev}}, \bibinfo {author} {\bibfnamefont {V.~Y.}\ \bibnamefont
  {Krivnov}}, \bibinfo {author} {\bibfnamefont {J.}~\bibnamefont {Schnack}},\
  and\ \bibinfo {author} {\bibfnamefont {J.}~\bibnamefont {Richter}},\
  }\bibfield  {title} {\emph {\bibinfo {title} {Magnetic field effects in
  frustrated ferromagnetic spin chains}},\ }\href
  {https://doi.org/10.1103/PhysRevB.101.054427} {\bibfield  {journal} {\bibinfo
   {journal} {Phys. Rev. B}\ }\textbf {\bibinfo {volume} {101}},\ \bibinfo
  {pages} {054427} (\bibinfo {year} {2020})}\BibitemShut {NoStop}%
\bibitem [{\citenamefont {Rausch}\ and\ \citenamefont
  {Karrasch}(2025)}]{rau25}%
  \BibitemOpen
  \bibfield  {author} {\bibinfo {author} {\bibfnamefont {R.}~\bibnamefont
  {Rausch}}\ and\ \bibinfo {author} {\bibfnamefont {C.}~\bibnamefont
  {Karrasch}},\ }\bibfield  {title} {\emph {\bibinfo {title} {Noncollinear
  phase of the antiferromagnetic sawtooth chain}},\ }\href
  {https://doi.org/10.1103/PhysRevB.111.045154} {\bibfield  {journal} {\bibinfo
   {journal} {Phys. Rev. B}\ }\textbf {\bibinfo {volume} {111}},\ \bibinfo
  {pages} {045154} (\bibinfo {year} {2025})}\BibitemShut {NoStop}%
\bibitem [{\citenamefont {Rausch}\ \emph {et~al.}(2023)\citenamefont {Rausch},
  \citenamefont {Peschke}, \citenamefont {Plorin}, \citenamefont {Schnack},\
  and\ \citenamefont {Karrasch}}]{schnack23a}%
  \BibitemOpen
  \bibfield  {author} {\bibinfo {author} {\bibfnamefont {R.}~\bibnamefont
  {Rausch}}, \bibinfo {author} {\bibfnamefont {M.}~\bibnamefont {Peschke}},
  \bibinfo {author} {\bibfnamefont {C.}~\bibnamefont {Plorin}}, \bibinfo
  {author} {\bibfnamefont {J.}~\bibnamefont {Schnack}},\ and\ \bibinfo {author}
  {\bibfnamefont {C.}~\bibnamefont {Karrasch}},\ }\bibfield  {title} {\emph
  {\bibinfo {title} {Quantum spin spiral ground state of the ferrimagnetic
  sawtooth chain}},\ }\href {https://doi.org/10.21468/SciPostPhys.14.3.052}
  {\bibfield  {journal} {\bibinfo  {journal} {SciPost Phys.}\ }\textbf
  {\bibinfo {volume} {14}},\ \bibinfo {pages} {052} (\bibinfo {year}
  {2023})}\BibitemShut {NoStop}%
\bibitem [{\citenamefont {Ohanyan}(2009)}]{oha09}%
  \BibitemOpen
  \bibfield  {author} {\bibinfo {author} {\bibfnamefont {V.}~\bibnamefont
  {Ohanyan}},\ }\bibfield  {title} {\emph {\bibinfo {title} {Antiferromagnetic
  sawtooth chain with {Heisenberg} and {Ising} bonds}},\ }\href
  {https://doi.org/10.5488/CMP.12.3.343} {\bibfield  {journal} {\bibinfo
  {journal} {Condens. Matter Phys.}\ }\textbf {\bibinfo {volume} {12}},\
  \bibinfo {pages} {343} (\bibinfo {year} {2009})}\BibitemShut {NoStop}%
\bibitem [{\citenamefont {Bellucci}\ and\ \citenamefont
  {Ohanyan}(2010)}]{bel10}%
  \BibitemOpen
  \bibfield  {author} {\bibinfo {author} {\bibfnamefont {S.}~\bibnamefont
  {Bellucci}}\ and\ \bibinfo {author} {\bibfnamefont {V.}~\bibnamefont
  {Ohanyan}},\ }\bibfield  {title} {\emph {\bibinfo {title} {Lattice
  distortions in a sawtooth chain with {Heisenberg} and {Ising} bonds}},\
  }\href {https://doi.org/10.1140/epjb/e2010-00146-x} {\bibfield  {journal}
  {\bibinfo  {journal} {Eur. Phys. J. B}\ }\textbf {\bibinfo {volume} {75}},\
  \bibinfo {pages} {531} (\bibinfo {year} {2010})}\BibitemShut {NoStop}%
\bibitem [{\citenamefont {Bellucci}\ and\ \citenamefont
  {Ohanyan}(2013)}]{bel13}%
  \BibitemOpen
  \bibfield  {author} {\bibinfo {author} {\bibfnamefont {S.}~\bibnamefont
  {Bellucci}}\ and\ \bibinfo {author} {\bibfnamefont {V.}~\bibnamefont
  {Ohanyan}},\ }\bibfield  {title} {\emph {\bibinfo {title} {Magnetocaloric
  effect in a sawtooth chain with competing interactions}},\ }\href
  {https://doi.org/10.1140/epjb/e2013-40466-8} {\bibfield  {journal} {\bibinfo
  {journal} {Eur. Phys. J. B}\ }\textbf {\bibinfo {volume} {86}},\ \bibinfo
  {pages} {446} (\bibinfo {year} {2013})}\BibitemShut {NoStop}%
\bibitem [{\citenamefont {Schulenburg}\ \emph {et~al.}(2002)\citenamefont
  {Schulenburg}, \citenamefont {Honecker}, \citenamefont {Schnack},
  \citenamefont {Richter},\ and\ \citenamefont {Schmidt}}]{schul02}%
  \BibitemOpen
  \bibfield  {author} {\bibinfo {author} {\bibfnamefont {J.}~\bibnamefont
  {Schulenburg}}, \bibinfo {author} {\bibfnamefont {A.}~\bibnamefont
  {Honecker}}, \bibinfo {author} {\bibfnamefont {J.}~\bibnamefont {Schnack}},
  \bibinfo {author} {\bibfnamefont {J.}~\bibnamefont {Richter}},\ and\ \bibinfo
  {author} {\bibfnamefont {H.-J.}\ \bibnamefont {Schmidt}},\ }\bibfield
  {title} {\emph {\bibinfo {title} {Macroscopic magnetization jumps due to
  independent magnons in frustrated quantum spin lattices}},\ }\href
  {https://doi.org/10.1103/PhysRevLett.88.167207} {\bibfield  {journal}
  {\bibinfo  {journal} {Phys. Rev. Lett.}\ }\textbf {\bibinfo {volume} {88}},\
  \bibinfo {pages} {167207} (\bibinfo {year} {2002})}\BibitemShut {NoStop}%
\bibitem [{\citenamefont {Richter}\ \emph
  {et~al.}(2004{\natexlab{a}})\citenamefont {Richter}, \citenamefont
  {Schulenburg}, \citenamefont {Honecker}, \citenamefont {Schnack},\ and\
  \citenamefont {Schmidt}}]{rich04}%
  \BibitemOpen
  \bibfield  {author} {\bibinfo {author} {\bibfnamefont {J.}~\bibnamefont
  {Richter}}, \bibinfo {author} {\bibfnamefont {J.}~\bibnamefont
  {Schulenburg}}, \bibinfo {author} {\bibfnamefont {A.}~\bibnamefont
  {Honecker}}, \bibinfo {author} {\bibfnamefont {J.}~\bibnamefont {Schnack}},\
  and\ \bibinfo {author} {\bibfnamefont {H.-J.}\ \bibnamefont {Schmidt}},\
  }\bibfield  {title} {\emph {\bibinfo {title} {Exact eigenstates and
  macroscopic magnetization jumps in strongly frustrated spin lattices}},\
  }\href {https://doi.org/10.1088/0953-8984/16/11/010} {\bibfield  {journal}
  {\bibinfo  {journal} {J. Phys.: Condens. Matter}\ }\textbf {\bibinfo {volume}
  {16}},\ \bibinfo {pages} {S779} (\bibinfo {year}
  {2004}{\natexlab{a}})}\BibitemShut {NoStop}%
\bibitem [{\citenamefont {Richter}(2005)}]{rich05}%
  \BibitemOpen
  \bibfield  {author} {\bibinfo {author} {\bibfnamefont {J.}~\bibnamefont
  {Richter}},\ }\bibfield  {title} {\emph {\bibinfo {title} {Localized-magnon
  states in strongly frustrated quantum spin lattices}},\ }\href
  {https://doi.org/10.1063/1.2008130} {\bibfield  {journal} {\bibinfo
  {journal} {Low Temp. Phys.}\ }\textbf {\bibinfo {volume} {31}},\ \bibinfo
  {pages} {695} (\bibinfo {year} {2005})}\BibitemShut {NoStop}%
\bibitem [{\citenamefont {Zhitomirsky}\ and\ \citenamefont
  {Tsunetsugu}(2004)}]{zhi04}%
  \BibitemOpen
  \bibfield  {author} {\bibinfo {author} {\bibfnamefont {M.~E.}\ \bibnamefont
  {Zhitomirsky}}\ and\ \bibinfo {author} {\bibfnamefont {H.}~\bibnamefont
  {Tsunetsugu}},\ }\bibfield  {title} {\emph {\bibinfo {title} {Exact
  low-temperature behavior of a kagome antiferromagnet at high fields}},\
  }\href {https://doi.org/10.1103/PhysRevB.70.100403} {\bibfield  {journal}
  {\bibinfo  {journal} {Phys. Rev. B}\ }\textbf {\bibinfo {volume} {70}},\
  \bibinfo {pages} {100403(R)} (\bibinfo {year} {2004})}\BibitemShut {NoStop}%
\bibitem [{\citenamefont {Derzhko}\ and\ \citenamefont
  {Richter}(2004)}]{der04}%
  \BibitemOpen
  \bibfield  {author} {\bibinfo {author} {\bibfnamefont {O.}~\bibnamefont
  {Derzhko}}\ and\ \bibinfo {author} {\bibfnamefont {J.}~\bibnamefont
  {Richter}},\ }\bibfield  {title} {\emph {\bibinfo {title} {Finite
  low-temperature entropy of some strongly frustrated quantum spin lattices in
  a magnetic field}},\ }\href {https://doi.org/10.1103/PhysRevB.70.104415}
  {\bibfield  {journal} {\bibinfo  {journal} {Phys. Rev. B}\ }\textbf {\bibinfo
  {volume} {70}},\ \bibinfo {pages} {104415} (\bibinfo {year}
  {2004})}\BibitemShut {NoStop}%
\bibitem [{\citenamefont {Derzhko}\ and\ \citenamefont
  {Richter}(2006)}]{der06}%
  \BibitemOpen
  \bibfield  {author} {\bibinfo {author} {\bibfnamefont {O.}~\bibnamefont
  {Derzhko}}\ and\ \bibinfo {author} {\bibfnamefont {J.}~\bibnamefont
  {Richter}},\ }\bibfield  {title} {\emph {\bibinfo {title} {Universal
  low-temperature behavior of frustrated quantum antiferromagnets in the
  vicinity of the saturation field}},\ }\href
  {https://doi.org/10.1140/epjb/e2006-00104-1} {\bibfield  {journal} {\bibinfo
  {journal} {Eur. Phys. J. B}\ }\textbf {\bibinfo {volume} {52}},\ \bibinfo
  {pages} {23} (\bibinfo {year} {2006})}\BibitemShut {NoStop}%
\bibitem [{\citenamefont {Derzhko}\ \emph {et~al.}(2007)\citenamefont
  {Derzhko}, \citenamefont {Richter}, \citenamefont {Honecker},\ and\
  \citenamefont {Schmidt}}]{der07}%
  \BibitemOpen
  \bibfield  {author} {\bibinfo {author} {\bibfnamefont {O.}~\bibnamefont
  {Derzhko}}, \bibinfo {author} {\bibfnamefont {J.}~\bibnamefont {Richter}},
  \bibinfo {author} {\bibfnamefont {A.}~\bibnamefont {Honecker}},\ and\
  \bibinfo {author} {\bibfnamefont {H.-J.}\ \bibnamefont {Schmidt}},\
  }\bibfield  {title} {\emph {\bibinfo {title} {Universal properties of highly
  frustrated quantum magnets in strong magnetic fields}},\ }\href
  {https://doi.org/10.1063/1.2780165} {\bibfield  {journal} {\bibinfo
  {journal} {Low Temp. Phys.}\ }\textbf {\bibinfo {volume} {33}},\ \bibinfo
  {pages} {745} (\bibinfo {year} {2007})}\BibitemShut {NoStop}%
\bibitem [{\citenamefont {Richter}\ \emph {et~al.}(2008)\citenamefont
  {Richter}, \citenamefont {Derzhko},\ and\ \citenamefont {Honecker}}]{rich08}%
  \BibitemOpen
  \bibfield  {author} {\bibinfo {author} {\bibfnamefont {J.}~\bibnamefont
  {Richter}}, \bibinfo {author} {\bibfnamefont {O.}~\bibnamefont {Derzhko}},\
  and\ \bibinfo {author} {\bibfnamefont {A.}~\bibnamefont {Honecker}},\
  }\bibfield  {title} {\emph {\bibinfo {title} {Thermodynamics of highly
  frustrated spin models}},\ }\href {https://doi.org/10.1142/S0217979208048023}
  {\bibfield  {journal} {\bibinfo  {journal} {Int. J. Mod. Phys. B}\ }\textbf
  {\bibinfo {volume} {22}},\ \bibinfo {pages} {4418} (\bibinfo {year}
  {2008})}\BibitemShut {NoStop}%
\bibitem [{\citenamefont {Derzhko}\ \emph {et~al.}(2015)\citenamefont
  {Derzhko}, \citenamefont {Richter},\ and\ \citenamefont
  {Maksymenko}}]{der15}%
  \BibitemOpen
  \bibfield  {author} {\bibinfo {author} {\bibfnamefont {O.}~\bibnamefont
  {Derzhko}}, \bibinfo {author} {\bibfnamefont {J.}~\bibnamefont {Richter}},\
  and\ \bibinfo {author} {\bibfnamefont {M.}~\bibnamefont {Maksymenko}},\
  }\bibfield  {title} {\emph {\bibinfo {title} {Strongly correlated flat-band
  systems: {The} route from {Heisenberg} spins to {Hubbard} electrons}},\
  }\href {https://doi.org/10.1142/S0217979215300078} {\bibfield  {journal}
  {\bibinfo  {journal} {Int. J. Mod. Phys. B}\ }\textbf {\bibinfo {volume}
  {29}},\ \bibinfo {pages} {1530007} (\bibinfo {year} {2015})}\BibitemShut
  {NoStop}%
\bibitem [{\citenamefont {Metavitsiadis}\ \emph {et~al.}(2020)\citenamefont
  {Metavitsiadis}, \citenamefont {Psaroudaki},\ and\ \citenamefont
  {Brenig}}]{met20}%
  \BibitemOpen
  \bibfield  {author} {\bibinfo {author} {\bibfnamefont {A.}~\bibnamefont
  {Metavitsiadis}}, \bibinfo {author} {\bibfnamefont {C.}~\bibnamefont
  {Psaroudaki}},\ and\ \bibinfo {author} {\bibfnamefont {W.}~\bibnamefont
  {Brenig}},\ }\bibfield  {title} {\emph {\bibinfo {title} {Enhancement of
  magnetization plateaus in low-dimensional spin systems}},\ }\href
  {https://doi.org/10.1103/PhysRevB.101.235143} {\bibfield  {journal} {\bibinfo
   {journal} {Phys. Rev. B}\ }\textbf {\bibinfo {volume} {101}},\ \bibinfo
  {pages} {235143} (\bibinfo {year} {2020})}\BibitemShut {NoStop}%
\bibitem [{\citenamefont {Derzhko}\ \emph {et~al.}(2020)\citenamefont
  {Derzhko}, \citenamefont {Schnack}, \citenamefont {Dmitriev}, \citenamefont
  {Krivnov},\ and\ \citenamefont {Richter}}]{der20}%
  \BibitemOpen
  \bibfield  {author} {\bibinfo {author} {\bibfnamefont {O.}~\bibnamefont
  {Derzhko}}, \bibinfo {author} {\bibfnamefont {J.}~\bibnamefont {Schnack}},
  \bibinfo {author} {\bibfnamefont {D.~V.}\ \bibnamefont {Dmitriev}}, \bibinfo
  {author} {\bibfnamefont {V.~Y.}\ \bibnamefont {Krivnov}},\ and\ \bibinfo
  {author} {\bibfnamefont {J.}~\bibnamefont {Richter}},\ }\bibfield  {title}
  {\emph {\bibinfo {title} {Flat-band physics in the spin-1/2 sawtooth
  chain}},\ }\href {https://doi.org/10.1140/epjb/e2020-10224-1} {\bibfield
  {journal} {\bibinfo  {journal} {Eur. Phys. J. B}\ }\textbf {\bibinfo {volume}
  {93}},\ \bibinfo {pages} {161} (\bibinfo {year} {2020})}\BibitemShut
  {NoStop}%
\bibitem [{\citenamefont {Richter}\ \emph {et~al.}(2020)\citenamefont
  {Richter}, \citenamefont {Schulenburg}, \citenamefont {Dmitriev},
  \citenamefont {Krivnov},\ and\ \citenamefont {Schnack}}]{CMP20}%
  \BibitemOpen
  \bibfield  {author} {\bibinfo {author} {\bibfnamefont {J.}~\bibnamefont
  {Richter}}, \bibinfo {author} {\bibfnamefont {J.}~\bibnamefont
  {Schulenburg}}, \bibinfo {author} {\bibfnamefont {D.~V.}\ \bibnamefont
  {Dmitriev}}, \bibinfo {author} {\bibfnamefont {V.~Y.}\ \bibnamefont
  {Krivnov}},\ and\ \bibinfo {author} {\bibfnamefont {J.}~\bibnamefont
  {Schnack}},\ }\bibfield  {title} {\emph {\bibinfo {title} {Anomalous
  thermodynamics of a quantum spin system with large residual entropy}},\
  }\href {https://doi.org/10.5488/CMP.23.43713} {\bibfield  {journal} {\bibinfo
   {journal} {Condens. Matter Phys.}\ }\textbf {\bibinfo {volume} {23}},\
  \bibinfo {pages} {43713} (\bibinfo {year} {2020})}\BibitemShut {NoStop}%
\bibitem [{\citenamefont {Acevedo}\ \emph {et~al.}(2020)\citenamefont
  {Acevedo}, \citenamefont {Pujol},\ and\ \citenamefont {Lamas}}]{ace20}%
  \BibitemOpen
  \bibfield  {author} {\bibinfo {author} {\bibfnamefont {S.}~\bibnamefont
  {Acevedo}}, \bibinfo {author} {\bibfnamefont {P.}~\bibnamefont {Pujol}},\
  and\ \bibinfo {author} {\bibfnamefont {C.~A.}\ \bibnamefont {Lamas}},\
  }\bibfield  {title} {\emph {\bibinfo {title} {Current jumps in flat-band
  ladders with {Dzyaloshinskii-Moriya} interactions}},\ }\href
  {https://doi.org/10.1103/PhysRevB.102.195139} {\bibfield  {journal} {\bibinfo
   {journal} {Phys. Rev. B}\ }\textbf {\bibinfo {volume} {102}},\ \bibinfo
  {pages} {195139} (\bibinfo {year} {2020})}\BibitemShut {NoStop}%
\bibitem [{\citenamefont {Richter}\ \emph {et~al.}(2022)\citenamefont
  {Richter}, \citenamefont {Ohanyan}, \citenamefont {Schulenburg},\ and\
  \citenamefont {Schnack}}]{SC_KNB}%
  \BibitemOpen
  \bibfield  {author} {\bibinfo {author} {\bibfnamefont {J.}~\bibnamefont
  {Richter}}, \bibinfo {author} {\bibfnamefont {V.}~\bibnamefont {Ohanyan}},
  \bibinfo {author} {\bibfnamefont {J.}~\bibnamefont {Schulenburg}},\ and\
  \bibinfo {author} {\bibfnamefont {J.}~\bibnamefont {Schnack}},\ }\bibfield
  {title} {\emph {\bibinfo {title} {Magnetoelectric properties of frustrated
  quantum spin systems}},\ }\href {https://doi.org/10.1103/PhysRevB.105.054420}
  {\bibfield  {journal} {\bibinfo  {journal} {Phys. Rev. B}\ }\textbf {\bibinfo
  {volume} {105}},\ \bibinfo {pages} {054420} (\bibinfo {year}
  {2022})}\BibitemShut {NoStop}%
\bibitem [{\citenamefont {Johannesmann}\ \emph {et~al.}(2023)\citenamefont
  {Johannesmann}, \citenamefont {Eckseler}, \citenamefont {Schlüter},\ and\
  \citenamefont {Schnack}}]{schnack23b}%
  \BibitemOpen
  \bibfield  {author} {\bibinfo {author} {\bibfnamefont {F.}~\bibnamefont
  {Johannesmann}}, \bibinfo {author} {\bibfnamefont {J.}~\bibnamefont
  {Eckseler}}, \bibinfo {author} {\bibfnamefont {H.}~\bibnamefont
  {Schlüter}},\ and\ \bibinfo {author} {\bibfnamefont {J.}~\bibnamefont
  {Schnack}},\ }\bibfield  {title} {\emph {\bibinfo {title} {Nonergodic
  one-magnon magnetization dynamics of the antiferromagnetic delta chain}},\
  }\href {https://doi.org/10.1103/PhysRevB.108.064304} {\bibfield  {journal}
  {\bibinfo  {journal} {Phys. Rev. B}\ }\textbf {\bibinfo {volume} {108}},\
  \bibinfo {pages} {064304} (\bibinfo {year} {2023})}\BibitemShut {NoStop}%
\bibitem [{\citenamefont {Reichert}\ \emph {et~al.}(2024)\citenamefont
  {Reichert}, \citenamefont {Schlüter}, \citenamefont {Heitmann},
  \citenamefont {Richter}, \citenamefont {Rausch},\ and\ \citenamefont
  {Schnack}}]{schnack24}%
  \BibitemOpen
  \bibfield  {author} {\bibinfo {author} {\bibfnamefont {N.}~\bibnamefont
  {Reichert}}, \bibinfo {author} {\bibfnamefont {H.}~\bibnamefont {Schlüter}},
  \bibinfo {author} {\bibfnamefont {T.}~\bibnamefont {Heitmann}}, \bibinfo
  {author} {\bibfnamefont {J.}~\bibnamefont {Richter}}, \bibinfo {author}
  {\bibfnamefont {R.}~\bibnamefont {Rausch}},\ and\ \bibinfo {author}
  {\bibfnamefont {J.}~\bibnamefont {Schnack}},\ }\bibfield  {title} {\emph
  {\bibinfo {title} {Magneto- and barocaloric properties of the
  ferro-antiferromagnetic sawtooth chain}},\ }\href
  {https://doi.org/10.1515/zna-2023-0267} {\bibfield  {journal} {\bibinfo
  {journal} {Z. Naturforsch. A}\ }\textbf {\bibinfo {volume} {79}},\ \bibinfo
  {pages} {283} (\bibinfo {year} {2024})}\BibitemShut {NoStop}%
\bibitem [{\citenamefont {Eckseler}\ and\ \citenamefont
  {Schnack}(2025)}]{schnack25}%
  \BibitemOpen
  \bibfield  {author} {\bibinfo {author} {\bibfnamefont {J.}~\bibnamefont
  {Eckseler}}\ and\ \bibinfo {author} {\bibfnamefont {J.}~\bibnamefont
  {Schnack}},\ }\bibfield  {title} {\emph {\bibinfo {title} {Permanent
  oscillations and solitary wave behavior in flatband {Heisenberg} quantum spin
  systems}},\ }\href {https://doi.org/10.1103/PhysRevResearch.7.013178}
  {\bibfield  {journal} {\bibinfo  {journal} {Phys. Rev. Res.}\ }\textbf
  {\bibinfo {volume} {7}},\ \bibinfo {pages} {013178} (\bibinfo {year}
  {2025})}\BibitemShut {NoStop}%
\bibitem [{\citenamefont {Zhitomirsky}\ and\ \citenamefont
  {Tsunetsugu}(2005)}]{zhi05}%
  \BibitemOpen
  \bibfield  {author} {\bibinfo {author} {\bibfnamefont {M.~E.}\ \bibnamefont
  {Zhitomirsky}}\ and\ \bibinfo {author} {\bibfnamefont {H.}~\bibnamefont
  {Tsunetsugu}},\ }\bibfield  {title} {\emph {\bibinfo {title} {Magnon
  localization and low-temperature properties of frustrated spin systems}},\
  }\href {https://doi.org/10.1143/PTPS.160.361} {\bibfield  {journal} {\bibinfo
   {journal} {Prog. Theor. Phys. Suppl.}\ }\textbf {\bibinfo {volume} {160}},\
  \bibinfo {pages} {361} (\bibinfo {year} {2005})}\BibitemShut {NoStop}%
\bibitem [{\citenamefont {Schnack}\ \emph {et~al.}(2020)\citenamefont
  {Schnack}, \citenamefont {Schulenburg}, \citenamefont {Honecker},\ and\
  \citenamefont {Richter}}]{prl2020}%
  \BibitemOpen
  \bibfield  {author} {\bibinfo {author} {\bibfnamefont {J.}~\bibnamefont
  {Schnack}}, \bibinfo {author} {\bibfnamefont {J.}~\bibnamefont
  {Schulenburg}}, \bibinfo {author} {\bibfnamefont {A.}~\bibnamefont
  {Honecker}},\ and\ \bibinfo {author} {\bibfnamefont {J.}~\bibnamefont
  {Richter}},\ }\bibfield  {title} {\emph {\bibinfo {title} {Magnon
  crystallization in the kagome lattice antiferromagnet}},\ }\href
  {https://doi.org/10.1103/PhysRevLett.125.117207} {\bibfield  {journal}
  {\bibinfo  {journal} {Phys. Rev. Lett.}\ }\textbf {\bibinfo {volume} {125}},\
  \bibinfo {pages} {117207} (\bibinfo {year} {2020})}\BibitemShut {NoStop}%
\bibitem [{\citenamefont {Richter}\ \emph
  {et~al.}(2004{\natexlab{b}})\citenamefont {Richter}, \citenamefont
  {Derzhko},\ and\ \citenamefont {Schulenburg}}]{spin-peierls}%
  \BibitemOpen
  \bibfield  {author} {\bibinfo {author} {\bibfnamefont {J.}~\bibnamefont
  {Richter}}, \bibinfo {author} {\bibfnamefont {O.}~\bibnamefont {Derzhko}},\
  and\ \bibinfo {author} {\bibfnamefont {J.}~\bibnamefont {Schulenburg}},\
  }\bibfield  {title} {\emph {\bibinfo {title} {Magnetic-field-induced
  {spin-Peierls} instability in strongly frustrated quantum spin lattices}},\
  }\href {https://doi.org/10.1103/PhysRevLett.93.107206} {\bibfield  {journal}
  {\bibinfo  {journal} {Phys. Rev. Lett.}\ }\textbf {\bibinfo {volume} {93}},\
  \bibinfo {pages} {107206} (\bibinfo {year} {2004}{\natexlab{b}})}\BibitemShut
  {NoStop}%
\bibitem [{\citenamefont {Zhitomirsky}\ and\ \citenamefont
  {Honecker}(2004)}]{zhi04h}%
  \BibitemOpen
  \bibfield  {author} {\bibinfo {author} {\bibfnamefont {M.~E.}\ \bibnamefont
  {Zhitomirsky}}\ and\ \bibinfo {author} {\bibfnamefont {A.}~\bibnamefont
  {Honecker}},\ }\bibfield  {title} {\emph {\bibinfo {title} {Magnetocaloric
  effect in one-dimensional antiferromagnets}},\ }\bibfield  {journal}
  {\bibinfo  {journal} {J. Stat. Mech.}\ }\textbf {\bibinfo {volume}
  {P07012}},\ \href {https://doi.org/10.1088/1742-5468/2004/07/P07012}
  {10.1088/1742-5468/2004/07/P07012} (\bibinfo {year} {2004})\BibitemShut
  {NoStop}%
\bibitem [{\citenamefont {Schmidt}\ \emph {et~al.}(2006)\citenamefont
  {Schmidt}, \citenamefont {Richter},\ and\ \citenamefont {Moessner}}]{sch06}%
  \BibitemOpen
  \bibfield  {author} {\bibinfo {author} {\bibfnamefont {H.-J.}\ \bibnamefont
  {Schmidt}}, \bibinfo {author} {\bibfnamefont {J.}~\bibnamefont {Richter}},\
  and\ \bibinfo {author} {\bibfnamefont {R.}~\bibnamefont {Moessner}},\
  }\bibfield  {title} {\emph {\bibinfo {title} {Excitations in strongly
  frustrated spin lattices}},\ }\href
  {https://doi.org/10.1088/0305-4470/39/34/S31} {\bibfield  {journal} {\bibinfo
   {journal} {J. Phys. A: Math. Gen.}\ }\textbf {\bibinfo {volume} {39}},\
  \bibinfo {pages} {10673} (\bibinfo {year} {2006})}\BibitemShut {NoStop}%
\bibitem [{\citenamefont {Heinze}\ \emph {et~al.}(2021)\citenamefont {Heinze},
  \citenamefont {Jeschke}, \citenamefont {Mazin}, \citenamefont
  {Metavitsiadis}, \citenamefont {Reehuis}, \citenamefont {Feyerherm},
  \citenamefont {Hoffmann}, \citenamefont {Bartkowiak}, \citenamefont
  {Prokhnenko}, \citenamefont {Wolter}, \citenamefont {Ding}, \citenamefont
  {Zapf}, \citenamefont {Corvalán~Moya}, \citenamefont {Weickert},
  \citenamefont {Jaime}, \citenamefont {Rule}, \citenamefont {Menzel},
  \citenamefont {Valentí}, \citenamefont {Brenig},\ and\ \citenamefont
  {Süllow}}]{ata21}%
  \BibitemOpen
  \bibfield  {author} {\bibinfo {author} {\bibfnamefont {L.}~\bibnamefont
  {Heinze}}, \bibinfo {author} {\bibfnamefont {H.~O.}\ \bibnamefont {Jeschke}},
  \bibinfo {author} {\bibfnamefont {I.~I.}\ \bibnamefont {Mazin}}, \bibinfo
  {author} {\bibfnamefont {A.}~\bibnamefont {Metavitsiadis}}, \bibinfo {author}
  {\bibfnamefont {M.}~\bibnamefont {Reehuis}}, \bibinfo {author} {\bibfnamefont
  {R.}~\bibnamefont {Feyerherm}}, \bibinfo {author} {\bibfnamefont {J.-U.}\
  \bibnamefont {Hoffmann}}, \bibinfo {author} {\bibfnamefont {M.}~\bibnamefont
  {Bartkowiak}}, \bibinfo {author} {\bibfnamefont {O.}~\bibnamefont
  {Prokhnenko}}, \bibinfo {author} {\bibfnamefont {A.~U.~B.}\ \bibnamefont
  {Wolter}}, \bibinfo {author} {\bibfnamefont {X.}~\bibnamefont {Ding}},
  \bibinfo {author} {\bibfnamefont {V.~S.}\ \bibnamefont {Zapf}}, \bibinfo
  {author} {\bibfnamefont {C.}~\bibnamefont {Corvalán~Moya}}, \bibinfo
  {author} {\bibfnamefont {F.}~\bibnamefont {Weickert}}, \bibinfo {author}
  {\bibfnamefont {M.}~\bibnamefont {Jaime}}, \bibinfo {author} {\bibfnamefont
  {K.~C.}\ \bibnamefont {Rule}}, \bibinfo {author} {\bibfnamefont
  {D.}~\bibnamefont {Menzel}}, \bibinfo {author} {\bibfnamefont
  {R.}~\bibnamefont {Valentí}}, \bibinfo {author} {\bibfnamefont
  {W.}~\bibnamefont {Brenig}},\ and\ \bibinfo {author} {\bibfnamefont
  {S.}~\bibnamefont {Süllow}},\ }\bibfield  {title} {\emph {\bibinfo {title}
  {Magnetization process of atacamite: {A} case of weakly coupled {$S$ $=$
  $1/2$} sawtooth chains}},\ }\href
  {https://doi.org/10.1103/PhysRevLett.126.207201} {\bibfield  {journal}
  {\bibinfo  {journal} {Phys. Rev. Lett.}\ }\textbf {\bibinfo {volume} {126}},\
  \bibinfo {pages} {207201} (\bibinfo {year} {2021})}\BibitemShut {NoStop}%
\bibitem [{\citenamefont {Heinze}\ \emph {et~al.}(2025)\citenamefont {Heinze},
  \citenamefont {Kotte}, \citenamefont {Rausch}, \citenamefont {Demuer},
  \citenamefont {Luther}, \citenamefont {Feyerherm}, \citenamefont {Ammerlaan},
  \citenamefont {Zeitler}, \citenamefont {Gorbunov}, \citenamefont {Uhlarz},
  \citenamefont {Rule}, \citenamefont {Wolter}, \citenamefont {Kühne},
  \citenamefont {Wosnitza}, \citenamefont {Karrasch},\ and\ \citenamefont
  {Süllow}}]{ata25}%
  \BibitemOpen
  \bibfield  {author} {\bibinfo {author} {\bibfnamefont {L.}~\bibnamefont
  {Heinze}}, \bibinfo {author} {\bibfnamefont {T.}~\bibnamefont {Kotte}},
  \bibinfo {author} {\bibfnamefont {R.}~\bibnamefont {Rausch}}, \bibinfo
  {author} {\bibfnamefont {A.}~\bibnamefont {Demuer}}, \bibinfo {author}
  {\bibfnamefont {S.}~\bibnamefont {Luther}}, \bibinfo {author} {\bibfnamefont
  {R.}~\bibnamefont {Feyerherm}}, \bibinfo {author} {\bibfnamefont {E.~L.
  Q.~N.}\ \bibnamefont {Ammerlaan}}, \bibinfo {author} {\bibfnamefont
  {U.}~\bibnamefont {Zeitler}}, \bibinfo {author} {\bibfnamefont {D.~I.}\
  \bibnamefont {Gorbunov}}, \bibinfo {author} {\bibfnamefont {M.}~\bibnamefont
  {Uhlarz}}, \bibinfo {author} {\bibfnamefont {K.~C.}\ \bibnamefont {Rule}},
  \bibinfo {author} {\bibfnamefont {A.~U.~B.}\ \bibnamefont {Wolter}}, \bibinfo
  {author} {\bibfnamefont {H.}~\bibnamefont {Kühne}}, \bibinfo {author}
  {\bibfnamefont {J.}~\bibnamefont {Wosnitza}}, \bibinfo {author}
  {\bibfnamefont {C.}~\bibnamefont {Karrasch}},\ and\ \bibinfo {author}
  {\bibfnamefont {S.}~\bibnamefont {Süllow}},\ }\bibfield  {title} {\emph
  {\bibinfo {title} {Atacamite {Cu$_2$Cl(OH)$_3$} in high magnetic fields:
  {Quantum} criticality and dimensional reduction of a sawtooth-chain
  compound}},\ }\href {https://doi.org/10.1103/PhysRevLett.134.216701}
  {\bibfield  {journal} {\bibinfo  {journal} {Phys. Rev. Lett.}\ }\textbf
  {\bibinfo {volume} {134}},\ \bibinfo {pages} {216701} (\bibinfo {year}
  {2025})}\BibitemShut {NoStop}%
\bibitem [{\citenamefont {Allen}\ \emph {et~al.}(2025)\citenamefont {Allen},
  \citenamefont {Heinze}, \citenamefont {Mole}, \citenamefont {Süllow},
  \citenamefont {Janson}, \citenamefont {Nishimoto}, \citenamefont {Lewis},\
  and\ \citenamefont {Rule}}]{ata25b}%
  \BibitemOpen
  \bibfield  {author} {\bibinfo {author} {\bibfnamefont {J.~L.}\ \bibnamefont
  {Allen}}, \bibinfo {author} {\bibfnamefont {L.}~\bibnamefont {Heinze}},
  \bibinfo {author} {\bibfnamefont {R.~A.}\ \bibnamefont {Mole}}, \bibinfo
  {author} {\bibfnamefont {S.}~\bibnamefont {Süllow}}, \bibinfo {author}
  {\bibfnamefont {O.}~\bibnamefont {Janson}}, \bibinfo {author} {\bibfnamefont
  {S.}~\bibnamefont {Nishimoto}}, \bibinfo {author} {\bibfnamefont {R.~A.}\
  \bibnamefont {Lewis}},\ and\ \bibinfo {author} {\bibfnamefont {K.~C.}\
  \bibnamefont {Rule}},\ }\href {https://doi.org/10.48550/arXiv.2508.02201}
  {\bibinfo {title} {The magnetic ground state of atacamite {Cu$_2$Cl(OH)$_3$}:
  {The} crucial role of frustrated zigzag chains revealed by inelastic neutron
  scattering}} (\bibinfo {year} {2025}),\ \Eprint
  {https://arxiv.org/abs/2508.02201} {arXiv:2508.02201} \BibitemShut {NoStop}%
\bibitem [{\citenamefont {Kikuchi}\ \emph {et~al.}(2011)\citenamefont
  {Kikuchi}, \citenamefont {Fujii}, \citenamefont {Takahashi}, \citenamefont
  {Azuma}, \citenamefont {Shimakawa}, \citenamefont {Taniguchi}, \citenamefont
  {Matsuo},\ and\ \citenamefont {Kindo}}]{euch}%
  \BibitemOpen
  \bibfield  {author} {\bibinfo {author} {\bibfnamefont {H.}~\bibnamefont
  {Kikuchi}}, \bibinfo {author} {\bibfnamefont {Y.}~\bibnamefont {Fujii}},
  \bibinfo {author} {\bibfnamefont {D.}~\bibnamefont {Takahashi}}, \bibinfo
  {author} {\bibfnamefont {M.}~\bibnamefont {Azuma}}, \bibinfo {author}
  {\bibfnamefont {Y.}~\bibnamefont {Shimakawa}}, \bibinfo {author}
  {\bibfnamefont {T.}~\bibnamefont {Taniguchi}}, \bibinfo {author}
  {\bibfnamefont {A.}~\bibnamefont {Matsuo}},\ and\ \bibinfo {author}
  {\bibfnamefont {K.}~\bibnamefont {Kindo}},\ }\bibfield  {title} {\emph
  {\bibinfo {title} {Experimental realization of a {$S$ $=$ $1/2$} sawtooth
  chain}},\ }\href {https://doi.org/10.1088/1742-6596/320/1/012045} {\bibfield
  {journal} {\bibinfo  {journal} {J. Phys.: Conf. Ser.}\ }\textbf {\bibinfo
  {volume} {320}},\ \bibinfo {pages} {012045} (\bibinfo {year}
  {2011})}\BibitemShut {NoStop}%
\bibitem [{\citenamefont {Garlea}\ \emph {et~al.}(2014)\citenamefont {Garlea},
  \citenamefont {Sanjeewa}, \citenamefont {McGuire}, \citenamefont {Kumar},
  \citenamefont {Sulejmanovic}, \citenamefont {He},\ and\ \citenamefont
  {Hwu}}]{Fe3+}%
  \BibitemOpen
  \bibfield  {author} {\bibinfo {author} {\bibfnamefont {V.~O.}\ \bibnamefont
  {Garlea}}, \bibinfo {author} {\bibfnamefont {L.~D.}\ \bibnamefont
  {Sanjeewa}}, \bibinfo {author} {\bibfnamefont {M.~A.}\ \bibnamefont
  {McGuire}}, \bibinfo {author} {\bibfnamefont {P.}~\bibnamefont {Kumar}},
  \bibinfo {author} {\bibfnamefont {D.}~\bibnamefont {Sulejmanovic}}, \bibinfo
  {author} {\bibfnamefont {J.}~\bibnamefont {He}},\ and\ \bibinfo {author}
  {\bibfnamefont {S.-J.}\ \bibnamefont {Hwu}},\ }\bibfield  {title} {\emph
  {\bibinfo {title} {Complex magnetic behavior of the sawtooth {Fe} chains in
  {Rb$_2$Fe$_2$O(AsO$_4$)$_2$}}},\ }\href
  {https://doi.org/10.1103/PhysRevB.89.014426} {\bibfield  {journal} {\bibinfo
  {journal} {Phys. Rev. B}\ }\textbf {\bibinfo {volume} {89}},\ \bibinfo
  {pages} {014426} (\bibinfo {year} {2014})}\BibitemShut {NoStop}%
\bibitem [{\citenamefont {Gnezdilov}\ \emph {et~al.}(2019)\citenamefont
  {Gnezdilov}, \citenamefont {Pashkevich}, \citenamefont {Kurnosov},
  \citenamefont {Zhuravlev}, \citenamefont {Wulferding}, \citenamefont
  {Lemmens}, \citenamefont {Menzel}, \citenamefont {Kozlyakova}, \citenamefont
  {Akhrorov}, \citenamefont {Kuznetsova}, \citenamefont {Berdonosov},
  \citenamefont {Dolgikh}, \citenamefont {Volkova},\ and\ \citenamefont
  {Vasiliev}}]{Fe3+2}%
  \BibitemOpen
  \bibfield  {author} {\bibinfo {author} {\bibfnamefont {V.~P.}\ \bibnamefont
  {Gnezdilov}}, \bibinfo {author} {\bibfnamefont {Y.~G.}\ \bibnamefont
  {Pashkevich}}, \bibinfo {author} {\bibfnamefont {V.~S.}\ \bibnamefont
  {Kurnosov}}, \bibinfo {author} {\bibfnamefont {O.~V.}\ \bibnamefont
  {Zhuravlev}}, \bibinfo {author} {\bibfnamefont {D.}~\bibnamefont
  {Wulferding}}, \bibinfo {author} {\bibfnamefont {P.}~\bibnamefont {Lemmens}},
  \bibinfo {author} {\bibfnamefont {D.}~\bibnamefont {Menzel}}, \bibinfo
  {author} {\bibfnamefont {E.~S.}\ \bibnamefont {Kozlyakova}}, \bibinfo
  {author} {\bibfnamefont {A.~Y.}\ \bibnamefont {Akhrorov}}, \bibinfo {author}
  {\bibfnamefont {E.~S.}\ \bibnamefont {Kuznetsova}}, \bibinfo {author}
  {\bibfnamefont {P.~S.}\ \bibnamefont {Berdonosov}}, \bibinfo {author}
  {\bibfnamefont {V.~A.}\ \bibnamefont {Dolgikh}}, \bibinfo {author}
  {\bibfnamefont {O.~S.}\ \bibnamefont {Volkova}},\ and\ \bibinfo {author}
  {\bibfnamefont {A.~N.}\ \bibnamefont {Vasiliev}},\ }\bibfield  {title} {\emph
  {\bibinfo {title} {Flat-band spin dynamics and phonon anomalies of the
  saw-tooth spin-chain system {Fe$_2$O(SeO$_3$)$_2$}}},\ }\href
  {https://doi.org/10.1103/PhysRevB.99.064413} {\bibfield  {journal} {\bibinfo
  {journal} {Phys. Rev. B}\ }\textbf {\bibinfo {volume} {99}},\ \bibinfo
  {pages} {064413} (\bibinfo {year} {2019})}\BibitemShut {NoStop}%
\bibitem [{\citenamefont {Nhalil}\ \emph {et~al.}(2019)\citenamefont {Nhalil},
  \citenamefont {Baral}, \citenamefont {Khamala}, \citenamefont {Cosio},
  \citenamefont {Singamaneni}, \citenamefont {Fitta}, \citenamefont {Antonio},
  \citenamefont {Gofryk}, \citenamefont {Zope}, \citenamefont {Baruah},
  \citenamefont {Saparov},\ and\ \citenamefont {Nair}}]{oliv1}%
  \BibitemOpen
  \bibfield  {author} {\bibinfo {author} {\bibfnamefont {H.}~\bibnamefont
  {Nhalil}}, \bibinfo {author} {\bibfnamefont {R.}~\bibnamefont {Baral}},
  \bibinfo {author} {\bibfnamefont {B.~O.}\ \bibnamefont {Khamala}}, \bibinfo
  {author} {\bibfnamefont {A.}~\bibnamefont {Cosio}}, \bibinfo {author}
  {\bibfnamefont {S.~R.}\ \bibnamefont {Singamaneni}}, \bibinfo {author}
  {\bibfnamefont {M.}~\bibnamefont {Fitta}}, \bibinfo {author} {\bibfnamefont
  {D.}~\bibnamefont {Antonio}}, \bibinfo {author} {\bibfnamefont
  {K.}~\bibnamefont {Gofryk}}, \bibinfo {author} {\bibfnamefont {R.~R.}\
  \bibnamefont {Zope}}, \bibinfo {author} {\bibfnamefont {T.}~\bibnamefont
  {Baruah}}, \bibinfo {author} {\bibfnamefont {B.}~\bibnamefont {Saparov}},\
  and\ \bibinfo {author} {\bibfnamefont {H.~S.}\ \bibnamefont {Nair}},\
  }\bibfield  {title} {\emph {\bibinfo {title} {Magnetic properties of a
  frustrated sawtooth compound}},\ }\href
  {https://doi.org/10.1103/PhysRevB.99.184434} {\bibfield  {journal} {\bibinfo
  {journal} {Phys. Rev. B}\ }\textbf {\bibinfo {volume} {99}},\ \bibinfo
  {pages} {184434} (\bibinfo {year} {2019})}\BibitemShut {NoStop}%
\bibitem [{\citenamefont {Pan}\ \emph {et~al.}(2023)\citenamefont {Pan},
  \citenamefont {Hu}, \citenamefont {Huang}, \citenamefont {Shi}, \citenamefont
  {Wang}, \citenamefont {Liu}, \citenamefont {Zhang}, \citenamefont {Xu},
  \citenamefont {Wang}, \citenamefont {Hao}, \citenamefont {Cheng},\ and\
  \citenamefont {Yao}}]{oliv2}%
  \BibitemOpen
  \bibfield  {author} {\bibinfo {author} {\bibfnamefont {F.}~\bibnamefont
  {Pan}}, \bibinfo {author} {\bibfnamefont {X.}~\bibnamefont {Hu}}, \bibinfo
  {author} {\bibfnamefont {J.}~\bibnamefont {Huang}}, \bibinfo {author}
  {\bibfnamefont {B.}~\bibnamefont {Shi}}, \bibinfo {author} {\bibfnamefont
  {J.}~\bibnamefont {Wang}}, \bibinfo {author} {\bibfnamefont {J.}~\bibnamefont
  {Liu}}, \bibinfo {author} {\bibfnamefont {H.}~\bibnamefont {Zhang}}, \bibinfo
  {author} {\bibfnamefont {D.}~\bibnamefont {Xu}}, \bibinfo {author}
  {\bibfnamefont {H.}~\bibnamefont {Wang}}, \bibinfo {author} {\bibfnamefont
  {L.}~\bibnamefont {Hao}}, \bibinfo {author} {\bibfnamefont {P.}~\bibnamefont
  {Cheng}},\ and\ \bibinfo {author} {\bibfnamefont {D.-X.}\ \bibnamefont
  {Yao}},\ }\bibfield  {title} {\emph {\bibinfo {title} {Magnetic properties of
  the spin-1/2 sawtooth chain with competing interactions}},\ }\href
  {https://doi.org/10.1103/PhysRevB.107.224423} {\bibfield  {journal} {\bibinfo
   {journal} {Phys. Rev. B}\ }\textbf {\bibinfo {volume} {107}},\ \bibinfo
  {pages} {224423} (\bibinfo {year} {2023})}\BibitemShut {NoStop}%
\bibitem [{\citenamefont {Mandujano}\ \emph
  {et~al.}(2023{\natexlab{a}})\citenamefont {Mandujano}, \citenamefont {Metta},
  \citenamefont {Barišić}, \citenamefont {Zhang}, \citenamefont {Tabiś},
  \citenamefont {Muniraju},\ and\ \citenamefont {Nair}}]{spir}%
  \BibitemOpen
  \bibfield  {author} {\bibinfo {author} {\bibfnamefont {H.~C.}\ \bibnamefont
  {Mandujano}}, \bibinfo {author} {\bibfnamefont {A.}~\bibnamefont {Metta}},
  \bibinfo {author} {\bibfnamefont {N.}~\bibnamefont {Barišić}}, \bibinfo
  {author} {\bibfnamefont {Q.}~\bibnamefont {Zhang}}, \bibinfo {author}
  {\bibfnamefont {W.}~\bibnamefont {Tabiś}}, \bibinfo {author} {\bibfnamefont
  {N.~K.~C.}\ \bibnamefont {Muniraju}},\ and\ \bibinfo {author} {\bibfnamefont
  {H.~S.}\ \bibnamefont {Nair}},\ }\bibfield  {title} {\emph {\bibinfo {title}
  {Magnetic structure and excitations in a frustrated sawtooth system}},\
  }\href {https://doi.org/10.1103/PhysRevMaterials.7.024422} {\bibfield
  {journal} {\bibinfo  {journal} {Phys. Rev. Materials}\ }\textbf {\bibinfo
  {volume} {7}},\ \bibinfo {pages} {024422} (\bibinfo {year}
  {2023}{\natexlab{a}})}\BibitemShut {NoStop}%
\bibitem [{\citenamefont {Oshima}\ \emph {et~al.}(2012)\citenamefont {Oshima},
  \citenamefont {Nojiri}, \citenamefont {Schnack}, \citenamefont {Kögerler},\
  and\ \citenamefont {Luban}}]{MoV}%
  \BibitemOpen
  \bibfield  {author} {\bibinfo {author} {\bibfnamefont {Y.}~\bibnamefont
  {Oshima}}, \bibinfo {author} {\bibfnamefont {H.}~\bibnamefont {Nojiri}},
  \bibinfo {author} {\bibfnamefont {J.}~\bibnamefont {Schnack}}, \bibinfo
  {author} {\bibfnamefont {P.}~\bibnamefont {Kögerler}},\ and\ \bibinfo
  {author} {\bibfnamefont {M.}~\bibnamefont {Luban}},\ }\bibfield  {title}
  {\emph {\bibinfo {title} {Magnetization plateaus in sawtooth spin systems}},\
  }\href {https://doi.org/10.1103/PhysRevB.85.024413} {\bibfield  {journal}
  {\bibinfo  {journal} {Phys. Rev. B}\ }\textbf {\bibinfo {volume} {85}},\
  \bibinfo {pages} {024413} (\bibinfo {year} {2012})}\BibitemShut {NoStop}%
\bibitem [{\citenamefont {Baniodeh}\ \emph {et~al.}(2018)\citenamefont
  {Baniodeh}, \citenamefont {Magnani}, \citenamefont {Lan}, \citenamefont
  {Buth}, \citenamefont {Anson}, \citenamefont {Richter}, \citenamefont
  {Affronte}, \citenamefont {Schnack},\ and\ \citenamefont {Powell}}]{FeGd}%
  \BibitemOpen
  \bibfield  {author} {\bibinfo {author} {\bibfnamefont {A.}~\bibnamefont
  {Baniodeh}}, \bibinfo {author} {\bibfnamefont {N.}~\bibnamefont {Magnani}},
  \bibinfo {author} {\bibfnamefont {Y.}~\bibnamefont {Lan}}, \bibinfo {author}
  {\bibfnamefont {G.}~\bibnamefont {Buth}}, \bibinfo {author} {\bibfnamefont
  {C.~E.}\ \bibnamefont {Anson}}, \bibinfo {author} {\bibfnamefont
  {J.}~\bibnamefont {Richter}}, \bibinfo {author} {\bibfnamefont
  {M.}~\bibnamefont {Affronte}}, \bibinfo {author} {\bibfnamefont
  {J.}~\bibnamefont {Schnack}},\ and\ \bibinfo {author} {\bibfnamefont {A.~K.}\
  \bibnamefont {Powell}},\ }\bibfield  {title} {\emph {\bibinfo {title} {High
  spin cycles: {Topping} the spin record for a single molecule verging on
  quantum criticality}},\ }\href {https://doi.org/10.1038/s41535-018-0082-7}
  {\bibfield  {journal} {\bibinfo  {journal} {npj Quantum Mater.}\ }\textbf
  {\bibinfo {volume} {3}},\ \bibinfo {pages} {10} (\bibinfo {year}
  {2018})}\BibitemShut {NoStop}%
\bibitem [{\citenamefont {Katsura}\ \emph {et~al.}(2005)\citenamefont
  {Katsura}, \citenamefont {Nagaosa},\ and\ \citenamefont {Balatsky}}]{KNB1}%
  \BibitemOpen
  \bibfield  {author} {\bibinfo {author} {\bibfnamefont {H.}~\bibnamefont
  {Katsura}}, \bibinfo {author} {\bibfnamefont {N.}~\bibnamefont {Nagaosa}},\
  and\ \bibinfo {author} {\bibfnamefont {A.~V.}\ \bibnamefont {Balatsky}},\
  }\bibfield  {title} {\emph {\bibinfo {title} {Spin current and
  magnetoelectric effect in noncollinear magnets}},\ }\href
  {https://doi.org/10.1103/PhysRevLett.95.057205} {\bibfield  {journal}
  {\bibinfo  {journal} {Phys. Rev. Lett.}\ }\textbf {\bibinfo {volume} {95}},\
  \bibinfo {pages} {057205} (\bibinfo {year} {2005})}\BibitemShut {NoStop}%
\bibitem [{\citenamefont {Jia}\ \emph {et~al.}(2006)\citenamefont {Jia},
  \citenamefont {Onoda}, \citenamefont {Nagaosa},\ and\ \citenamefont
  {Han}}]{KNB2}%
  \BibitemOpen
  \bibfield  {author} {\bibinfo {author} {\bibfnamefont {C.}~\bibnamefont
  {Jia}}, \bibinfo {author} {\bibfnamefont {S.}~\bibnamefont {Onoda}}, \bibinfo
  {author} {\bibfnamefont {N.}~\bibnamefont {Nagaosa}},\ and\ \bibinfo {author}
  {\bibfnamefont {J.~H.}\ \bibnamefont {Han}},\ }\bibfield  {title} {\emph
  {\bibinfo {title} {Theory of electric polarization in multiferroic
  magnets}},\ }\href {https://doi.org/10.1103/PhysRevB.74.224444} {\bibfield
  {journal} {\bibinfo  {journal} {Phys. Rev. B}\ }\textbf {\bibinfo {volume}
  {74}},\ \bibinfo {pages} {224444} (\bibinfo {year} {2006})}\BibitemShut
  {NoStop}%
\bibitem [{\citenamefont {Solovyev}\ \emph {et~al.}(2021)\citenamefont
  {Solovyev}, \citenamefont {Ono},\ and\ \citenamefont {Nikolaev}}]{Sol21}%
  \BibitemOpen
  \bibfield  {author} {\bibinfo {author} {\bibfnamefont {I.~V.}\ \bibnamefont
  {Solovyev}}, \bibinfo {author} {\bibfnamefont {R.}~\bibnamefont {Ono}},\ and\
  \bibinfo {author} {\bibfnamefont {S.}~\bibnamefont {Nikolaev}},\ }\bibfield
  {title} {\emph {\bibinfo {title} {Spin-current mechanism of multiferroicity
  in frustrated magnets}},\ }\href
  {https://doi.org/10.1103/PhysRevLett.127.187601} {\bibfield  {journal}
  {\bibinfo  {journal} {Phys. Rev. Lett.}\ }\textbf {\bibinfo {volume} {127}},\
  \bibinfo {pages} {187601} (\bibinfo {year} {2021})}\BibitemShut {NoStop}%
\bibitem [{\citenamefont {Solovyev}(2025)}]{Sol25}%
  \BibitemOpen
  \bibfield  {author} {\bibinfo {author} {\bibfnamefont {I.~V.}\ \bibnamefont
  {Solovyev}},\ }\bibfield  {title} {\emph {\bibinfo {title} {Microscopic
  theory of magnetic interactions in complex oxides}},\ }\href
  {https://doi.org/10.3390/condmat10010021} {\bibfield  {journal} {\bibinfo
  {journal} {Condens. Matter}\ }\textbf {\bibinfo {volume} {10}},\ \bibinfo
  {pages} {21} (\bibinfo {year} {2025})}\BibitemShut {NoStop}%
\bibitem [{\citenamefont {Khomskii}(2009)}]{khom}%
  \BibitemOpen
  \bibfield  {author} {\bibinfo {author} {\bibfnamefont {D.}~\bibnamefont
  {Khomskii}},\ }\bibfield  {title} {\emph {\bibinfo {title} {Classifying
  multiferroics: {Mechanisms} and effects}},\ }\href
  {https://doi.org/10.1103/Physics.2.20} {\bibfield  {journal} {\bibinfo
  {journal} {Physics}\ }\textbf {\bibinfo {volume} {2}},\ \bibinfo {pages} {20}
  (\bibinfo {year} {2009})}\BibitemShut {NoStop}%
\bibitem [{\citenamefont {Cheong}\ and\ \citenamefont {Mostovoy}(2007)}]{MEE1}%
  \BibitemOpen
  \bibfield  {author} {\bibinfo {author} {\bibfnamefont {S.-W.}\ \bibnamefont
  {Cheong}}\ and\ \bibinfo {author} {\bibfnamefont {M.}~\bibnamefont
  {Mostovoy}},\ }\bibfield  {title} {\emph {\bibinfo {title} {Multiferroics: a
  magnetic twist for ferroelectricity}},\ }\href
  {https://doi.org/10.1038/nmat1804} {\bibfield  {journal} {\bibinfo  {journal}
  {Nature Materials}\ }\textbf {\bibinfo {volume} {6}},\ \bibinfo {pages} {13}
  (\bibinfo {year} {2007})}\BibitemShut {NoStop}%
\bibitem [{\citenamefont {Eerenstein}\ \emph {et~al.}(2006)\citenamefont
  {Eerenstein}, \citenamefont {Mathur},\ and\ \citenamefont {Scott}}]{MEE2}%
  \BibitemOpen
  \bibfield  {author} {\bibinfo {author} {\bibfnamefont {W.}~\bibnamefont
  {Eerenstein}}, \bibinfo {author} {\bibfnamefont {N.~D.}\ \bibnamefont
  {Mathur}},\ and\ \bibinfo {author} {\bibfnamefont {J.~F.}\ \bibnamefont
  {Scott}},\ }\bibfield  {title} {\emph {\bibinfo {title} {Multiferroic and
  magnetoelectric materials}},\ }\href {https://doi.org/10.1038/nature05023}
  {\bibfield  {journal} {\bibinfo  {journal} {Nature}\ }\textbf {\bibinfo
  {volume} {442}},\ \bibinfo {pages} {760} (\bibinfo {year}
  {2006})}\BibitemShut {NoStop}%
\bibitem [{\citenamefont {Tokura}\ and\ \citenamefont {Seki}(2010)}]{Tok10}%
  \BibitemOpen
  \bibfield  {author} {\bibinfo {author} {\bibfnamefont {Y.}~\bibnamefont
  {Tokura}}\ and\ \bibinfo {author} {\bibfnamefont {S.}~\bibnamefont {Seki}},\
  }\bibfield  {title} {\emph {\bibinfo {title} {Multiferroics with spiral spin
  orders}},\ }\href {https://doi.org/10.1002/adma.200901961} {\bibfield
  {journal} {\bibinfo  {journal} {Adv. Mater.}\ }\textbf {\bibinfo {volume}
  {22}},\ \bibinfo {pages} {1554} (\bibinfo {year} {2010})}\BibitemShut
  {NoStop}%
\bibitem [{\citenamefont {Tokura}\ \emph {et~al.}(2014)\citenamefont {Tokura},
  \citenamefont {Seki},\ and\ \citenamefont {Nagaosa}}]{Tok14}%
  \BibitemOpen
  \bibfield  {author} {\bibinfo {author} {\bibfnamefont {Y.}~\bibnamefont
  {Tokura}}, \bibinfo {author} {\bibfnamefont {S.}~\bibnamefont {Seki}},\ and\
  \bibinfo {author} {\bibfnamefont {N.}~\bibnamefont {Nagaosa}},\ }\bibfield
  {title} {\emph {\bibinfo {title} {Multiferroics of spin origin}},\ }\href
  {https://doi.org/10.1088/0034-4885/77/7/076501} {\bibfield  {journal}
  {\bibinfo  {journal} {Rep. Prog. Phys.}\ }\textbf {\bibinfo {volume} {77}},\
  \bibinfo {pages} {076501} (\bibinfo {year} {2014})}\BibitemShut {NoStop}%
\bibitem [{\citenamefont {Brockmann}\ \emph {et~al.}(2013)\citenamefont
  {Brockmann}, \citenamefont {Klümper},\ and\ \citenamefont
  {Ohanyan}}]{bro13}%
  \BibitemOpen
  \bibfield  {author} {\bibinfo {author} {\bibfnamefont {M.}~\bibnamefont
  {Brockmann}}, \bibinfo {author} {\bibfnamefont {A.}~\bibnamefont
  {Klümper}},\ and\ \bibinfo {author} {\bibfnamefont {V.}~\bibnamefont
  {Ohanyan}},\ }\bibfield  {title} {\emph {\bibinfo {title} {Ground-state phase
  diagram of a spin-$1/2$ {Heisenberg} sawtooth chain}},\ }\href
  {https://doi.org/10.1103/PhysRevB.87.054407} {\bibfield  {journal} {\bibinfo
  {journal} {Phys. Rev. B}\ }\textbf {\bibinfo {volume} {87}},\ \bibinfo
  {pages} {054407} (\bibinfo {year} {2013})}\BibitemShut {NoStop}%
\bibitem [{\citenamefont {Baran}\ \emph {et~al.}(2018)\citenamefont {Baran},
  \citenamefont {Ohanyan},\ and\ \citenamefont {Verkholyak}}]{baran18}%
  \BibitemOpen
  \bibfield  {author} {\bibinfo {author} {\bibfnamefont {O.}~\bibnamefont
  {Baran}}, \bibinfo {author} {\bibfnamefont {V.}~\bibnamefont {Ohanyan}},\
  and\ \bibinfo {author} {\bibfnamefont {T.}~\bibnamefont {Verkholyak}},\
  }\bibfield  {title} {\emph {\bibinfo {title} {Magnetocaloric effect in a
  spin-$1/2$ {Heisenberg} sawtooth chain}},\ }\href
  {https://doi.org/10.1103/PhysRevB.98.064415} {\bibfield  {journal} {\bibinfo
  {journal} {Phys. Rev. B}\ }\textbf {\bibinfo {volume} {98}},\ \bibinfo
  {pages} {064415} (\bibinfo {year} {2018})}\BibitemShut {NoStop}%
\bibitem [{\citenamefont {Thakur}\ and\ \citenamefont
  {Durganandini}(2018)}]{thakur18}%
  \BibitemOpen
  \bibfield  {author} {\bibinfo {author} {\bibfnamefont {P.}~\bibnamefont
  {Thakur}}\ and\ \bibinfo {author} {\bibfnamefont {P.}~\bibnamefont
  {Durganandini}},\ }\bibfield  {title} {\emph {\bibinfo {title} {Thermodynamic
  properties of the spin-$1/2$ {Heisenberg} sawtooth chain}},\ }\href
  {https://doi.org/10.1103/PhysRevB.97.064413} {\bibfield  {journal} {\bibinfo
  {journal} {Phys. Rev. B}\ }\textbf {\bibinfo {volume} {97}},\ \bibinfo
  {pages} {064413} (\bibinfo {year} {2018})}\BibitemShut {NoStop}%
\bibitem [{\citenamefont {Yi}\ \emph {et~al.}(2019)\citenamefont {Yi},
  \citenamefont {You}, \citenamefont {Wu},\ and\ \citenamefont {Oleś}}]{XYZ}%
  \BibitemOpen
  \bibfield  {author} {\bibinfo {author} {\bibfnamefont {T.-C.}\ \bibnamefont
  {Yi}}, \bibinfo {author} {\bibfnamefont {W.-L.}\ \bibnamefont {You}},
  \bibinfo {author} {\bibfnamefont {N.}~\bibnamefont {Wu}},\ and\ \bibinfo
  {author} {\bibfnamefont {A.~M.}\ \bibnamefont {Oleś}},\ }\bibfield  {title}
  {\emph {\bibinfo {title} {Competing interactions in a spin-$1/2$ {Heisenberg}
  sawtooth chain}},\ }\href {https://doi.org/10.1103/PhysRevB.100.024423}
  {\bibfield  {journal} {\bibinfo  {journal} {Phys. Rev. B}\ }\textbf {\bibinfo
  {volume} {100}},\ \bibinfo {pages} {024423} (\bibinfo {year}
  {2019})}\BibitemShut {NoStop}%
\bibitem [{\citenamefont {Ohanyan}(2020)}]{oha20}%
  \BibitemOpen
  \bibfield  {author} {\bibinfo {author} {\bibfnamefont {V.}~\bibnamefont
  {Ohanyan}},\ }\bibfield  {title} {\emph {\bibinfo {title} {Magnetization and
  phase transitions of the sawtooth chain}},\ }\href
  {https://doi.org/10.5488/CMP.23.43704} {\bibfield  {journal} {\bibinfo
  {journal} {Condens. Matter Phys.}\ }\textbf {\bibinfo {volume} {23}},\
  \bibinfo {pages} {43704} (\bibinfo {year} {2020})}\BibitemShut {NoStop}%
\bibitem [{\citenamefont {Menchyshyn}\ \emph {et~al.}(2015)\citenamefont
  {Menchyshyn}, \citenamefont {Ohanyan}, \citenamefont {Verkholyak},
  \citenamefont {Krokhmalskii},\ and\ \citenamefont {Derzhko}}]{mench15}%
  \BibitemOpen
  \bibfield  {author} {\bibinfo {author} {\bibfnamefont {O.}~\bibnamefont
  {Menchyshyn}}, \bibinfo {author} {\bibfnamefont {V.}~\bibnamefont {Ohanyan}},
  \bibinfo {author} {\bibfnamefont {T.}~\bibnamefont {Verkholyak}}, \bibinfo
  {author} {\bibfnamefont {T.}~\bibnamefont {Krokhmalskii}},\ and\ \bibinfo
  {author} {\bibfnamefont {O.}~\bibnamefont {Derzhko}},\ }\bibfield  {title}
  {\emph {\bibinfo {title} {Thermodynamics of a spin-$1/2$ {Heisenberg}
  sawtooth chain}},\ }\href {https://doi.org/10.1103/PhysRevB.92.184427}
  {\bibfield  {journal} {\bibinfo  {journal} {Phys. Rev. B}\ }\textbf {\bibinfo
  {volume} {92}},\ \bibinfo {pages} {184427} (\bibinfo {year}
  {2015})}\BibitemShut {NoStop}%
\bibitem [{\citenamefont {Sznajd}(2018)}]{sznajd18}%
  \BibitemOpen
  \bibfield  {author} {\bibinfo {author} {\bibfnamefont {J.}~\bibnamefont
  {Sznajd}},\ }\bibfield  {title} {\emph {\bibinfo {title} {Magnetic properties
  of frustrated {Heisenberg} chains}},\ }\href
  {https://doi.org/10.1103/PhysRevB.97.214410} {\bibfield  {journal} {\bibinfo
  {journal} {Phys. Rev. B}\ }\textbf {\bibinfo {volume} {97}},\ \bibinfo
  {pages} {214410} (\bibinfo {year} {2018})}\BibitemShut {NoStop}%
\bibitem [{\citenamefont {Sznajd}(2019)}]{sznajd19}%
  \BibitemOpen
  \bibfield  {author} {\bibinfo {author} {\bibfnamefont {J.}~\bibnamefont
  {Sznajd}},\ }\bibfield  {title} {\emph {\bibinfo {title} {Magnetization
  process in frustrated {Heisenberg} chains}},\ }\href
  {https://doi.org/10.1016/j.jmmm.2019.02.030} {\bibfield  {journal} {\bibinfo
  {journal} {J. Magn. Magn. Mater.}\ }\textbf {\bibinfo {volume} {479}},\
  \bibinfo {pages} {254} (\bibinfo {year} {2019})}\BibitemShut {NoStop}%
\bibitem [{\citenamefont {You}\ \emph {et~al.}(2014)\citenamefont {You},
  \citenamefont {Liu}, \citenamefont {Horsch},\ and\ \citenamefont
  {Oleś}}]{oles}%
  \BibitemOpen
  \bibfield  {author} {\bibinfo {author} {\bibfnamefont {W.-L.}\ \bibnamefont
  {You}}, \bibinfo {author} {\bibfnamefont {G.-H.}\ \bibnamefont {Liu}},
  \bibinfo {author} {\bibfnamefont {P.}~\bibnamefont {Horsch}},\ and\ \bibinfo
  {author} {\bibfnamefont {A.~M.}\ \bibnamefont {Oleś}},\ }\bibfield  {title}
  {\emph {\bibinfo {title} {Phase diagram of the spin-$1/2$ {Heisenberg}
  sawtooth chain with competing interactions}},\ }\href
  {https://doi.org/10.1103/PhysRevB.90.094413} {\bibfield  {journal} {\bibinfo
  {journal} {Phys. Rev. B}\ }\textbf {\bibinfo {volume} {90}},\ \bibinfo
  {pages} {094413} (\bibinfo {year} {2014})}\BibitemShut {NoStop}%
\bibitem [{\citenamefont {Strečka}\ \emph {et~al.}(2020)\citenamefont
  {Strečka}, \citenamefont {Gálisová},\ and\ \citenamefont
  {Verkholyak}}]{stre20}%
  \BibitemOpen
  \bibfield  {author} {\bibinfo {author} {\bibfnamefont {J.}~\bibnamefont
  {Strečka}}, \bibinfo {author} {\bibfnamefont {L.}~\bibnamefont
  {Gálisová}},\ and\ \bibinfo {author} {\bibfnamefont {T.}~\bibnamefont
  {Verkholyak}},\ }\bibfield  {title} {\emph {\bibinfo {title} {Exactly
  solvable model of a mixed-spin {Ising}–{Heisenberg} sawtooth chain}},\
  }\href {https://doi.org/10.1103/PhysRevE.101.012103} {\bibfield  {journal}
  {\bibinfo  {journal} {Phys. Rev. E}\ }\textbf {\bibinfo {volume} {101}},\
  \bibinfo {pages} {012103} (\bibinfo {year} {2020})}\BibitemShut {NoStop}%
\bibitem [{\citenamefont {Cabra}\ \emph {et~al.}(2019)\citenamefont {Cabra},
  \citenamefont {Dobry}, \citenamefont {Gazza},\ and\ \citenamefont
  {Rossini}}]{cabra19}%
  \BibitemOpen
  \bibfield  {author} {\bibinfo {author} {\bibfnamefont {D.~C.}\ \bibnamefont
  {Cabra}}, \bibinfo {author} {\bibfnamefont {A.~O.}\ \bibnamefont {Dobry}},
  \bibinfo {author} {\bibfnamefont {C.~J.}\ \bibnamefont {Gazza}},\ and\
  \bibinfo {author} {\bibfnamefont {G.~L.}\ \bibnamefont {Rossini}},\
  }\bibfield  {title} {\emph {\bibinfo {title} {Magnetization plateaus in
  frustrated {Heisenberg} chains}},\ }\href
  {https://doi.org/10.1103/PhysRevB.100.161111} {\bibfield  {journal} {\bibinfo
   {journal} {Phys. Rev. B}\ }\textbf {\bibinfo {volume} {100}},\ \bibinfo
  {pages} {161111(R)} (\bibinfo {year} {2019})}\BibitemShut {NoStop}%
\bibitem [{\citenamefont {Baran}(2021)}]{bar21}%
  \BibitemOpen
  \bibfield  {author} {\bibinfo {author} {\bibfnamefont {O.}~\bibnamefont
  {Baran}},\ }\bibfield  {title} {\emph {\bibinfo {title} {Energy flux effect
  in the one-dimensional spin-1/2 xx model of magnetoelectric. lagrange
  multiplier method}},\ }\href
  {https://ujp.bitp.kiev.ua/index.php/ujp/article/view/2021032} {\bibfield
  {journal} {\bibinfo  {journal} {Ukr. J. Phys.}\ }\textbf {\bibinfo {volume}
  {66}},\ \bibinfo {pages} {890} (\bibinfo {year} {2021})}\BibitemShut
  {NoStop}%
\bibitem [{\citenamefont {Dmitriev}\ and\ \citenamefont
  {Krivnov}(2016{\natexlab{b}})}]{J1J2J3}%
  \BibitemOpen
  \bibfield  {author} {\bibinfo {author} {\bibfnamefont {D.~V.}\ \bibnamefont
  {Dmitriev}}\ and\ \bibinfo {author} {\bibfnamefont {V.~Y.}\ \bibnamefont
  {Krivnov}},\ }\bibfield  {title} {\emph {\bibinfo {title} {Thermodynamics of
  a frustrated spin-$1/2$ {Heisenberg} chain with competing interactions}},\
  }\href {https://doi.org/10.1088/0953-8984/30/38/385803} {\bibfield  {journal}
  {\bibinfo  {journal} {J. Phys.: Condens. Matter}\ }\textbf {\bibinfo {volume}
  {30}},\ \bibinfo {pages} {385803} (\bibinfo {year}
  {2016}{\natexlab{b}})}\BibitemShut {NoStop}%
\bibitem [{\citenamefont {Weisstein}(2025{\natexlab{a}})}]{AlGem}%
  \BibitemOpen
  \bibfield  {author} {\bibinfo {author} {\bibfnamefont {E.~W.}\ \bibnamefont
  {Weisstein}},\ }\href {https://mathworld.wolfram.com/QuarticCurve.html}
  {\bibinfo {title} {Quartic curve}},\ \bibinfo {howpublished} {From
  MathWorld--A Wolfram Resource} (\bibinfo {year}
  {2025}{\natexlab{a}})\BibitemShut {NoStop}%
\bibitem [{\citenamefont {Weisstein}(2025{\natexlab{b}})}]{cruc}%
  \BibitemOpen
  \bibfield  {author} {\bibinfo {author} {\bibfnamefont {E.~W.}\ \bibnamefont
  {Weisstein}},\ }\href {https://mathworld.wolfram.com/Cruciform.html}
  {\bibinfo {title} {Cruciform}},\ \bibinfo {howpublished} {From MathWorld--A
  Wolfram Resource} (\bibinfo {year} {2025}{\natexlab{b}})\BibitemShut
  {NoStop}%
\bibitem [{\citenamefont {Bergholtz}\ and\ \citenamefont {Liu}(2013)}]{ber13}%
  \BibitemOpen
  \bibfield  {author} {\bibinfo {author} {\bibfnamefont {E.~J.}\ \bibnamefont
  {Bergholtz}}\ and\ \bibinfo {author} {\bibfnamefont {Z.}~\bibnamefont
  {Liu}},\ }\bibfield  {title} {\emph {\bibinfo {title} {Topological flat band
  models and fractional {Chern} insulators}},\ }\href
  {https://doi.org/10.1142/S021797921330017X} {\bibfield  {journal} {\bibinfo
  {journal} {Int. J. Mod. Phys. B}\ }\textbf {\bibinfo {volume} {27}},\
  \bibinfo {pages} {1330017} (\bibinfo {year} {2013})}\BibitemShut {NoStop}%
\bibitem [{\citenamefont {Parameswaran}\ \emph {et~al.}(2013)\citenamefont
  {Parameswaran}, \citenamefont {Roy},\ and\ \citenamefont {Sondhi}}]{par13}%
  \BibitemOpen
  \bibfield  {author} {\bibinfo {author} {\bibfnamefont {S.~A.}\ \bibnamefont
  {Parameswaran}}, \bibinfo {author} {\bibfnamefont {R.}~\bibnamefont {Roy}},\
  and\ \bibinfo {author} {\bibfnamefont {S.~L.}\ \bibnamefont {Sondhi}},\
  }\bibfield  {title} {\emph {\bibinfo {title} {Fractional quantum {Hall}
  physics in topological flat bands}},\ }\href
  {https://doi.org/10.1016/j.crhy.2013.04.003} {\bibfield  {journal} {\bibinfo
  {journal} {Comptes Rendus Phys.}\ }\textbf {\bibinfo {volume} {14}},\
  \bibinfo {pages} {816} (\bibinfo {year} {2013})}\BibitemShut {NoStop}%
\bibitem [{\citenamefont {Leykam}\ \emph {et~al.}(2013)\citenamefont {Leykam},
  \citenamefont {Flach}, \citenamefont {Bahat-Treidel},\ and\ \citenamefont
  {Desyatnikov}}]{ley13}%
  \BibitemOpen
  \bibfield  {author} {\bibinfo {author} {\bibfnamefont {D.}~\bibnamefont
  {Leykam}}, \bibinfo {author} {\bibfnamefont {S.}~\bibnamefont {Flach}},
  \bibinfo {author} {\bibfnamefont {O.}~\bibnamefont {Bahat-Treidel}},\ and\
  \bibinfo {author} {\bibfnamefont {A.~S.}\ \bibnamefont {Desyatnikov}},\
  }\bibfield  {title} {\emph {\bibinfo {title} {Flat band states: {Disorder}
  and nonlinearity}},\ }\href {https://doi.org/10.1103/PhysRevB.88.224203}
  {\bibfield  {journal} {\bibinfo  {journal} {Phys. Rev. B}\ }\textbf {\bibinfo
  {volume} {88}},\ \bibinfo {pages} {224203} (\bibinfo {year}
  {2013})}\BibitemShut {NoStop}%
\bibitem [{\citenamefont {Leykam}\ \emph {et~al.}(2018)\citenamefont {Leykam},
  \citenamefont {Andreanov},\ and\ \citenamefont {Flach}}]{ley18}%
  \BibitemOpen
  \bibfield  {author} {\bibinfo {author} {\bibfnamefont {D.}~\bibnamefont
  {Leykam}}, \bibinfo {author} {\bibfnamefont {A.}~\bibnamefont {Andreanov}},\
  and\ \bibinfo {author} {\bibfnamefont {S.}~\bibnamefont {Flach}},\ }\bibfield
   {title} {\emph {\bibinfo {title} {Artificial flat band systems: {From}
  lattice models to experiments}},\ }\href
  {https://doi.org/10.1080/23746149.2018.1473052} {\bibfield  {journal}
  {\bibinfo  {journal} {Adv. Phys.: X}\ }\textbf {\bibinfo {volume} {3}},\
  \bibinfo {pages} {1473052} (\bibinfo {year} {2018})}\BibitemShut {NoStop}%
\bibitem [{\citenamefont {Bae}\ \emph {et~al.}(2023)\citenamefont {Bae},
  \citenamefont {Sedrakyan},\ and\ \citenamefont {Maltl}}]{bae23}%
  \BibitemOpen
  \bibfield  {author} {\bibinfo {author} {\bibfnamefont {J.-H.}\ \bibnamefont
  {Bae}}, \bibinfo {author} {\bibfnamefont {T.}~\bibnamefont {Sedrakyan}},\
  and\ \bibinfo {author} {\bibfnamefont {S.}~\bibnamefont {Maltl}},\ }\bibfield
   {title} {\emph {\bibinfo {title} {Flat bands and correlated phases in
  quantum spin systems}},\ }\href
  {https://doi.org/10.21468/SciPostPhys.15.4.139} {\bibfield  {journal}
  {\bibinfo  {journal} {SciPost Phys.}\ }\textbf {\bibinfo {volume} {15}},\
  \bibinfo {pages} {139} (\bibinfo {year} {2023})}\BibitemShut {NoStop}%
\bibitem [{\citenamefont {Lee}\ \emph {et~al.}(2024)\citenamefont {Lee},
  \citenamefont {Andreanov}, \citenamefont {Sedrakyan},\ and\ \citenamefont
  {Flach}}]{lee24}%
  \BibitemOpen
  \bibfield  {author} {\bibinfo {author} {\bibfnamefont {S.}~\bibnamefont
  {Lee}}, \bibinfo {author} {\bibfnamefont {A.}~\bibnamefont {Andreanov}},
  \bibinfo {author} {\bibfnamefont {T.}~\bibnamefont {Sedrakyan}},\ and\
  \bibinfo {author} {\bibfnamefont {S.}~\bibnamefont {Flach}},\ }\bibfield
  {title} {\emph {\bibinfo {title} {Trapping hard-core bosons in flat-band
  lattices}},\ }\href {https://doi.org/10.1103/PhysRevB.109.245137} {\bibfield
  {journal} {\bibinfo  {journal} {Phys. Rev. B}\ }\textbf {\bibinfo {volume}
  {109}},\ \bibinfo {pages} {245137} (\bibinfo {year} {2024})}\BibitemShut
  {NoStop}%
\bibitem [{\citenamefont {Mallick}\ and\ \citenamefont
  {Andreanov}(2025)}]{mall25}%
  \BibitemOpen
  \bibfield  {author} {\bibinfo {author} {\bibfnamefont {A.}~\bibnamefont
  {Mallick}}\ and\ \bibinfo {author} {\bibfnamefont {A.}~\bibnamefont
  {Andreanov}},\ }\bibfield  {title} {\emph {\bibinfo {title} {Flat bands in
  tight-binding lattices with anisotropic potentials}},\ }\href
  {https://doi.org/10.1103/PhysRevB.111.014201} {\bibfield  {journal} {\bibinfo
   {journal} {Phys. Rev. B}\ }\textbf {\bibinfo {volume} {111}},\ \bibinfo
  {pages} {014201} (\bibinfo {year} {2025})}\BibitemShut {NoStop}%
\bibitem [{\citenamefont {Sutherland}(1986)}]{suth86}%
  \BibitemOpen
  \bibfield  {author} {\bibinfo {author} {\bibfnamefont {B.}~\bibnamefont
  {Sutherland}},\ }\bibfield  {title} {\emph {\bibinfo {title} {Localization of
  electronic wave functions due to local topology}},\ }\href
  {https://doi.org/10.1103/PhysRevB.34.5208} {\bibfield  {journal} {\bibinfo
  {journal} {Phys. Rev. B}\ }\textbf {\bibinfo {volume} {34}},\ \bibinfo
  {pages} {5208} (\bibinfo {year} {1986})}\BibitemShut {NoStop}%
\bibitem [{\citenamefont {Mielke}(1991)}]{mielke}%
  \BibitemOpen
  \bibfield  {author} {\bibinfo {author} {\bibfnamefont {A.}~\bibnamefont
  {Mielke}},\ }\bibfield  {title} {\emph {\bibinfo {title} {Ferromagnetism in
  the {Hubbard} model on line graphs and further considerations}},\ }\href
  {https://doi.org/10.1088/0305-4470/24/2/011} {\bibfield  {journal} {\bibinfo
  {journal} {J. Phys. A: Math. Gen.}\ }\textbf {\bibinfo {volume} {24}},\
  \bibinfo {pages} {L73} (\bibinfo {year} {1991})}\BibitemShut {NoStop}%
\bibitem [{\citenamefont {Tasaki}(1992)}]{tas92}%
  \BibitemOpen
  \bibfield  {author} {\bibinfo {author} {\bibfnamefont {H.}~\bibnamefont
  {Tasaki}},\ }\bibfield  {title} {\emph {\bibinfo {title} {Ferromagnetism in
  the {Hubbard} models with degenerate single-electron ground states}},\ }\href
  {https://doi.org/10.1103/PhysRevLett.69.1608} {\bibfield  {journal} {\bibinfo
   {journal} {Phys. Rev. Lett.}\ }\textbf {\bibinfo {volume} {69}},\ \bibinfo
  {pages} {1608} (\bibinfo {year} {1992})}\BibitemShut {NoStop}%
\bibitem [{\citenamefont {Mielke}\ and\ \citenamefont
  {Tasaki}(1993)}]{mielke93}%
  \BibitemOpen
  \bibfield  {author} {\bibinfo {author} {\bibfnamefont {A.}~\bibnamefont
  {Mielke}}\ and\ \bibinfo {author} {\bibfnamefont {H.}~\bibnamefont
  {Tasaki}},\ }\bibfield  {title} {\emph {\bibinfo {title} {Ferromagnetism in
  the {Hubbard} model}},\ }\href {https://doi.org/10.1007/BF02108079}
  {\bibfield  {journal} {\bibinfo  {journal} {Commun. Math. Phys.}\ }\textbf
  {\bibinfo {volume} {158}},\ \bibinfo {pages} {341} (\bibinfo {year}
  {1993})}\BibitemShut {NoStop}%
\bibitem [{\citenamefont {Tasaki}(1998)}]{tas98}%
  \BibitemOpen
  \bibfield  {author} {\bibinfo {author} {\bibfnamefont {H.}~\bibnamefont
  {Tasaki}},\ }\bibfield  {title} {\emph {\bibinfo {title} {From {Nagaoka's}
  ferromagnetism to flat-band ferromagnetism and beyond: {An} introduction to
  ferromagnetism in the {Hubbard} model}},\ }\href
  {https://doi.org/10.1143/PTP.99.489} {\bibfield  {journal} {\bibinfo
  {journal} {Prog. Theor. Phys.}\ }\textbf {\bibinfo {volume} {99}},\ \bibinfo
  {pages} {489} (\bibinfo {year} {1998})}\BibitemShut {NoStop}%
\bibitem [{\citenamefont {Derzhko}\ \emph {et~al.}(2010)\citenamefont
  {Derzhko}, \citenamefont {Richter}, \citenamefont {Honecker}, \citenamefont
  {Maksymenko},\ and\ \citenamefont {Moessner}}]{Moes10}%
  \BibitemOpen
  \bibfield  {author} {\bibinfo {author} {\bibfnamefont {O.}~\bibnamefont
  {Derzhko}}, \bibinfo {author} {\bibfnamefont {J.}~\bibnamefont {Richter}},
  \bibinfo {author} {\bibfnamefont {A.}~\bibnamefont {Honecker}}, \bibinfo
  {author} {\bibfnamefont {M.}~\bibnamefont {Maksymenko}},\ and\ \bibinfo
  {author} {\bibfnamefont {R.}~\bibnamefont {Moessner}},\ }\bibfield  {title}
  {\emph {\bibinfo {title} {Low-temperature properties of the {Hubbard} model
  on highly frustrated one-dimensional lattices}},\ }\href
  {https://doi.org/10.1103/PhysRevB.81.014421} {\bibfield  {journal} {\bibinfo
  {journal} {Phys. Rev. B}\ }\textbf {\bibinfo {volume} {81}},\ \bibinfo
  {pages} {014421} (\bibinfo {year} {2010})}\BibitemShut {NoStop}%
\bibitem [{\citenamefont {Maksymenko}\ \emph {et~al.}(2012)\citenamefont
  {Maksymenko}, \citenamefont {Honecker}, \citenamefont {Moessner},
  \citenamefont {Richter},\ and\ \citenamefont {Derzhko}}]{Moes12}%
  \BibitemOpen
  \bibfield  {author} {\bibinfo {author} {\bibfnamefont {M.}~\bibnamefont
  {Maksymenko}}, \bibinfo {author} {\bibfnamefont {A.}~\bibnamefont
  {Honecker}}, \bibinfo {author} {\bibfnamefont {R.}~\bibnamefont {Moessner}},
  \bibinfo {author} {\bibfnamefont {J.}~\bibnamefont {Richter}},\ and\ \bibinfo
  {author} {\bibfnamefont {O.}~\bibnamefont {Derzhko}},\ }\bibfield  {title}
  {\emph {\bibinfo {title} {Flat-band ferromagnetism as a {Pauli-correlated}
  percolation problem}},\ }\href
  {https://doi.org/10.1103/PhysRevLett.109.096404} {\bibfield  {journal}
  {\bibinfo  {journal} {Phys. Rev. Lett.}\ }\textbf {\bibinfo {volume} {109}},\
  \bibinfo {pages} {096404} (\bibinfo {year} {2012})}\BibitemShut {NoStop}%
\bibitem [{\citenamefont {Leykam}\ and\ \citenamefont {Flach}(2018)}]{Ley18a}%
  \BibitemOpen
  \bibfield  {author} {\bibinfo {author} {\bibfnamefont {D.}~\bibnamefont
  {Leykam}}\ and\ \bibinfo {author} {\bibfnamefont {S.}~\bibnamefont {Flach}},\
  }\bibfield  {title} {\emph {\bibinfo {title} {Photonic flatbands}},\ }\href
  {https://doi.org/10.1063/1.5034365} {\bibfield  {journal} {\bibinfo
  {journal} {APL Photonics}\ }\textbf {\bibinfo {volume} {3}},\ \bibinfo
  {pages} {070901} (\bibinfo {year} {2018})}\BibitemShut {NoStop}%
\bibitem [{\citenamefont {Yang}\ \emph {et~al.}(2024)\citenamefont {Yang},
  \citenamefont {Li}, \citenamefont {Yang}, \citenamefont {Xie}, \citenamefont
  {Zhang}, \citenamefont {Yuan}, \citenamefont {Cai}, \citenamefont {Wang},\
  and\ \citenamefont {Gao}}]{nat24}%
  \BibitemOpen
  \bibfield  {author} {\bibinfo {author} {\bibfnamefont {J.}~\bibnamefont
  {Yang}}, \bibinfo {author} {\bibfnamefont {Y.}~\bibnamefont {Li}}, \bibinfo
  {author} {\bibfnamefont {Y.}~\bibnamefont {Yang}}, \bibinfo {author}
  {\bibfnamefont {X.}~\bibnamefont {Xie}}, \bibinfo {author} {\bibfnamefont
  {Z.}~\bibnamefont {Zhang}}, \bibinfo {author} {\bibfnamefont
  {J.}~\bibnamefont {Yuan}}, \bibinfo {author} {\bibfnamefont {H.}~\bibnamefont
  {Cai}}, \bibinfo {author} {\bibfnamefont {D.-W.}\ \bibnamefont {Wang}},\ and\
  \bibinfo {author} {\bibfnamefont {F.}~\bibnamefont {Gao}},\ }\bibfield
  {title} {\emph {\bibinfo {title} {Realization of all-band-flat photonic
  lattices}},\ }\href {https://doi.org/10.1038/s41467-024-45710-8} {\bibfield
  {journal} {\bibinfo  {journal} {Nat. Commun.}\ }\textbf {\bibinfo {volume}
  {15}},\ \bibinfo {pages} {1484} (\bibinfo {year} {2024})}\BibitemShut
  {NoStop}%
\bibitem [{\citenamefont {Eyvazi}\ \emph {et~al.}(2025)\citenamefont {Eyvazi},
  \citenamefont {Mamonov}, \citenamefont {Heilmann}, \citenamefont {Cuerda},\
  and\ \citenamefont {Törmä}}]{phot25}%
  \BibitemOpen
  \bibfield  {author} {\bibinfo {author} {\bibfnamefont {S.}~\bibnamefont
  {Eyvazi}}, \bibinfo {author} {\bibfnamefont {E.~A.}\ \bibnamefont {Mamonov}},
  \bibinfo {author} {\bibfnamefont {R.}~\bibnamefont {Heilmann}}, \bibinfo
  {author} {\bibfnamefont {J.}~\bibnamefont {Cuerda}},\ and\ \bibinfo {author}
  {\bibfnamefont {P.}~\bibnamefont {Törmä}},\ }\bibfield  {title} {\emph
  {\bibinfo {title} {Flat-band lasing in silicon waveguide-integrated
  metasurfaces}},\ }\href {https://doi.org/10.1021/acsphotonics.4c01902}
  {\bibfield  {journal} {\bibinfo  {journal} {ACS Photonics}\ }\textbf
  {\bibinfo {volume} {12}},\ \bibinfo {pages} {1570} (\bibinfo {year}
  {2025})}\BibitemShut {NoStop}%
\bibitem [{\citenamefont {Taie}\ \emph {et~al.}(2015)\citenamefont {Taie},
  \citenamefont {Ozawa}, \citenamefont {Ichinose}, \citenamefont {Nishio},
  \citenamefont {Nakajima},\ and\ \citenamefont {Takahashi}}]{Taie15}%
  \BibitemOpen
  \bibfield  {author} {\bibinfo {author} {\bibfnamefont {S.}~\bibnamefont
  {Taie}}, \bibinfo {author} {\bibfnamefont {H.}~\bibnamefont {Ozawa}},
  \bibinfo {author} {\bibfnamefont {T.}~\bibnamefont {Ichinose}}, \bibinfo
  {author} {\bibfnamefont {T.}~\bibnamefont {Nishio}}, \bibinfo {author}
  {\bibfnamefont {S.}~\bibnamefont {Nakajima}},\ and\ \bibinfo {author}
  {\bibfnamefont {Y.}~\bibnamefont {Takahashi}},\ }\bibfield  {title} {\emph
  {\bibinfo {title} {Coherent driving and freezing of bosonic matter wave in an
  optical {Lieb} lattice}},\ }\href {https://doi.org/10.1126/sciadv.1500854}
  {\bibfield  {journal} {\bibinfo  {journal} {Sci. Adv.}\ }\textbf {\bibinfo
  {volume} {1}},\ \bibinfo {pages} {e1500854} (\bibinfo {year}
  {2015})}\BibitemShut {NoStop}%
\bibitem [{\citenamefont {Ozawa}\ \emph {et~al.}(2017)\citenamefont {Ozawa},
  \citenamefont {Taie}, \citenamefont {Ichinose},\ and\ \citenamefont
  {Takahashi}}]{Ozawa17}%
  \BibitemOpen
  \bibfield  {author} {\bibinfo {author} {\bibfnamefont {H.}~\bibnamefont
  {Ozawa}}, \bibinfo {author} {\bibfnamefont {S.}~\bibnamefont {Taie}},
  \bibinfo {author} {\bibfnamefont {T.}~\bibnamefont {Ichinose}},\ and\
  \bibinfo {author} {\bibfnamefont {Y.}~\bibnamefont {Takahashi}},\ }\bibfield
  {title} {\emph {\bibinfo {title} {Interaction-driven shift and distortion of
  a flat band in an optical {Lieb} lattice}},\ }\href
  {https://doi.org/10.1103/PhysRevLett.118.175301} {\bibfield  {journal}
  {\bibinfo  {journal} {Phys. Rev. Lett.}\ }\textbf {\bibinfo {volume} {118}},\
  \bibinfo {pages} {175301} (\bibinfo {year} {2017})}\BibitemShut {NoStop}%
\bibitem [{\citenamefont {Taie}\ \emph {et~al.}(2020)\citenamefont {Taie},
  \citenamefont {Ichinose}, \citenamefont {Ozawa},\ and\ \citenamefont
  {Takahashi}}]{Taie20}%
  \BibitemOpen
  \bibfield  {author} {\bibinfo {author} {\bibfnamefont {S.}~\bibnamefont
  {Taie}}, \bibinfo {author} {\bibfnamefont {T.}~\bibnamefont {Ichinose}},
  \bibinfo {author} {\bibfnamefont {H.}~\bibnamefont {Ozawa}},\ and\ \bibinfo
  {author} {\bibfnamefont {Y.}~\bibnamefont {Takahashi}},\ }\bibfield  {title}
  {\emph {\bibinfo {title} {Spatial adiabatic passage of massive quantum
  particles in an optical {Lieb} lattice}},\ }\href
  {https://doi.org/10.1038/s41467-020-17645-9} {\bibfield  {journal} {\bibinfo
  {journal} {Nat. Commun.}\ }\textbf {\bibinfo {volume} {11}},\ \bibinfo
  {pages} {1} (\bibinfo {year} {2020})}\BibitemShut {NoStop}%
\bibitem [{\citenamefont {Jiang}\ \emph {et~al.}(2023)\citenamefont {Jiang},
  \citenamefont {Hsieh}, \citenamefont {Jones}, \citenamefont {Majchrzak},
  \citenamefont {Chakradhar}, \citenamefont {Watanabe}, \citenamefont
  {Taniguchi}, \citenamefont {Miwa}, \citenamefont {Chen},\ and\ \citenamefont
  {Ulstrup}}]{gr23}%
  \BibitemOpen
  \bibfield  {author} {\bibinfo {author} {\bibfnamefont {Z.}~\bibnamefont
  {Jiang}}, \bibinfo {author} {\bibfnamefont {K.}~\bibnamefont {Hsieh}},
  \bibinfo {author} {\bibfnamefont {A.~J.~H.}\ \bibnamefont {Jones}}, \bibinfo
  {author} {\bibfnamefont {P.}~\bibnamefont {Majchrzak}}, \bibinfo {author}
  {\bibfnamefont {S.}~\bibnamefont {Chakradhar}}, \bibinfo {author}
  {\bibfnamefont {K.}~\bibnamefont {Watanabe}}, \bibinfo {author}
  {\bibfnamefont {T.}~\bibnamefont {Taniguchi}}, \bibinfo {author}
  {\bibfnamefont {J.~A.}\ \bibnamefont {Miwa}}, \bibinfo {author}
  {\bibfnamefont {Y.~P.}\ \bibnamefont {Chen}},\ and\ \bibinfo {author}
  {\bibfnamefont {S.}~\bibnamefont {Ulstrup}},\ }\bibfield  {title} {\emph
  {\bibinfo {title} {Revealing flat bands and hybridization gaps in a twisted
  bilayer graphene device with microarpes}},\ }\href
  {https://doi.org/10.1088/2053-1583/acd8af} {\bibfield  {journal} {\bibinfo
  {journal} {2D Mater.}\ }\textbf {\bibinfo {volume} {10}},\ \bibinfo {pages}
  {045027} (\bibinfo {year} {2023})}\BibitemShut {NoStop}%
\bibitem [{\citenamefont {Wang}\ \emph {et~al.}(2024)\citenamefont {Wang},
  \citenamefont {Ma}, \citenamefont {Chen}, \citenamefont {Wu}, \citenamefont
  {Xu}, \citenamefont {Dai}, \citenamefont {Zhu}, \citenamefont {Ren},
  \citenamefont {Gao},\ and\ \citenamefont {Lin}}]{gr24}%
  \BibitemOpen
  \bibfield  {author} {\bibinfo {author} {\bibfnamefont {Z.-Y.}\ \bibnamefont
  {Wang}}, \bibinfo {author} {\bibfnamefont {J.-J.}\ \bibnamefont {Ma}},
  \bibinfo {author} {\bibfnamefont {Q.}~\bibnamefont {Chen}}, \bibinfo {author}
  {\bibfnamefont {K.}~\bibnamefont {Wu}}, \bibinfo {author} {\bibfnamefont
  {S.}~\bibnamefont {Xu}}, \bibinfo {author} {\bibfnamefont {Q.}~\bibnamefont
  {Dai}}, \bibinfo {author} {\bibfnamefont {Z.}~\bibnamefont {Zhu}}, \bibinfo
  {author} {\bibfnamefont {J.}~\bibnamefont {Ren}}, \bibinfo {author}
  {\bibfnamefont {H.-J.}\ \bibnamefont {Gao}},\ and\ \bibinfo {author}
  {\bibfnamefont {X.}~\bibnamefont {Lin}},\ }\bibfield  {title} {\emph
  {\bibinfo {title} {Visualizing localized nematic states in twisted double
  bilayer graphene}},\ }\href {https://doi.org/10.1039/D4NR02816A} {\bibfield
  {journal} {\bibinfo  {journal} {Nanoscale}\ }\textbf {\bibinfo {volume}
  {16}},\ \bibinfo {pages} {18852} (\bibinfo {year} {2024})}\BibitemShut
  {NoStop}%
\bibitem [{\citenamefont {Abanin}\ \emph {et~al.}(2019)\citenamefont {Abanin},
  \citenamefont {Altman}, \citenamefont {Bloch},\ and\ \citenamefont
  {Serbyn}}]{MBL}%
  \BibitemOpen
  \bibfield  {author} {\bibinfo {author} {\bibfnamefont {D.~A.}\ \bibnamefont
  {Abanin}}, \bibinfo {author} {\bibfnamefont {E.}~\bibnamefont {Altman}},
  \bibinfo {author} {\bibfnamefont {I.}~\bibnamefont {Bloch}},\ and\ \bibinfo
  {author} {\bibfnamefont {M.}~\bibnamefont {Serbyn}},\ }\bibfield  {title}
  {\emph {\bibinfo {title} {Colloquium: {Many}-body localization,
  thermalization, and entanglement}},\ }\href
  {https://doi.org/10.1103/RevModPhys.91.021001} {\bibfield  {journal}
  {\bibinfo  {journal} {Rev. Mod. Phys.}\ }\textbf {\bibinfo {volume} {91}},\
  \bibinfo {pages} {021001} (\bibinfo {year} {2019})}\BibitemShut {NoStop}%
\bibitem [{\citenamefont {Holstein}\ and\ \citenamefont
  {Primakoff}(1940)}]{HP}%
  \BibitemOpen
  \bibfield  {author} {\bibinfo {author} {\bibfnamefont {T.}~\bibnamefont
  {Holstein}}\ and\ \bibinfo {author} {\bibfnamefont {H.}~\bibnamefont
  {Primakoff}},\ }\bibfield  {title} {\emph {\bibinfo {title} {Field dependence
  of the intrinsic domain magnetization of a ferromagnet}},\ }\href
  {https://doi.org/10.1103/PhysRev.58.1098} {\bibfield  {journal} {\bibinfo
  {journal} {Phys. Rev.}\ }\textbf {\bibinfo {volume} {58}},\ \bibinfo {pages}
  {1098} (\bibinfo {year} {1940})}\BibitemShut {NoStop}%
\bibitem [{\citenamefont {Dyson}(1956{\natexlab{a}})}]{Dys1}%
  \BibitemOpen
  \bibfield  {author} {\bibinfo {author} {\bibfnamefont {F.~J.}\ \bibnamefont
  {Dyson}},\ }\bibfield  {title} {\emph {\bibinfo {title} {General theory of
  spin-wave interactions}},\ }\href {https://doi.org/10.1103/PhysRev.102.1217}
  {\bibfield  {journal} {\bibinfo  {journal} {Phys. Rev.}\ }\textbf {\bibinfo
  {volume} {102}},\ \bibinfo {pages} {1217} (\bibinfo {year}
  {1956}{\natexlab{a}})}\BibitemShut {NoStop}%
\bibitem [{\citenamefont {Dyson}(1956{\natexlab{b}})}]{Dys2}%
  \BibitemOpen
  \bibfield  {author} {\bibinfo {author} {\bibfnamefont {F.~J.}\ \bibnamefont
  {Dyson}},\ }\bibfield  {title} {\emph {\bibinfo {title} {Thermodynamic
  behavior of an ideal ferromagnet}},\ }\href
  {https://doi.org/10.1103/PhysRev.102.1230} {\bibfield  {journal} {\bibinfo
  {journal} {Phys. Rev.}\ }\textbf {\bibinfo {volume} {102}},\ \bibinfo {pages}
  {1230} (\bibinfo {year} {1956}{\natexlab{b}})}\BibitemShut {NoStop}%
\bibitem [{\citenamefont {Maleev}(1958)}]{Mal}%
  \BibitemOpen
  \bibfield  {author} {\bibinfo {author} {\bibfnamefont {S.~V.}\ \bibnamefont
  {Maleev}},\ }\bibfield  {title} {\emph {\bibinfo {title} {Scattering of slow
  neutrons by ferromagnets}},\ }\href
  {https://www.jetp.ras.ru/cgi-bin/e/index/e/6/4/p776?a=list} {\bibfield
  {journal} {\bibinfo  {journal} {Sov. Phys. JETP}\ }\textbf {\bibinfo {volume}
  {6}},\ \bibinfo {pages} {776} (\bibinfo {year} {1958})}\BibitemShut {NoStop}%
\bibitem [{\citenamefont {Gong}\ \emph {et~al.}(2015)\citenamefont {Gong},
  \citenamefont {Qian}, \citenamefont {Yan}, \citenamefont {Scarola},\ and\
  \citenamefont {Zhang}}]{DM_OL}%
  \BibitemOpen
  \bibfield  {author} {\bibinfo {author} {\bibfnamefont {M.}~\bibnamefont
  {Gong}}, \bibinfo {author} {\bibfnamefont {Y.}~\bibnamefont {Qian}}, \bibinfo
  {author} {\bibfnamefont {M.}~\bibnamefont {Yan}}, \bibinfo {author}
  {\bibfnamefont {V.~W.}\ \bibnamefont {Scarola}},\ and\ \bibinfo {author}
  {\bibfnamefont {C.}~\bibnamefont {Zhang}},\ }\bibfield  {title} {\emph
  {\bibinfo {title} {{Dzyaloshinskii-Moriya} interaction and spiral order in
  spin-orbit coupled optical lattices}},\ }\href
  {https://doi.org/10.1038/srep10050} {\bibfield  {journal} {\bibinfo
  {journal} {Sci. Rep.}\ }\textbf {\bibinfo {volume} {5}},\ \bibinfo {pages}
  {10050} (\bibinfo {year} {2015})}\BibitemShut {NoStop}%
\bibitem [{\citenamefont {Kunimi}\ \emph {et~al.}(2024)\citenamefont {Kunimi},
  \citenamefont {Tomita}, \citenamefont {Katsura},\ and\ \citenamefont
  {Kato}}]{Ryd}%
  \BibitemOpen
  \bibfield  {author} {\bibinfo {author} {\bibfnamefont {M.}~\bibnamefont
  {Kunimi}}, \bibinfo {author} {\bibfnamefont {T.}~\bibnamefont {Tomita}},
  \bibinfo {author} {\bibfnamefont {H.}~\bibnamefont {Katsura}},\ and\ \bibinfo
  {author} {\bibfnamefont {Y.}~\bibnamefont {Kato}},\ }\bibfield  {title}
  {\emph {\bibinfo {title} {Proposal for simulating quantum spin models with
  the {Dzyaloshinskii-Moriya} interaction using {Rydberg} atoms and the
  construction of asymptotic quantum many-body scar states}},\ }\href
  {https://doi.org/10.1103/PhysRevA.110.043312} {\bibfield  {journal} {\bibinfo
   {journal} {Phys. Rev. A}\ }\textbf {\bibinfo {volume} {110}},\ \bibinfo
  {pages} {043312} (\bibinfo {year} {2024})}\BibitemShut {NoStop}%
\bibitem [{\citenamefont {Mandujano}\ \emph
  {et~al.}(2023{\natexlab{b}})\citenamefont {Mandujano}, \citenamefont {Metta},
  \citenamefont {Barišić}, \citenamefont {Zhang}, \citenamefont {Tabiś},
  \citenamefont {Muniraju},\ and\ \citenamefont {Nair}}]{BeCr}%
  \BibitemOpen
  \bibfield  {author} {\bibinfo {author} {\bibfnamefont {H.~C.}\ \bibnamefont
  {Mandujano}}, \bibinfo {author} {\bibfnamefont {A.}~\bibnamefont {Metta}},
  \bibinfo {author} {\bibfnamefont {N.}~\bibnamefont {Barišić}}, \bibinfo
  {author} {\bibfnamefont {Q.}~\bibnamefont {Zhang}}, \bibinfo {author}
  {\bibfnamefont {W.}~\bibnamefont {Tabiś}}, \bibinfo {author} {\bibfnamefont
  {N.~K.~C.}\ \bibnamefont {Muniraju}},\ and\ \bibinfo {author} {\bibfnamefont
  {H.~S.}\ \bibnamefont {Nair}},\ }\bibfield  {title} {\emph {\bibinfo {title}
  {Magnetic structure and excitations in a frustrated sawtooth system}},\
  }\href {https://doi.org/10.1103/PhysRevMaterials.7.024422} {\bibfield
  {journal} {\bibinfo  {journal} {Phys. Rev. Materials}\ }\textbf {\bibinfo
  {volume} {7}},\ \bibinfo {pages} {024422} (\bibinfo {year}
  {2023}{\natexlab{b}})}\BibitemShut {NoStop}%
\bibitem [{\citenamefont {Kocsis}\ \emph {et~al.}(2021)\citenamefont {Kocsis},
  \citenamefont {Tokunaga}, \citenamefont {Tokura},\ and\ \citenamefont
  {Taguchi}}]{koc21}%
  \BibitemOpen
  \bibfield  {author} {\bibinfo {author} {\bibfnamefont {V.}~\bibnamefont
  {Kocsis}}, \bibinfo {author} {\bibfnamefont {Y.}~\bibnamefont {Tokunaga}},
  \bibinfo {author} {\bibfnamefont {Y.}~\bibnamefont {Tokura}},\ and\ \bibinfo
  {author} {\bibfnamefont {Y.}~\bibnamefont {Taguchi}},\ }\bibfield  {title}
  {\emph {\bibinfo {title} {Switching of antiferromagnetic states in
  {LiCoPO$_4$} as investigated via the magnetoelectric effect}},\ }\href
  {https://doi.org/10.1103/PhysRevB.104.054426} {\bibfield  {journal} {\bibinfo
   {journal} {Phys. Rev. B}\ }\textbf {\bibinfo {volume} {104}},\ \bibinfo
  {pages} {054426} (\bibinfo {year} {2021})}\BibitemShut {NoStop}%
\bibitem [{\citenamefont {Ukleev}\ \emph {et~al.}(2025)\citenamefont {Ukleev},
  \citenamefont {Magrez},\ and\ \citenamefont {White}}]{ukl25}%
  \BibitemOpen
  \bibfield  {author} {\bibinfo {author} {\bibfnamefont {V.}~\bibnamefont
  {Ukleev}}, \bibinfo {author} {\bibfnamefont {A.}~\bibnamefont {Magrez}},\
  and\ \bibinfo {author} {\bibfnamefont {J.~S.}\ \bibnamefont {White}},\
  }\bibfield  {title} {\emph {\bibinfo {title} {Magnetoelectric and magnetic
  properties of frustrated materials}},\ }\href
  {https://doi.org/10.1063/5.0201234} {\bibfield  {journal} {\bibinfo
  {journal} {J. Appl. Phys.}\ }\textbf {\bibinfo {volume} {138}},\ \bibinfo
  {pages} {103905} (\bibinfo {year} {2025})}\BibitemShut {NoStop}%
\end{thebibliography}

%

\end{document}